\documentclass[aps,prc,twocolumn,showpacs,superscriptaddress,longbibliography,floatfix,10pt]{revtex4-2}
\usepackage{CJK}
\usepackage{indentfirst}
\usepackage{bm}
\usepackage{dcolumn}
\usepackage{amssymb}
\usepackage{amsmath}
\usepackage{graphicx}
\usepackage{dcolumn}
\usepackage{bm}
\usepackage{xcolor}
\usepackage{fix-cm}
\usepackage{mathptmx}
\usepackage[T1]{fontenc}
\usepackage{epstopdf}
\usepackage{longtable}
\usepackage[colorlinks,citecolor=blue,linkcolor=blue,anchorcolor=blue,filecolor=blue,urlcolor=blue]{hyperref}
\usepackage{multirow}
\makeatletter
\def\NAT@def@citea{\def\@citea{\NAT@separator}}
\makeatother

\begin{document}	
	\hyphenpenalty=5000
	\tolerance=2000
	
	\begin{CJK*}{UTF8}{}
		\title{
			Accuracy of the mean-field theory in describing ground-state properties of light nuclei}
		
		\author{Yu-Ting Rong}
		\email[Corresponding author:~]{rongyuting@gxnu.edu.cn}
		\affiliation{Guangxi Key Laboratory of Nuclear Physics and Technology, Guangxi Normal University, Guilin, 541004, China}
	
		\date{\today}
		
		\begin{abstract}

The relativistic mean-field model, augmented with three types of center-of-mass corrections and two types of rotational corrections, is employed to investigate the ground-state properties of helium, beryllium, and carbon isotopes. 
The efficacy of the mean-field approach in describing the binding energies, quadrupole deformations, root-mean-square charge radii, root-mean-square matter radii, and neutron skins of these light nuclei is evaluated. 
By averaging the binding energies obtained from six selected effective interactions, a mass-dependent behavior of the mean-field approximation is elucidated. 
The findings from radii reveal that, unlike in heavy nuclei, the exchange terms of the center-of-mass correction play an indispensable role in accurately describing the radii of light nuclei.
The mean-field approximation, when augmented with center-of-mass and rotational corrections, effectively reproduces the energies and radii of light nuclei. However, due to the absence of many-body correlations between valence neutrons, the mean-field approximation falls short in describing the deformations and shell evolutions of the helium and beryllium isotopic chains.

		\end{abstract}
		
		
		\maketitle
	\end{CJK*}
	
\section{INTRODUCTION}\label{sec1}

Light nuclei have attracted considerable attention due to their unique properties. One of these properties is the cluster phenomenon~\cite{Freer2018_RMP90-035004,Bijker2020_PPNP110-103735,Bishop2019_PRC100-034320}, such as the presence of two-alpha cluster structure in $^8$Be, the Hoyle state in $^{12}$C~\cite{Epelbaum2012_PRL109-252501,DellAquila2017_PRL119-132501,Shen2021_EPJA57-276}, and the tetrahedral shape in~$^{16}$O~\cite{Robson1982_PRC25-1108,Bijker2014_PRL112-152501,Halcrow2020_JPCS1643-012136}. Additionally, with the development of rare-isotope beam facilities, the exploration of the drip line~\cite{Pfuetzner2012_RMP84-567,Ahn2019_PRL123-212501} associated with the isospin limitation and the halo phenomena in extremely neutron- or proton-rich nuclei~\cite{Tanihata2013_PPNP68-215} has become a hot topic. Another important aspect is the shell evolution, where both experimental and theoretical investigations have suggested the possible existence of the magic number 6 in some semi-magic unstable light nuclei. Notably, there have been proposals for subshell closures in $^8$He~\cite{Skaza2006_PRC73-044301,Otsuka2001_PRL87-082502} and $^{14}$C~\cite{Sorlin2008_PPNP61-602,Tran2018_NutureCommu9-1594}. Conversely, the traditional magic number 8 is significantly compromised in $^{12}$Be~\cite{Imai2009_PLB673-179,Krieger2012_PRL108-142501,Pain2006_PRL96-032502}.

Since a light nucleus consists of few nucleons, a real Hamiltonian
for it can be constructed with bare nucleon-nucleon interactions obtained from scattering. 
Together with the improvements in computer performance, {\textit{ab initio}} methods are being utilized to study exotic nuclear properties in light nuclei from first principle, and these methods are now being extended to heavier regions as well~\cite{Freer2018_RMP90-035004} .
While density functional theories (DFTs) ~\cite{Schunck2019_EDF-Nuclei} using universal effective interactions have been successful in describing many nuclear phenomena for nuclei with mass numbers $A\geq 16$, it is commonly thought that DFTs are not suitable for light nuclei. This is due to the mean-field approximation used in DFTs, which erases the few-body correlations between nucleons and poses significant difficulties for accurately describing properties such as binding energy, charge radius, neutron skin, and surface thickness in the light mass region.

Covariant density functional theories (CDFTs), which take Lorentz symmetry into account, provide microscopic frameworks for a global description of atomic nuclei ~\cite{Meng2006_PPNP57-470,Meng2016_RDFNS}. These approaches have been extended to study light nuclei ~\cite{Meng1996_PRL77-3963,Zhu1994_PLB328-1,Arumugam2005_PRC71-064308,Lu2011_PRC84-014328,Tang2013_CPL30-012101,Tang2013_CPL30-012101,Sun2018_PLB785-530,Yang2021_PRL126-082501,Wang2022_CTP74-015303}, with $^{11}$Li serving as a typical example \cite{Meng1996_PRL77-3963,Zhu1994_PLB328-1}. In Ref.~\cite{Meng1996_PRL77-3963}, pairing and continuum effects were considered, and both the binding energies and radii of the isotopic chain from $^6$Li to $^{11}$Li, as well as the halo structure of $^{11}$Li, were successfully reproduced. 
Additionally, the $\alpha$-clustering and halo structures in beryllium and boron nuclei, along with several prominent cluster structures in both the ground and intrinsic excited states of $\alpha$ nuclei from $^{12}$C to $^{32}$S, were well-described by relativistic mean-field (RMF) calculations~\cite{Arumugam2005_PRC71-064308}.
Recently, halo structures in $^{22}$C and $^{17}$B were also well described by the deformed relativistic Hartree-Bogoliubov model in continuum (DRHBc) model~\cite{Sun2018_PLB785-530,Yang2021_PRL126-082501}. The triangular shape in $^{12}$C was studied through parity and angular momentum projections based on the multidimensionally constrained relativistic Hartree-Bogoliubov (p-MDCRHB) model~\cite{Wang2022_CTP74-015303}. 
These works suggest that results for light nuclei calculated with CDFTs and their corrections appear to be reliable, at least for bulk properties such as binding energies, densities, and single-particle levels.  If this is indeed the case, comparing {\textit{ab initio}} calculations with CDFTs will help us understand the connections between the two methods and even build new-generation DFTs with realistic nuclear forces, such as the relativistic chiral nucleon-nucleon interactions~\cite{Lu2022_PRL128-142002,Ren2018_ChinPhysC42-14103}. 

Therefore, in this work, my aim is to systematically study the bulk properties of nuclei with proton number $Z<8$ using various CDFTs. The primary objective is to evaluate the ability of these approaches to accurately describe the ground-state properties of light nuclei. Specifically, I will investigate whether the effective interactions, which are constrained by the properties of heavy nuclei, can provide reliable predictions for the ground-state properties of light nuclei within the CDFT frameworks. Moreover, I will examine the ability of CDFTs to accurately describe the properties of $^4$He, which serves as the starting nucleus for $\alpha$ cluster structures. I will also investigate whether CDFTs can reproduce the observed shell evolutions in these isotopes.

To do this, I will investigate even-even nuclei in the light mass region, utilizing three distinct effective interactions: nonlinear meson exchange (NL-ME), density-dependent meson exchange (DD-ME), and density-dependent point coupling (DD-PC). The calculations are based on the multidimensional constrained relativistic Hartree-Bogoliubov (MDCRHB) model~~\cite{Zhao2017_PRC95-014320,Zhou2016_PS91-063008}, accounting for both center-of-mass and rotational corrections. For the first time, I compute the root-mean-square (rms) radii including full microscopic center-of-mass correction, with exchange terms previously deemed negligible in the $Z>8$ mass region~\cite{Long2004_PRC69-034319}.
In Sec.~\ref{sec2}, I provide a brief overview of the MDCRHB model and the corrections applied in this study. In Sec.~\ref{sec3}, I present the calculated binding energies, radii, deformations, and potential energy curves for helium, beryllium, and carbon isotopes, along with detailed discussions of the results. Finally, in Sec.~\ref{sec4}, I summarize my findings.

\section{THEORETICAL FRAMEWORK}\label{sec2}
\subsection{Mean-field description}

In the present work, the RHB theory is employed to provide a unified description of the relativistic mean-field and the pairing correlations via the Bogoliubov transformation~\cite{Ring1980}. The RHB equation reads
\begin{equation}
	\label{eq:rhb}
	\int d^{3}\bm{r}^{\prime}
	\left( \begin{array}{cc} h-\lambda  &  \Delta                      \\
		-\Delta^{*}   & -h+\lambda \end{array}
	\right)
	\left( \begin{array}{c} U_{k} \\ V_{k} \end{array} \right)
	= E_{k}
	\left( \begin{array}{c} U_{k} \\ V_{k} \end{array} \right),
\end{equation}
where $h$ is the single-particle Hamiltonian, $\Delta$ is the pairing field, $\lambda$ is the Fermi energy, $E_k$ is the quasiparticle energy, and $(U_k,V_k)^T$ is the wave function.
The single-particle Hamiltonian 
\begin{equation}
	h=\alpha\cdot p+\beta[m+S(r)]+V(r)+\Sigma_R(r),
\end{equation}
consists of the kinetic energy term, the scalar potential $S(r)$, the vector potential $V(r)$, and the rearrangement potential $\Sigma_R(r)$. $m$ denotes the mass of nucleon.
For meson-exchange interactions,
\begin{equation}
	\begin{aligned}
		&S(r)=g_\sigma \sigma, \\
		&V(r)=g_\omega \omega_0+g_\rho \rho_0 \cdot \tau_3+e\dfrac{1-\tau_3}{2}A_0, \\
		&\Sigma_R(r)=\dfrac{\partial g_\sigma}{\partial \rho_V} \rho_S \sigma + \dfrac{\partial g_\omega}{\partial \rho_V} \rho_V \omega_0 + \dfrac{\partial g_\rho}{\partial \rho_V} \rho_V\tau_3 \rho_0,
	\end{aligned}  
\end{equation} 
where $g_\sigma, g_\omega$, and $g_\rho$ are coupling constants of $\sigma, \omega_0$ and $\rho_0$ meson fields, $A_0$ is the time-like component of the Coulomb field, $e$ is the charge unit for protons, $\rho_S$ and $\rho_V$ are isoscalar and isovector densities, respectively. 
For point-coupling interactions,
\begin{equation}
	\begin{aligned}
		S(r)=&\alpha_S\rho_S+\alpha_{TS}\rho_{TS}\tau_3 + \beta_S	\rho_S^2 +\gamma_S\rho_S^3 \\
		&+\delta_S\Delta \rho_S +\delta_{TS}\Delta \rho_{TS}\tau_3, \\
		V(r)=&\alpha_V\rho_V+\alpha_{TV}\rho_V\tau_3 +\gamma_V\rho_V^3 \\
		&+\delta_V\Delta\rho_V+\delta_{TV}\Delta \rho_{TV}\tau_3 +e\dfrac{1-\tau_3}{2}A_0, \\
		\Sigma_R(r)=&\dfrac{1}{2} \dfrac{\partial \alpha_S}{\partial \rho_V} \rho_S^2 
		+ \dfrac{1}{2} \dfrac{\partial \alpha_V}{\partial \rho_V} \rho_V^2 +\dfrac{1}{2} \dfrac{\partial \alpha_{TV}}{\partial \rho_{V}} \rho_{TV}^2,
	\end{aligned}  
\end{equation}
where $\alpha_S,~\alpha_V,~\alpha_{TS}, ~\alpha_{TV}, ~\beta_S, ~\gamma_S, ~\gamma_V, ~\delta_S, ~\delta_V, ~\delta_{TS}$, and $\delta_{TV}$ are coupling constants for different channels, $\rho_{TS}$ and $\rho_{TV}$ are time-like components of isoscalar current
and time-like components of isovector current, respectively.

The pairing field reads 
\begin{equation}
	\begin{aligned}
		&\Delta(\bm{r}_{1}\sigma_{1},\bm{r}_{2}\sigma_{2}) 
		=  \int d^{3}\bm{r}_{1}^{\prime} d^{3}\bm{r}_{2}^{\prime} 
		\sum_{\sigma_{1}^{\prime}\sigma_{2}^{\prime}}  \\
		&	V(\bm{r}_{1}         \sigma_{1},          \bm{r}_{2}         \sigma_{2},
		\bm{r}_{1}^{\prime}\sigma_{1}^{\prime}, \bm{r}_{2}^{\prime}\sigma_{2}^{\prime}) 
		\kappa 
		(\bm{r}_{1}^{\prime}\sigma_{1}^{\prime}, 
		\bm{r}_{2}^{\prime}\sigma_{2}^{\prime}),
	\end{aligned}
\end{equation}
where 
$V$ is the effective pairing interaction and $\kappa$ is the pairing tensor. 
In this work,
I use a separable pairing force of finite
range with pairing strength $G = 728$ MeV$\cdot$fm$^3$ and effective range of the pairing force $a = 0.644$ fm~\cite{Tian2009_PLB676-44}. 

The deformations with $V_4$ symmetry are allowed in the MDCRHB model. The deformation parameter $\beta_{\lambda\mu}$ is determined by 
\begin{equation}
	\beta_{\lambda\mu}=\dfrac{4\pi}{3A R^\lambda}Q_{\lambda\mu},
\end{equation}
where $R$ is the radius of the nucleus, $A$ is the number of nucleons, and $Q_{\lambda\mu}$ is the intrinsic multipole moment. $Q_{\lambda\mu}$ is calculated from the vector density by
\begin{equation}
	Q_{\lambda\mu}=\int d^3r\rho_V(\bm{r}) r^\lambda Y_{\lambda\mu}(\Omega),
\end{equation}
where $Y_{\lambda\mu}$ is the spherical harmonic.

\subsection{Corrections}

I consider both the center-of-mass and rotational corrections for the calculated binding energies. The resulting binding energy is 
\begin{equation}
	E_{\rm B}=-E_{\rm MF}-E_{\rm c.m.}-E_{\rm rot.}.
\end{equation}
where $E_{\rm MF},~E_{\rm c.m.}$, and $E_{\rm rot.}$ denote the energies of mean field, center-of-mass correction and rotational correction, respectively.  
The center-of-mass energy can be evaluated
analytically from the harmonic oscillator (HO) states. Using the usual parameterisation of the oscillator constant from
the Nilsson model one obtains an estimate as~\cite{Bender2000_EPJA7-467}
\begin{equation}\label{eq:E_HO}
	E_{\rm c.m.}^{\rm HO}=-\dfrac{3}{4} \cdot 41 A^{-1/3}~\rm MeV,
\end{equation}
where the harmonic oscillator's energy $\hbar\omega=41 A^{-1/3}$ MeV is adopted.
The other way is the so call microscopic method, in which the center-of-mass correction can be given by calculating the change in binding energy from projection-after-variation in first-order approximation as follow
\begin{equation}\label{eq:E_mic}
	E_{\rm c.m.}^{\rm mic}=-\dfrac{\langle P_{\rm c.m.}^2\rangle }{2mA},
\end{equation}
with 
\begin{equation}\label{eq:cm-P-square}	
	\begin{aligned}
		\langle P_{\rm c.m.}^2\rangle=&
		\sum_i \upsilon_i^2 p_{ii}^2-\sum_{i,j}\upsilon_i^2\upsilon_j^2 p_{ij}p_{ij}^* \\
		& +\sum_{i,j}\upsilon_iu_i\upsilon_ju_j p_{ij}p_{\bar{i}\bar{j}},
	\end{aligned}
\end{equation}
where $i$ and $j$ denote the actual quasiparticle states. $\upsilon_i$ is the occupation probability. $p_{ii}^2$ is the expectation value of the square of the quasiparticle momentum operator, and $p_{ij}$ is the off-diagonal matrix element of the quasiparticle momentum operator. They are called direct and exchange terms later. The rotational energy correction is calculated by
\begin{equation}\label{eq:Erot}
	E_{\rm rot.}=-\dfrac{1}{2} \sum_{k=1}^3 \dfrac{\langle J_k^2\rangle}{I_k},
\end{equation}
where $k$ denotes the axis of rotation, $J_k$ denotes the component of the angular momentum
in the body-fixed frame of a nucleus. The 
moment of inertia $I_k$ is a linear combination of Inglis-Belyaev formula and the moment of inertia of rigid rotor, i.e., $I_k=0.8I_k^{\rm IB}+0.2I_k^{\rm rigid}$, with the Inglis-Belyaev formula~\cite{Inglis1956_PR103-1786,Beliaev1961_NP24-322}
\begin{equation}
	I_k^{\rm IB}=\sum_{i,j} \dfrac{(u_i\upsilon_j-\upsilon_i u_j)^2}{E_i+E_j} |\langle i|J_k|j\rangle|^2.
\end{equation}

Similar to the energy, the radius calculated from RHB model needs corrections, too. With center-of-mass correction, the square of the radius is estimated after RHB calculation by
\begin{equation}\label{eq:R-cm-corr}
	R^2=R_{\rm MF}^2-R_{\rm c.m.}^2.
\end{equation} 
where $R_{\rm MF}$ and $R$ are rms radius before and after center-of-mass correction, and $R_{\rm c.m.}$ is the rms radius of the center of mass.
I compare the radius calculated from HO approximation by
\begin{equation}\label{eq:Radius-corr-HO}
	(R_{\rm c.m.}^{\rm HO})^2=\dfrac{3\hbar^2}{2m A\cdot 41A^{-1/3}},
\end{equation}
and from mean-field expectation values by
\begin{equation}\label{eq:cm-R-square}
	\begin{aligned}
		(R_{\rm c.m.}^{\rm mic})^2=&
		\sum_i \upsilon_i^2(r^2)_{ii}-\sum_{i,j}\upsilon_i^2\upsilon_j^2 r_{ij}r_{ij}^* \\
		& +\sum_{i,j}\upsilon_i u_i\upsilon_j u_j r_{ij}r_{\bar{i}\bar{j}}.
	\end{aligned}
\end{equation}
Note that the second and third terms in Eq.~ (\ref{eq:cm-R-square}) are called the exchange terms, and usually omitted because of their fairly small effects in the mean-field models~\cite{Long2004_PRC69-034319}. I take these terms into account in this work to investigate the information given by them in light nuclei.
The above formulas are employed for both proton radius ($R_p$) and neutron radius ($R_n$).
The charge radius ($R_c$) is obtained from the proton radius by~\cite{Sugahara1994_NPA579-557}
\begin{equation}\label{eq:rc_from_rp}
	R_{c}^2=R_p^2+(0.862~{\rm fm})^2-(0.336~{\rm fm})^2 N/Z,
\end{equation}
in which the proton and neutron spin-orbit contributions to the charge radius~\cite{Horowitz2012_PRC86-045503,Kurasawa2019_PTEP2019-113D01} are neglected.


In the above calculations, the ground
states are obtained by applying the variational principle with a Bogoliubov vacuum, and the effects of fundamental translational
invariance and rotational symmetry are estimated approximately. However, wave functions and observables with certain symmetries cannot be achieved. 
The standard way to restore the broken
symmetries and calculate observables with good quantum numbers is through the projection-after-variation (PAV)
technique~\cite{Bender2003_RMP75-121,Niksic2011_PPNP66-519,Egido2016_PS91-073003,Robledo2018_JPG46-013001,Sheikh2021_JPG48-123001,Sun2021_SciBulletin66-2072}.
Recently, a projected multidimensionally-constrained relativistic Hartree-Bogoliubov (p-MDCRHB) model has been developed by incorporating the parity and angular momentum projections into the MDCRHB model. 
{\color{black}In this model, 
both the triaxial and octupole shapes are allowed. The wave function with a certain angular momentum $J$ and parity $\pi$ is obtained by
\begin{equation}
	|\Psi_{\alpha,q}^{JM\pi}\rangle=\sum_K f_\alpha^{JK\pi}\hat{P}^{J}_{MK}\hat{P}^\pi |\Phi(q)\rangle,
\end{equation}
where $K$ represents angular momentum projection onto $z$-axis in the intrinsic frame, $f_\alpha^{JK\pi}$ is the weight
function, and $q$ represents a collection of the deformation parameters. The operator $\hat{P}_{MK}^J$ projects out the component with angular momentum $J$ and its projection $M$ from the deformed mean-field wave function $|\Phi(q)\rangle$, and $\hat{P}^\pi$ is the parity projection operator.} 
To approximately restore the average proton and neutron numbers, two correction
terms are added to the Hamiltonian kernel ${\cal H}$ as in Ref.~\cite{Yao2010_PRC81-044311}. Finally, the weight function $f_\alpha^{JK\pi}$ and the eigenvalue $E_\alpha^{J\pi}$ are obtained by solving the generalized eigenvalue equation~\cite{Ring1980,Yao2009_PRC79-044312}
\begin{equation}
		\sum_{K'}\left\{{\cal H'}_{KK'}^{J\pi}(q;q)-E_{\alpha}^{J\pi} {\cal N}_{KK'}^{J\pi}(q;q)\right\}f_{\alpha}^{JK'\pi}=0,
\end{equation}
where ${\cal N}$ is the norm kernel, and ${\cal H'}$ is the Hamiltonian kernel with particle number correction.

This model has been used to study the low-lying states related to exotic nuclear shapes, such as the triangular shape associated with three-$\alpha$ configuration in $^{12}$C~\cite{Wang2022_CTP74-015303} and the octupole correlations in $^{96}$Zr~\cite{Rong2023_PLB840-137896}. {\color{black}In this work, I restrict the calculations to axial and reflection symmetry and perform angular momentum projection after variation to discuss the cluster structure and shell evolution in the light mass region. For simplicity, the configuration mixing associated with shape fluctuation is beyond the scope of this work.}

\section{RESULTS AND DISCUSSIONS}\label{sec3}

The DFTs are well-established for studying the properties of heavy nuclei, while they are thought to be difficult to describe light nuclei due to the deficiency of many-body correlations. 
However, as mentioned above, previous studies also found that proper treatments of the corrections based on the DFTs can help us reproduce the structures of light nuclei successfully.  
In this work, the ground-state properties are systematically calculated in the CDFT framework, with the center-of-mass and rotational corrections considered. 
The accuracy of the mean-field theory in describing the ground-state properties of light nuclei is investigated by comparing the calculated bulk properties with the corresponding experimental data and results from other models. 
To estimate the uncertainty from input parameter sets, I consider eight representative of them, which are classified into three types: (1) the nonlinear meson-exchange (NL-ME) interactions, including NLSH~\cite{Sharma1993_PLB312-377}, TM1~\cite{Sugahara1994_NPA579-557},
NL5(A)~\cite{Agbemava2019_PRC99-014318}, and PK1~\cite{Long2004_PRC69-034319}; (2) the density-dependent meson-exchange (DD-ME) interactions, including PKDD~\cite{Long2004_PRC69-034319}, DD-ME2~\cite{Lalazissis2005_PRC71-024312}, and DD-LZ1~\cite{Wei2020_ChinPhysC44-074107}; and (3) the density-dependent point-coupling (DD-PC) interaction DD-PC1~\cite{Niksic2008_PRC78-034318}. These interactions have been demonstrated to accurately describe the ground-state properties of heavy nuclei. 

\subsection{Binding Energies}

\begin{figure}[htbp]
	\centering
	\includegraphics[width=0.46\textwidth]{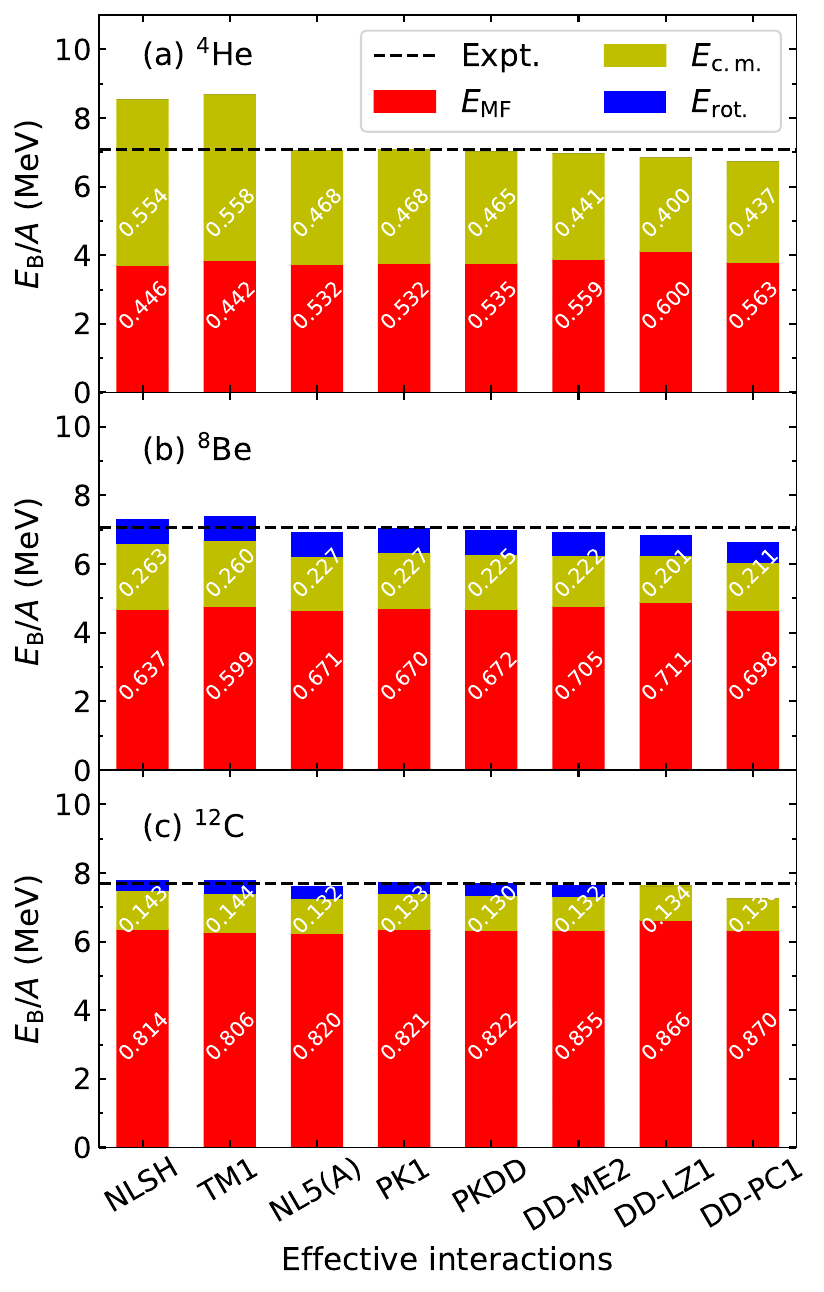}
	\caption{(Color online) Energies per nucleon $E_{\rm B}/A$ values for (a) $^4$He, (b) $^8$Be, and (c) $^{12}$C calculated by the MDCRHB model with eight selected parameter sets. The black dashed lines denote the experimental $E_{\rm B}/A$ values taken from AME2020~\cite{Wang2021_ChinPhysC45-030003}. The values in the columns are the ratios of the mean-field and center-of-mass energies to the total calculated energies, i.e., $-E_{\rm MF}/E_{\rm B}$ and $-E_{\rm c.m.}/E_{\rm B}$.}
	\label{fig:E-A-He-Be-C}
\end{figure}

I first test the validity of those parameter sets by calculating the binding energies of $^4$He, $^8$Be, and $^{12}$C using the eight parameter sets mentioned above. 
For each one, the corresponding center-of-mass energy correction $E_{\rm c.m.}$ is determined by the one used to fit the parameter set, i.e.,  Eq.~(\ref{eq:E_HO}) is adopted for
NLSH and TM1, but Eq.~(\ref{eq:E_mic}) is adopted for others.
For deformed nuclei, the rotational energy correction $E_{\rm rot.}$ is non-zero and calculated by Eq.~(\ref{eq:Erot}), in which the mean-field wave functions are used to calculate the expectation values.
The calculated energies per nucleon $E_{\rm B}/A$ values are shown in Fig.~\ref{fig:E-A-He-Be-C}.
The numbers in the bars are the ratios of the mean-field and center-of-mass energies to the total energy calculated by the MDCRHB model with each parameter set, respectively. 
The black dashed lines are experimental $E_{\rm B}/A$ values taken from AME2020~\cite{Wang2021_ChinPhysC45-030003}.

In Fig.~\ref{fig:E-A-He-Be-C}(a), the mean-field (center-of-mass) energy contributes about 53\% (47\%) to the total energy of $^4$He, 
and the rotational energy correction is zero.
The $E_{\rm B}/A$ values calculated by the NLSH and TM1 parameter sets overestimate the experimental value for about 1 MeV, while those calculated by the other parameter sets compare well with the experimental value. 
This indicates that the phenomenological mass-dependent formula for center-of-mass correction determined from the properties of heavy nuclei is not suitable for extending to light nuclei. However, with microscopic center-of-mass correction, this overestimation problem can be prevented, as suggested by Long et al.~\cite{Long2004_PRC69-034319}.

For $^8$Be in Fig.~\ref{fig:E-A-He-Be-C}(b), the mean-field energies contribute to $59.9\%-71.1\%$ of the $E_{\rm B}/A$ values, which are larger than those in $^4$He and become the dominant part of the binding energy. The center-of-mass energy correction accounts for more than 20\% of the $E_{\rm B}/A$, and that for the rotational correction is about 10\%. This result infers that for deformed light nuclei, not only is the center-of-mass energy correction essential for determinating the calculated binding energy, but the rotational correction is also indispensable. 
For the heavier nucleus, $^{12}$C, the mean-field, center-of-mass, and rotational motions take up about 82\%, 13\%, and 5\% of the $E_{\rm B}/A$ values, respectively, except for the results calculated with DD-LZ1 and DD-PC1, where the nucleus is spherical and there is no rotational correction.
Comparing the results obtained from different parameter sets, the $E_{\rm B}/A$ values obtained by using PK1 and PKDD are very close to the experimental values for all these three nuclei, while NL5(A), DD-ME2, DD-LZ1, and DD-PC1 systematically underestimate the experimental values by a few hundred keV.  
In conclusion, the binding erergies of these alpha nuclei are reproduced with CDFTs when both microscopic center-of-mass and rotational corrections are taken into consideration.

Next, I extend these calculations to even-even nuclei in helium, beryllium and carbon isotopes in order to evaluate the effectiveness of mean-field models in accurately describing the binding energies of light nuclei carefully. 
According to the above discussions, only the parameter sets fitted with microscopic center-of-mass correction are used in the following discussions. The obtained $E_{\rm B}/A$ values of $^{4,6,8}$He, $^{6,8,10,12,14}$Be and $^{10,12,14,16,18,20}$C are listed in table~\ref{tab:1}. In table~\ref{tab:1}, all the calculated binding energies reproduce the experimental values with high accuracy. The largest deviation of the $E_{\rm B}/A$ values is 0.416 MeV, calculated by DD-PC1 in $^{8}$Be.

\begin{figure*}[htbp]
	\centering
	\includegraphics[width=1.0\textwidth]{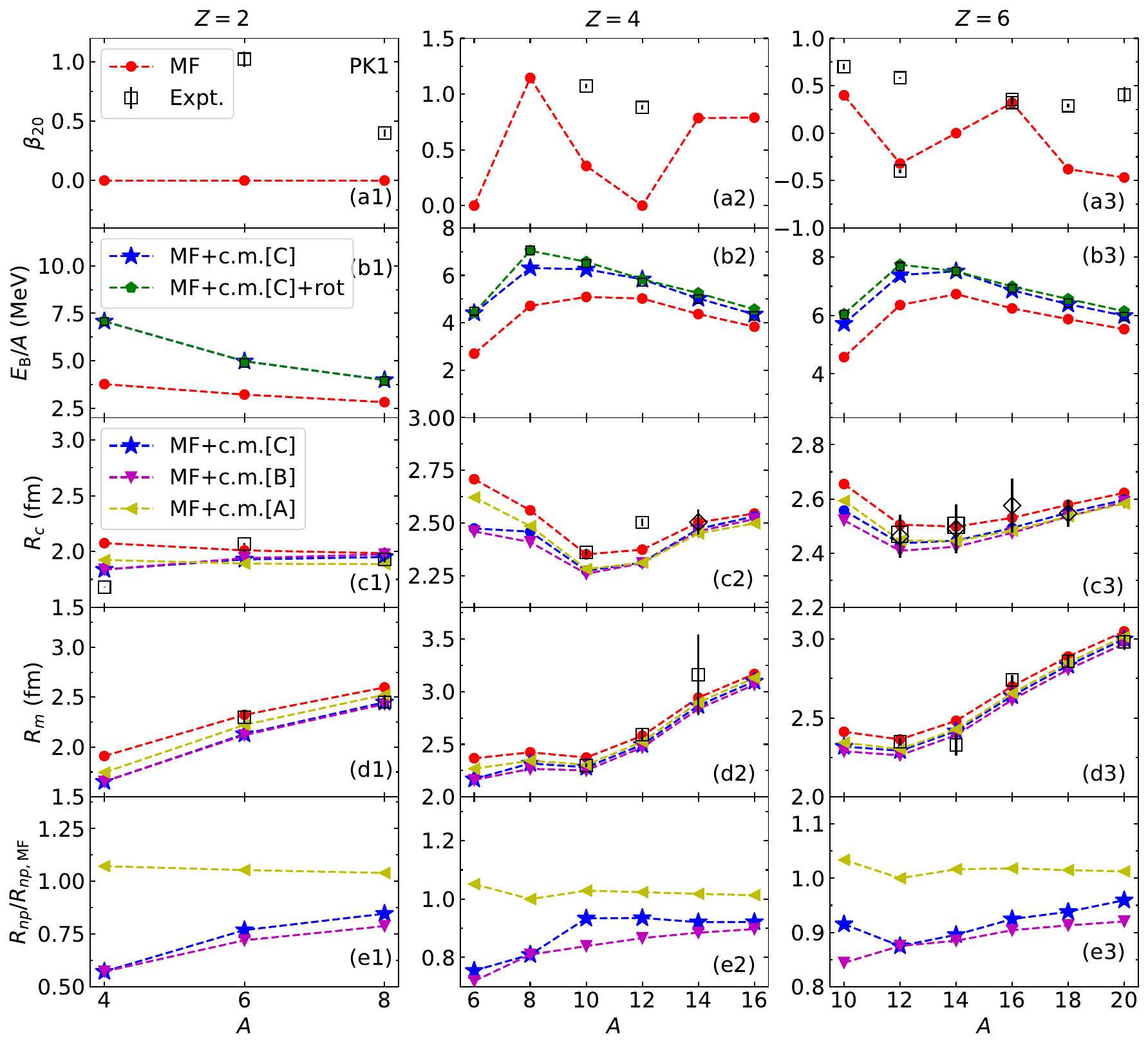}
	\caption{(Color online) Ground-state properties for helium ($Z=2$), beryllium ($Z=4$), and carbon ($Z=6$) isotopes. The evolutions of (a) the quadrupole deformation parameter $\beta_{20}$, (b) the energy per nucleon $E_{\rm B}/A$, (c) the root-mean-square (rms) charge radius $R_c$, (d) the rms mass radius $R_m$, and (e) the neutron skin ratio between pure mean-field and mean-field with corrections, $R_{np}/R_{np,{\rm MF}}$, calculated with the PK1 effective interaction are shown. For center-of-mass (c.m.) correction, "c.m.[A]" is corrected by Eq.~(\ref{eq:E_HO}) for energies and Eq.~(\ref{eq:Radius-corr-HO}) for radii; "c.m.[B]" is corrected by Eq.~(\ref{eq:E_mic}) for energies and Eq.~(\ref{eq:R-cm-corr}) for radii with only direct term in  Eq.~(\ref{eq:cm-R-square}); "c.m.[C]" is the same as "c.m.[B]" but with exchange terms. The green dashed line in (b) represents the result with rotational energy correction. The measured $\beta_{20}$, $E_{\rm B}/A$, $R_c$,  and $R_m$ listed in table~\ref{tab:1} are denoted by open squares. 
	The $R_c$ values with open diamonds are derived from proton radii obtained from measurement of charge-changing cross sections~\cite{Terashima2014_PTEP2014-101D02,Tran2016_PRC94-064604,Tran2018_NutureCommu9-1594}
	 with Eq.~(\ref{eq:rc_from_rp}). }
	\label{fig:He-PK1-beta-E-radius}
\end{figure*}

As an example, the energies in the mean-field approximation and with corrections calculated with PK1 parameter set are clearly visible in Fig.~\ref{fig:He-PK1-beta-E-radius}(b1-b3), together with the corresponding quadrupole deformations in Fig.~\ref{fig:He-PK1-beta-E-radius}(a1-a3). In these figure, the $E_{\rm B}/A$ values are all well-reproduced with center-of-mass and rotational energy corrections, especially for the two-alpha cluster structure nucleus $^8$Be. Although $^8$Be does not have the largest $E_{\rm B}/A$ in this isotopic chain in the mean field, it is repaired after corrections, indicating that the correction terms bring cluster effect to the mean-field description to some extent.

\begin{table*}[htbp]
	\setlength\tabcolsep{10.0pt}{
		\caption{The calculated energy per nucleon $E_{\rm B}/A$, root-mean-square (rms) matter radius $R_m$, rms charge radius $R_c$, and quadrupole deformation parameter $\beta_{20}$ with relativistic Hartree-Bogoliubov model. For energies, the microscopic center-of-mass correction with full exchange terms are used. For radii, results with and without exchange terms ($R_f$ and $R_d$) are listed for comparison. The rotational energy correction calculated by Eq.~(\ref{eq:Erot}) is included in $E_{\rm B}/A$ when the nucleus is deformed. Experimental (Expt.) values of $E_{\rm B}/A$ and $\beta_{20}$ are taken from Refs.~\cite{Wang2021_ChinPhysC45-030003} and~\cite{Pritychenko2016_ADNDT107-1}, respectively, except otherwise noted. The experimental radii and the corresponding references are also listed.}
		\label{tab:1} 
		\begin{tabular}{llrrrrrr}
			\toprule
			Nucleus & Interaction & $E_{\rm B}/A$ (MeV)  & $R_{c,f}$ (fm) & $R_{c,d}$ (fm) & $R_{m,f}$ (fm) & $R_{m,d}$ (fm) & $\beta_{20}$ \\
			\hline
			$^4$He &   Expt.      &   7.074     &  \multicolumn{2}{c}{1.681(4)~\cite{Sick2008_PRC77-041302(R)} }   \\
			{}     &   NL5(A)     &   7.008     &  1.846 &  1.846    & 1.663  &  1.663   & 0.000 \\
			{}     &   PK1        &   7.076     &  1.838 &  1.838    & 1.654  &  1.654   & 0.000 \\
			{}     &   PKDD       &   7.017     &  1.848 &  1.848    & 1.665  &  1.665   & 0.000 \\
			{}     &   DD-ME2     &   6.951     &  1.894 &  1.894    & 1.716  &  1.716   & 0.000 \\
			{}     &   DD-LZ1     &   6.842     &  1.982 &  1.982    & 1.812  &  1.812   & 0.000 \\
			{}     &   DD-PC1     &   6.723     &  1.925 &  1.925    & 1.750  &  1.750   & 0.000 \\
			$^6$He  &   Expt.      &  4.879      &  \multicolumn{2}{c}{2.068(11)~\cite{Mueller2007_PRL99-252501} }
			                                                         & \multicolumn{2}{c}{2.30(7)~\cite{Alkhazov2002_NPA712-269} }
			                                                                             &  1.024(66)  \\
			{}     &   NL5(A)     &  4.941      &  1.935 &  1.947    &  2.129  & 2.118   &  0.000	\\
			{}     &    PK1       &  4.973      &  1.929 &  1.940    &  2.130  & 2.120   &  0.000 \\
			{}     &   PKDD       &  4.873      &  1.946 &  1.958    &  2.154  & 2.144   &  0.000 \\
			{}     &  DD-ME2      &  4.835      &  2.005 &  2.014    &  2.211  & 2.202   &  0.000  \\
			{}     &  DD-LZ1      &  4.858      &  2.100 &  2.109    &  2.295  & 2.286   &  0.000 \\
			{}     &  DD-PC1      &  4.892      &  1.997 &  2.013    &  2.171  & 2.157   &  0.000 \\
			$^8$He &   Expt.      &  3.925      &  \multicolumn{2}{c}{1.929(26)~\cite{Mueller2007_PRL99-252501} }
			& \multicolumn{2}{c}{2.45(7)~\cite{Alkhazov2002_NPA712-269}} 
			& 0.40(3)~\cite{Holl2021_PLB822-136710} \\
			{}     &   NL5(A)     &  4.022      &  1.953 &  1.977    &  2.429  & 2.429   & 0.000	\\
			{}     &    PK1       &  3.983      &  1.951 &  1.974    &  2.449  & 2.449   & 0.000  \\
			{}     &    PKDD      &  3.833      &  1.968 &  1.992    &  2.475  & 2.475   & 0.000 \\
			{}     &   DD-ME2     &  3.715      &  2.036 &  1.974    &  2.541  & 2.541   & 0.000 \\
			{}     &   DD-LZ1     &  3.695      &  2.130 &  2.150    &  2.610  & 2.610   & 0.000 \\
			{}     &   DD-PC1     &  3.950      &  2.010 &  2.039    &  2.448  & 2.448   & 0.000 \\
			$^6$Be &    Expt.     &  4.487     &  {}     &      &	{}       & {}    \\
			{}     &    NL5(A)    &  4.388     &  2.467  &  2.452    &  2.166  &  2.157  & 0.000 \\
			{}     &    PK1       &  4.413     &  2.474  &  2.460    &  2.169  &  2.161  & 0.000 \\
			{}     &    PKDD      &  4.319     &  2.500  &  2.486    &  2.194  &  2.186  & 0.000 \\
			{}     &    DD-ME2    &  4.297     &  2.553  &  2.541    &  2.252  &  2.246  & 0.000 \\
			{}     &    DD-LZ1    &  4.331     &  2.630  &  2.618    &  2.341  &  2.334  & 0.000 \\
			{}     &    DD-PC1    &  4.345     &  2.494  &  2.471    &  2.204  &  2.191  & 0.000 \\
			$^8$Be &    Expt.     &  7.062     &  {}     &           &  {}     &    & {}   \\
			{}     &    NL5(A)    &  6.931     &  2.481  &  2.431    &  2.342  &  2.290  & 1.175	\\
			{}     &    PK1       &  7.040     &  2.459  &  2.410    &  2.319  &  2.267  &  1.145 \\
			{}     &    PKDD      &  6.981     &  2.475  &  2.426    &  2.335  &  2.284  &  1.158 \\
			{}     &    DD-ME2    &  6.928     &  2.527  &  2.477    &  2.390  &  2.338  &  1.213 \\
			{}     &    DD-LZ1    &  6.852     &  2.615  &  2.563    &  2.482  &  2.428  &  1.307 \\
			{}     &    DD-PC1    &  6.646     &  2.581  &  2.531    &  2.448  &  2.396  &  1.263 \\
			$^{10}$Be  &    Expt.     &  6.498     & \multicolumn{2}{c}{2.361(17)~\cite{Krieger2012_PRL108-142501} }
			&   \multicolumn{2}{c}{2.30(2)~\cite{Tanihata1988_PLB206-592} }
			& 1.071($^{+26}_{-20}$)\\
			{}     &   NL5(A)     &  6.472     &  2.292  &	 2.279   &  2.297   &  2.268  & 0.353	\\
			{}     &    PK1       &  6.574     &  2.273  &  2.259   &  2.282   &  2.252  &  0.356 \\
			{}     &   PKDD       &  6.510     &  2.306  &  2.289   &  2.312   &  2.282  &  0.385 \\
			{}     &   DD-ME2     &  6.431     &  2.325  &  2.313   &  2.331   &  2.303  &  0.316  \\
			{}     &   DD-LZ1     &  6.374     &  2.268  &  2.261   &  2.270   &  2.246  &  0.001 \\
			{}     &   DD-PC1     &  6.230     &  2.319  &  2.311   &  2.321   &  2.294  &  0.033 \\
			$^{12}$Be &   Expt.      &  5.721     & \multicolumn{2}{c}{2.503(15)~\cite{Krieger2012_PRL108-142501} }
			& \multicolumn{2}{c}{2.59(6)~\cite{Tanihata1988_PLB206-592} }
			& 0.88($^{+24}_{-12}$)\\
			{}     &   NL5(A)     &  5.844	    &  2.321  & 2.319    &  2.490   &  2.464  & 0.000	\\
			{}     &   PK1        &  5.834     &  2.312  & 2.310    &  2.495   &  2.470  & 0.000 \\
			{}     &   PKDD       &  5.724     &  2.331  & 2.327    &  2.508   &  2.484  & 0.000 \\
			{}     &   DD-ME2     &  5.746     &  2.371  & 2.368    &  2.534   &  2.510  & 0.000 \\
			{}     &   DD-LZ1     &  5.821     &  2.392  & 2.389    &  2.554   &  2.531  & 0.000 \\
			{}     &   DD-PC1     &  5.932     &  2.368  & 2.365    &  2.511   &  2.485  & 0.000 \\
						$^{14}$Be &   Expt.      &  4.994     &  \multicolumn{2}{c}{}           &  \multicolumn{2}{c}{3.16(38)~\cite{Tanihata1988_PLB206-592}} 
			&  {}  \\
			{}     &   NL5(A)     &  5.284     &  2.488  &  2.475   &  2.853   & 2.280   & 0.789	\\
			{}     &   PK1        &  5.255     &  2.472  &  2.460   &  2.867   &  2.835  & 0.785 \\
			{}     &   PKDD       &  5.112     &  2.490  &  2.478   &  2.889   &  2.856  & 0.797 \\
			{}     &   DD-ME2     &  5.039     &  2.509  &  2.503   &  2.917   &  2.886  & 0.758 \\
			{}     &   DD-LZ1     &  4.911     &  2.413  &  2.419   &  2.896   &  2.871  & 0.123 \\
			{}     &   DD-PC1     &  5.195     &  2.503  &  2.459   &  2.853   &  2.821  & 0.730 \\
			\toprule
		\end{tabular}
	}
\end{table*}

\begin{table*}[htbp]
	\setlength\tabcolsep{10.0pt}{
		\label{tab:2} 
		\begin{tabular}{llrrrrrr}
			\toprule
			Nucleus & Interaction & $E_{\rm B}/A$ (MeV)  & $R_{c,f}$ (fm) &  $R_{c,d}$ (fm) & $R_{m,f}$ (fm) &  $R_{m,d}$ (fm) & $\beta_{20}$ \\
			\hline
			$^{10}$C&	 Expt.    &  6.032     &         &           &         &         & 0.701($^{+32}_{-34}$) \\
			{}      &	 NL5(A)   &  5.958     &  2.567  &  2.532    &  2.333  &  2.304  & 0.397	\\
			{}      & 	 PK1      &  6.056     &  2.557  &  2.521    &  2.318  &  2.288  & 0.398 \\
			{}      & 	 PKDD     &  5.997     &  2.583  &  2.546    &  2.348  &  2.317  & 0.429 \\
			{}      &	 DD-ME2   &  5.914     &  2.608  &  2.573    &  2.374  &  2.346  & 0.361 \\
			{}      &	 DD-LZ1   &  5.814     &  2.558  &  2.525    &  2.319  &  2.294  & 0.019 \\
			{}      &	 DD-PC1   &  5.713     &  2.577  &  2.541    &  2.345  &  2.318  & 0.066 \\
			$^{12}$C    &	 Expt.    &  7.680     & \multicolumn{2}{c}{2.4702(22)~\cite{Angeli2013_ADNDT99-69}}   
			&  \multicolumn{2}{c}{2.35(2)~\cite{Kanungo2016_PRL117-102501}} 
			& $-0.40(2)$ \cite{Yasue1983_NPA394-29}     \\
			{}      &	 NL5(A)   &  7.608     &  2.469  &  2.440   &  2.327  &  2.297  & $-0.350$	\\
			{}      &	 PK1      &  7.744     &  2.436  &  2.408   &  2.292  &  2.263  & $-0.320$ \\
			{}      &	 PKDD     &  7.700     &  2.467  &  2.439   &  2.325  &  2.296  & $-0.346$ \\
			{}      &	 DD-ME2   &  7.660     &  2.491  &  2.463   &  2.349  &  2.320  & $-0.347$ \\
			{}      &	 DD-LZ1   &  7.633     &  2.376  &  2.354   &  2.228  &  2.204  & 0.002 \\
			{}      &	 DD-PC1   &  7.269     &  2.481  &  2.453   &  2.340  &  2.312  & $0.000$ \\
			$^{14}$C&	 Expt.    &  7.520     & \multicolumn{2}{c}{2.5025(87)~\cite{Angeli2013_ADNDT99-69}}   
			& \multicolumn{2}{c}{2.33(7)~\cite{Kanungo2016_PRL117-102501}} 
			&  {}   \\
			{}      &	 NL5(A)   &  7.438     &  2.466  &  2.443    &  2.431  &  2.405  & 0.000	\\
			{}      &	 PK1      &  7.517     &  2.446  &  2.423    &  2.417  &  2.391  & 0.000 \\
			{}      &	 PKDD     &  7.484     &  2.463  &  2.440    &  2.427  &  2.402  & 0.000 \\
			{}      &	 DD-ME2   &  7.495     &  2.493  &  2.470    &  2.449  &  2.424  & 0.000 \\
			{}      &	 DD-LZ1   &  7.623     &  2.474  &  2.452    &  2.436  &  2.414  & 0.000 \\
			{}      &	 DD-PC1   &  7.562     &  2.521  &  2.497    &  2.474  &  2.448  & 0.000 \\
			$^{16}$C&	 Expt.    &  6.922     &  \multicolumn{2}{c}{}    
			& \multicolumn{2}{c}{2.74(3)~\cite{Kanungo2016_PRL117-102501}} 
			& 0.323(18) \\
			{}      &     	      &            &         &           &         &         & $0.356^{+0.25}_{-0.23}$~\cite{Jiang2020_PRC101-024601} \\
			{}      &	 NL5(A)   &  6.950     &  2.512  &  2.494    &  2.644  &  2.619  & 0.327	\\
			{}      &	 PK1      &  6.994     &  2.492  &  2.475    &  2.639  &  2.614  & 0.320 \\
			{}      &	 PKDD     &  6.938     &  2.512  &  2.494    &  2.653  &  2.628  & 0.326 \\
			{}      &	 DD-ME2   &  6.935     &  2.547  &  2.529    &  2.679  &  2.654  & 0.316 \\
			{}      &	 DD-LZ1   &  7.015     &  2.522  &  2.506    &  2.674  &  2.652  & 0.192 \\
			{}      &	 DD-PC1   &  7.049     &  2.566  &  2.548    &  2.671  &  2.645  & 0.324 \\
			$^{18}$C&	 Expt.    &  6.426     &  \multicolumn{2}{c}{}    
			&  \multicolumn{2}{c}{2.86(4)~\cite{Kanungo2016_PRL117-102501}}  
			& 0.289($^{+20}_{-13}$)  \\
			{}      &	 NL5(A)   &  6.553     &  2.569  &  2.555    &  2.828  & 2.802   & $-0.381$\\
			{}      &	 PK1      &  6.567     &  2.550  &  2.536    &  2.832  & 2.807   & $-0.380$ \\
			{}      &	 PKDD     &  6.479     &  2.566  &  2.552    &  2.842  & 2.816   & $-0.374$ \\
			{}      &	 DD-ME2   &  6.456     &  2.602  &  2.587    &  2.864  & 2.839   & $-0.372$ \\
			{}      &	 DD-LZ1   &  6.526     &  2.571  &  2.559    &  2.841  & 2.817   & $-0.318$ \\
			{}      &	 DD-PC1   &  6.580     &  2.615  &  2.601    &  2.837  & 2.810   & $-0.362$ \\
			$^{20}$C&	 Expt.    &  5.961     & \multicolumn{2}{c}{}    
			& \multicolumn{2}{c}{2.98(5)~\cite{Togano2016_PLB761-412}} 
			& 0.405($^{+89}_{-45}$)  \\
			{}      &	 NL5(A)   &  6.171     &  2.615  & 2.607     &  2.987  &  2.957  & $-0.468$	\\
			{}      &	 PK1      &  6.146     &  2.596  & 2.588     &  2.999  &  2.970  & $-0.468$ \\
			{}      &	 PKDD     &  6.045     &  2.612  & 2.604     &  3.001  &  2.972  & $-0.457$ \\
			{}      &	 DD-ME2   &  6.006     &  2.649  & 2.640     &  3.011  &  2.983  & $-0.457$ \\
			{}	    &    DD-LZ1   &  6.026     &  2.609  & 2.602     &  2.953  &  2.926  & $-0.413$ \\
			{}	    &    DD-PC1   &  6.177     &  2.664  & 2.656     &  2.985  &  2.956  & $-0.453$ \\
			\toprule
		\end{tabular}
	}
\end{table*}

\begin{figure}[htbp]
	\centering
	\includegraphics[width=0.46\textwidth]{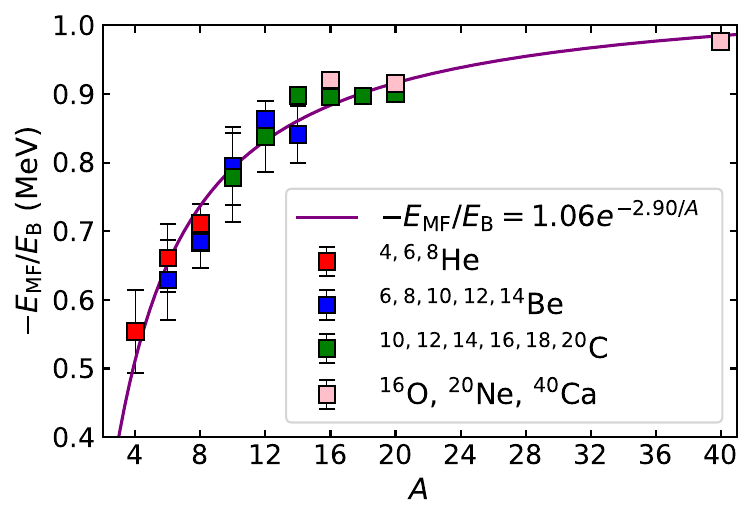}
	\caption{(Color online) The ratios of the mean-field energy to the binding energy, $-E_{\rm MF}/E_B$, as a function of the mass number $A$. The ratios are average values of the results calculated by using NL5(A), PK1, PKDD, DD-ME2, DD-LZ1, and DD-PC1. The corresponding errorbars are their root-mean-square deviations. The solid curve is fitted by all the average values in this figure.}
	\label{fig:Emf-Etot-function-A}
\end{figure}

The above results naturally brings a question: 
How does the percentage of the mean-field energy to the total energy in a nucleus depends on the mass number? To do this, I study the particle-number dependence by averaging the $E_{\rm MF}/E_{\rm B}$ ratios calculated from the six selected effective interactions NL5(A), PK1, PKDD, DD-ME2, DD-LZ1, and DD-PC1 for each nucleus. Meanwhile, the standard deviations of the ratios are also given as model uncertainties. Fig.~\ref{fig:Emf-Etot-function-A} displays the average $-E_{\rm MF}/E_{\rm B}$ values of the even-even helium, beryllium and carbon isotopes. Note that $^{16}$O, $^{20}$Ne, and $^{40}$Ca are added to constrain the particle-number dependence in the medium-mass region. It can be observed that the expected behavior, i.e., a favorable relation between the $-E_{\rm MF}/E_{\rm B}$ and the mass number in the light mass region until $A=40$ is obtained. For heavier nuclei, this dependence can be neglected. Given that the $-E_{\rm MF}/E_{\rm B}$ ratios for nuclei with the same mass number are slightly different, e.g., 0.795$\pm$0.057 for $^{10}$Be and 0.778$\pm$0.065 for $^{10}$C, one would find the isospin effect exists but can be neglected in the mass-number dependent behavior. To give a specific form of the mass-number dependent behavior, I fit all the energy ratios shown in Fig.~\ref{fig:Emf-Etot-function-A} and obtain the following relation 
\begin{equation}\label{eq:MF-A}
	-E_{\rm MF}/E_{\rm B}=1.06 e^{-2.90/A},
\end{equation}
where the mean-field energy ratios accumulate exponentially as the number of particles increases. Predicting by Eq.~(\ref{eq:MF-A}), when $A=42$, the mean-field energy is 99\% equals to the total energy.

\subsection{Radii}

Next, I discuss the radii. 
The radius is a quantity of great interest in describing a nucleus, providing information on deformations~\cite{Charlwood2009_PLB674-23}, exotic structures such as  halos~\cite{Mueller2007_PRL99-252501,Kanungo2016_PRL117-102501} and neutron/proton skins~\cite{Abrahamyan2012_PRL108-112502}, short-range correlations of nucleon-nucleon interactions~\cite{Weiss2019_PLB790-484}, and shell evolutions~\cite{Krieger2012_PRL108-142501}, etc.
Previous studies have shown that the inclusion of relativistic and center-of-mass corrections impacts the quality of energy density functionals optimized for charge radii data~\cite{Reinhard2021_PRC103-054310}.  
In this work, I focus on the treatment of the center-of-mass correction. The rms radius of the center of mass is calculated with HO approximation [Eq.~(\ref{eq:Radius-corr-HO})] or expectation values from RHB states with or without exchange terms [Eq.~(\ref{eq:cm-R-square})]. 

\begin{figure}[htbp]
	\centering
 	\includegraphics[width=0.46\textwidth]{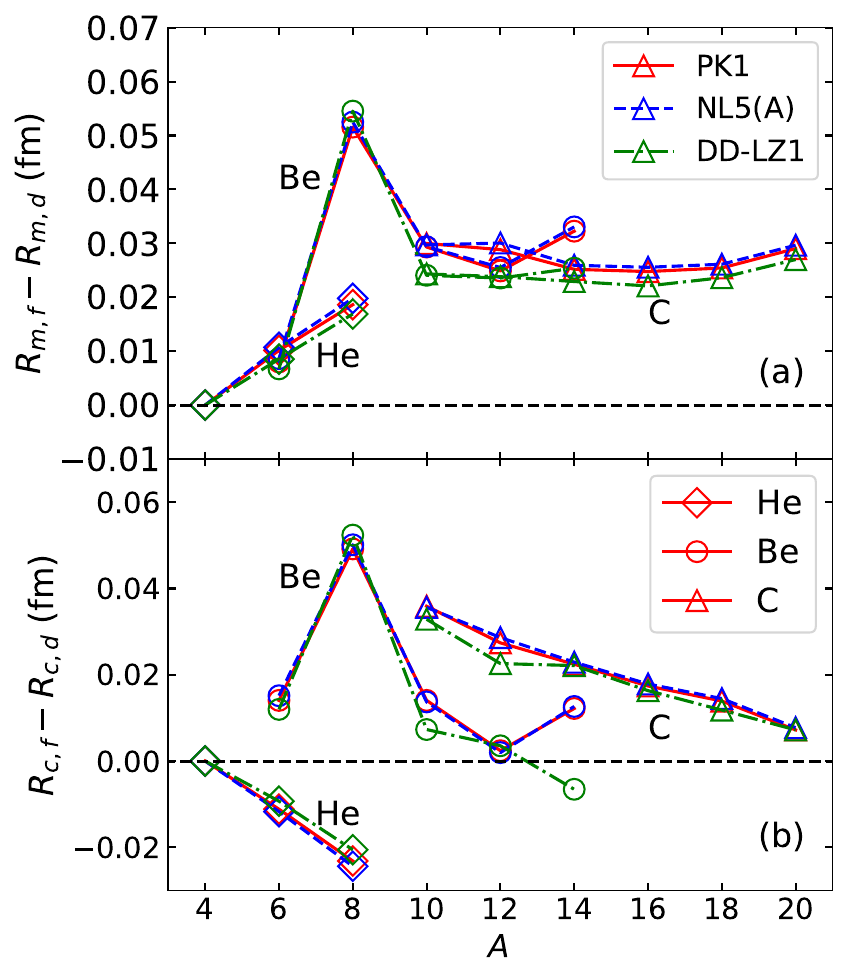}
	\caption{(Color online) The difference between the (a) rms matter radius $R_m$ and (b) rms charge radius $R_c$ calculated with microscopic center-of-mass correction with and without exchange terms (labeled by $R_{m,f}$ and $R_{m,d}$ for matter radius, respectively) for helium, beryllium, and carbon isotopes. The results are calculated with PK1 (red), NL5(A) (blue), and DD-LZ1 (green) parameter sets.}
	\label{fig:DRm_and_DRc}
\end{figure}

Table~\ref{tab:1} lists the radii  calculated by utilizing the microscopic center-of-mass correction with the six selected parameter sets. 
The radii corrected by Eq.~(\ref{eq:Radius-corr-HO}) are not listed because they are simple to calculate and independent of RHB wave functions. As can be seen from table~\ref{tab:1}, the exchange terms change both $R_c$ values and $R_m$ values from those with only the direct term, except for $^4$He.
To observe this effect clearly, I calculate the differences between the rms matter (charge) radii with and without exchange terms, i.e., $R_{m,f}-R_{m,d}$ ($R_{c,f}-R_{c,d}$), with PK1, NL5(A), and DD-LZ1 and show them in Fig.~\ref{fig:DRm_and_DRc}. It is evident that all the $R_{m,f}-R_{m,d}$ values are greater than or equal to zero, manifesting that the inclusion of the exchange terms increases $R_m$ value. However, $R_{c,f}-R_{c,d}<0$ is obtained for $^6$He and $^8$He with these three parameter sets and $^{14}$Be with DD-LZ1, meaning that $R_c$ may be reduced or increased depending on the choice of effective interactions for an individual nucleus. 
Overall, the differences range from 0.0 fm to 0.05 fm. Especially for $^{8}$Be, this effect introduces a two percent deviation to the calculated radius. 
Considering the unprecedented level of precision offered by new experimental techniques, which allows for the exploration of new physics and the elucidation of unclear observables~\cite{Miller2019_NP15-432,Adhikari2021_PRL126-172502}, the exchange terms cannot be ignored for these nuclei in the mean-field calculations.

According to the data in table~\ref{tab:1}, I plot the radii calculated with PK1 and different center-of-mass corrections, and compare them to the mean-field and measured values in Fig.~\ref{fig:He-PK1-beta-E-radius}(c-d).
In Fig.~\ref{fig:He-PK1-beta-E-radius}(c1), the mean-field $R_c$ values calculated with PK1 are around 2.0 fm and decrease as the neutron number increases. The curve including the center-of-mass correction with HO approximation shows a similar tendency in the isotopic chain as the mean-field result, but the corresponding $R_c$ values are smaller. This can be easily understood because this approximation only considers the mass-number dependence. However, with microscopic center-of-mass corrections, i.e., by using Eq.~(\ref{eq:Radius-corr-HO}), the isospin effect is considered, whether with or without exchange terms. For $^4$He, where the nucleons occupy only the $s$ single-particle level, the exchange terms are zero. For $^8$He, the reduction of the $R_c$ caused by the exchange terms cannot be neglected. Comparing these results with the experimental charge radii, the calculated value for $^4$He is larger, while the values for $^6$He and $^8$He are reasonably reproduced. 
Overall, the calculations with full microscopic correction perform better than the other three methods in describing the experimental curve, particularly for the $R_c$ values of $^4$He and $^8$He. 
However, a remaining issue is that the calculated relation $R_c(^6\rm{He})$< $R_c(^8\rm{He})$ contradicts the experimental observation~\cite{Mueller2007_PRL99-252501}. This inconsistency is associated with many-body correlations and will be discussed later from the perspective of deformation.

Regarding $R_m$ in Fig.~\ref{fig:He-PK1-beta-E-radius}(d1), It is clear that the treatment of the correction significantly affects the calculated radii in these nuclei. When compared with the data in Ref.~\cite{Alkhazov2002_NPA712-269}, which reveal $^6$He and $^8$He as halo nuclei, it is surprising that the calculated $R_m$ values for $^8$He are large enough, and the one with full microscopic center-of-mass correction is very close to the experimental value. 
one can further highlight the effects of different corrections by calculating the neutron skin ratio, i.e., $R_{np}/R_{np,{\rm MF}}$, where $R_{np}=R_n-R_p$ is the neutron skin thickness, and $R_n$ ($R_p$) is the rms neutron (proton) radius. In Fig.~\ref{fig:He-PK1-beta-E-radius}(e1), the ratios calculated by Eq.~(\ref{eq:cm-R-square}) are lower than 1.0, while those calculated by Eq.~(\ref{eq:Radius-corr-HO}) are larger than 1.0. This indicates that the microscopic methods reduce the neutron skins, but the HO approximation with energy fixed by the heavy nuclei increases them. 

\begin{figure}
	\centering
	\includegraphics[width=0.45\textwidth]{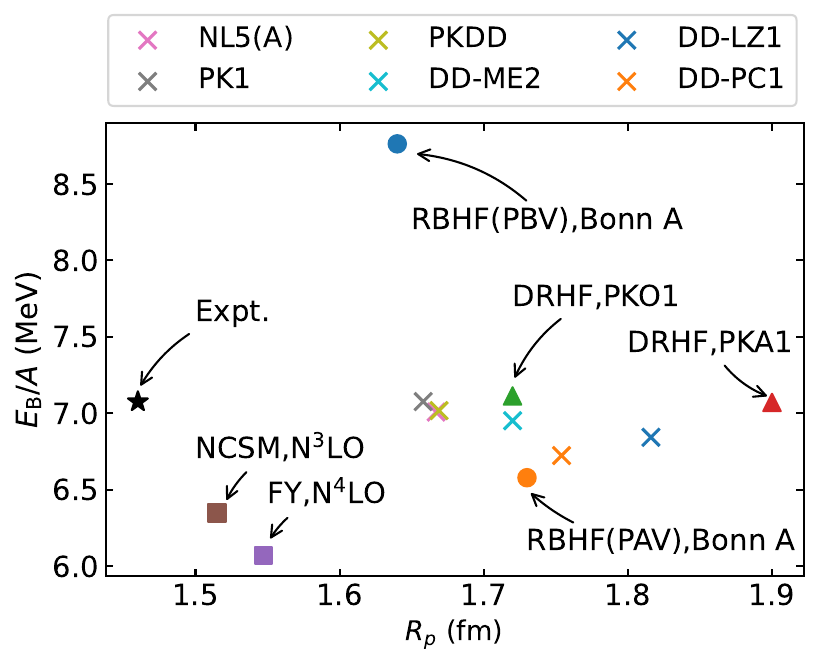}
	\caption{(Color online) Energies per nucleon $E_{\rm B}/A$ values for $^4$He as a function of the proton radius $R_p$ values calculated by the MDCRHB model with full microscopic center-of-mass (c.m.) correction, in comparison with RBHF model with Bonn A interaction \cite{Shen2017_PRC96-014316}, DRHF model with PKO1 \cite{Long2006_PLB640-150} and PKA1 \cite{Long2007_PRC76-034314} interactions, Faddeev-Yakubovsky (FY) equations with N$^4$LO \cite{Binder2016_PRC93-044002}, and no-core shell model (NCSM) with N$^3$LO NN potentials~\cite{Navratil2007_FS41-117}.}
	\label{fig:He4-E-Rp}
\end{figure}

Combining both binding energy and radius, in Fig.~\ref{fig:He4-E-Rp}, the $E_B/A$ values for $^4$He are shown as a function of it's proton radii $R_p$ values calculated with the six selected effective interactions, and compared with other models and the experimental point. It is noteworthy that all the calculated points are on the right side of the measured point. Calculated with effective interactions, the mean-field models, including MDCRHB model and DRHF model~\cite{Long2006_PLB640-150,Long2007_PRC76-034314}, provide proper description for the $E_{\rm B}/A$ but predict larger $R_p$ compared to the experimental value. In contrast, {\textit{ab initio}} methods employing realistic nuclear forces yield smaller deviations between the calculated and experimental $R_p$ values, but larger deviations for $E_{\rm B}/A$ values. 
The conjecture is that incorporating radii or densities into the fitting procedure may result in mean-field models that exhibit greater consistency with experimental measurements.

Then, let's turn to the beryllium isotopes. The $R_c$ values calculated with PK1 for beryllium isotopes are shown in Fig.~\ref{fig:He-PK1-beta-E-radius}(c2). Obviously, those with center-of-mass corrections are smaller than the corresponding $R_c$ values in the mean-field calculation, consistent with the increase in binding energy due to this correction. Furthermore, the microscopic type of center-of-mass correction with only a direct term drastically shrinks the $R_c$ values of $^{6}$Be and $^{8}$Be, while the inclusion of the exchange terms enlarges the value for $^{8}$Be notably. For heavier nuclei, the results with different center-of-mass corrections closely converge. When compared to the experimental $R_c$ values, the calculated results show a systematic shrinkage for $^{10}$Be and $^{12}$Be. 
This situation can be somewhat ameliorated by taking the beyond mean-field effects into consideration. The symmetry restoration, in particular, helps to achieve a larger $\beta_{20}$ from near spherical shape~\cite{Rodriguez-Guzman2000_PRC62-054319,Niksic2006_PRC73-034308} and increase the $R_c$ values of the studied nuclei. Detailed insights into this will be provided in the subsequent section. For $^{14}$Be, the $R_c$ obtained from PK1 closely aligns with the value derived from the charge-changing cross section measurement~\cite{Terashima2014_PTEP2014-101D02}. This could be associated with the large $\beta_{20}$ calculated with PK1. 
In Fig.~\ref{fig:He-PK1-beta-E-radius}(d2),
the correction effect for the $R_m$ is similar to that for $R_c$ in Fig.~\ref{fig:He-PK1-beta-E-radius}(c2), and the specifics are not revisited. 
The calculated $R_m$ values for $^{10}$Be and $^{12}$Be accurately reproduce the experimental values, and that for $^{14}$Be falls within the limits of the experimental error.
I examine the neutron skin ratios in Fig.~\ref{fig:He-PK1-beta-E-radius}(e2) and obtain the same conclusion as with helium isotopes: The microscopic center-of-mass correction diminishes the neutron skin when compared to the mean-field result, while the simple HO approximation yields an opposite tendency.

For carbon isotopes, the reduction of the charge and matter radii by employing center-of-mass correction from the mean-field values is still visible. As shown in Fig.~\ref{fig:He-PK1-beta-E-radius}(c3), the $R_c$ values calculated with full microscopic center-of-mass correction are very close to those calculated with HO approximation in this isotopic chain, and the same situation holds for $R_m$ values in Fig.~\ref{fig:He-PK1-beta-E-radius}(d3). This suggests that, from the perspective of radii, it is a valid approximation for these nuclei that the center-of-mass motion behaves as a harmonic oscillator vibration.
When compared with the experimental radii, the $R_c$ values in Fig.~\ref{fig:He-PK1-beta-E-radius}(c3) with corrections reproduce the experimental data well for $^{12}$C and $^{18}$C. Notably, the $R_c$ for $^{12}$C with PK1 is comparable to the measured values, in contrast to the value calculated with TMA in a previous study~\cite{Tran2018_NutureCommu9-1594}. However, those with pure mean-field calculations are closer to the results derived from measurements of charge-changing reactions~\cite{Tran2016_PRC94-064604}, rather than the corrected values.
As for $R_m$, the corrected values closely match the experimental data for $^{18}$C and $^{20}$C, but they surpass the experimental data for $^{12}$C and $^{16}$C. 
Experimental findings~\cite{Kanungo2016_PRL117-102501} indicate that the smallest $R_m$ in this isotopic chain is observed in $^{14}$C. Nevertheless, when calculated with PK1, the $R_c$ value for $^{14}$C is overestimated with all types of center-of-mass corrections, and the reduction of the radius in this nucleus throughout the evolution of the isotopic chain cannot be achieved.
In comparison with the results from DRHBc model~\cite{Sun2020_NPA1003-122011}, which adopts a spherical Dirac Woods-Saxon (WS) basis and considers continuum effects, results in this work align with theirs, except for $^{16}$C. The difference arises from the softness of the potential energy surface (PES) in the $\gamma=\arctan(\sqrt{2}\beta_{22}/\beta_{20})$ direction with triaxial degree of freedom, as demonstrated in Ref.~\cite{Lu2011_PRC84-014328}.
Additionally, Fig.~\ref{fig:He-PK1-beta-E-radius}(e3) underscores that the neutron skin can serve as a probe to differentiate between types of center-of-mass correction.

\subsection{Deformation and Shell Evolution}

The quadrupole deformations calculated with the six selected parameter sets for the studied light nuclei are illustrated in the last column of table~\ref{tab:1}. Apparently, for most of the studied nuclei, the deformation parameters calculated with different interactions are similar, except for some of those calculated with the point coupling interactions DD-PC1 and DD-LZ1, which tend to have spherical ground states rather than deformed ground states.

I then focus on the results calculated with PK1 as an example. $^{4,6,8}$He are spherical according to the MDCRHB model, implying that the valence neutrons are uniformly distributed around the surface of the alpha particle. This situation naturally leads to an increase in the charge and matter radii as the number of neutrons increases. However, charge radii extracted from the measured isotope shifts reveal a significant reduction in the charge radius from $^6$He to $^8$He~\cite{Mueller2007_PRL99-252501}. In Ref~\cite{Mueller2007_PRL99-252501}, the authors interpreted it as a change in the correlations of the excess neutrons: in $^6$He, the two neutrons are correlated so that on average they spend more time together on one side of the core rather than on opposite sides; 
While for $^8$He, the four excess neutrons are distributed in a more spherically symmetric fashion in the halo, resulting in less smearing of the charge in the core. The quadrupole deformation obtained from proton inelastic scattering also supports this picture~\cite{Mueller2007_PRL99-252501}.
In other words, both the charge radii and deformations from experiments reveal that the mean-field approximation misses the correction among the valence neutrons, which is essential in helium isotopes.  

Superficially, it is the zero quadrupole deformation that causes the inconsistent results with the experiment.
Suppose the experimental deformations can be reproduced by some corrections, then the calculated radius is modified by~\cite{Bohr1969} 
\begin{equation}\label{eq:Radius-deformed-nuclei}
	R^2=\left(1+\dfrac{5}{4\pi}\beta_2^2\right)R_{\rm sph}^2,
\end{equation}
where $R_{\rm sph}$ is the rms radius for a spherical nucleus. By incorporating the experimental $\beta_2$ into Eq.~(\ref{eq:Radius-deformed-nuclei}), the rms $R_c$ for $^6$He is calculated to be 2.296 fm, and that for $^8$He is 2.012 fm, based on the full microscopic center-of-mass correction. These values are larger than the measured values of 2.068 fm and 1.929 fm, respectively, and the trend of $R_c(^6\rm{He})$> $R_c(^8\rm{He})$ > $R_c(^4\rm{He})$ is reproduced.

\begin{figure*}[htbp]
	\centering
	\includegraphics[width=0.9\textwidth]{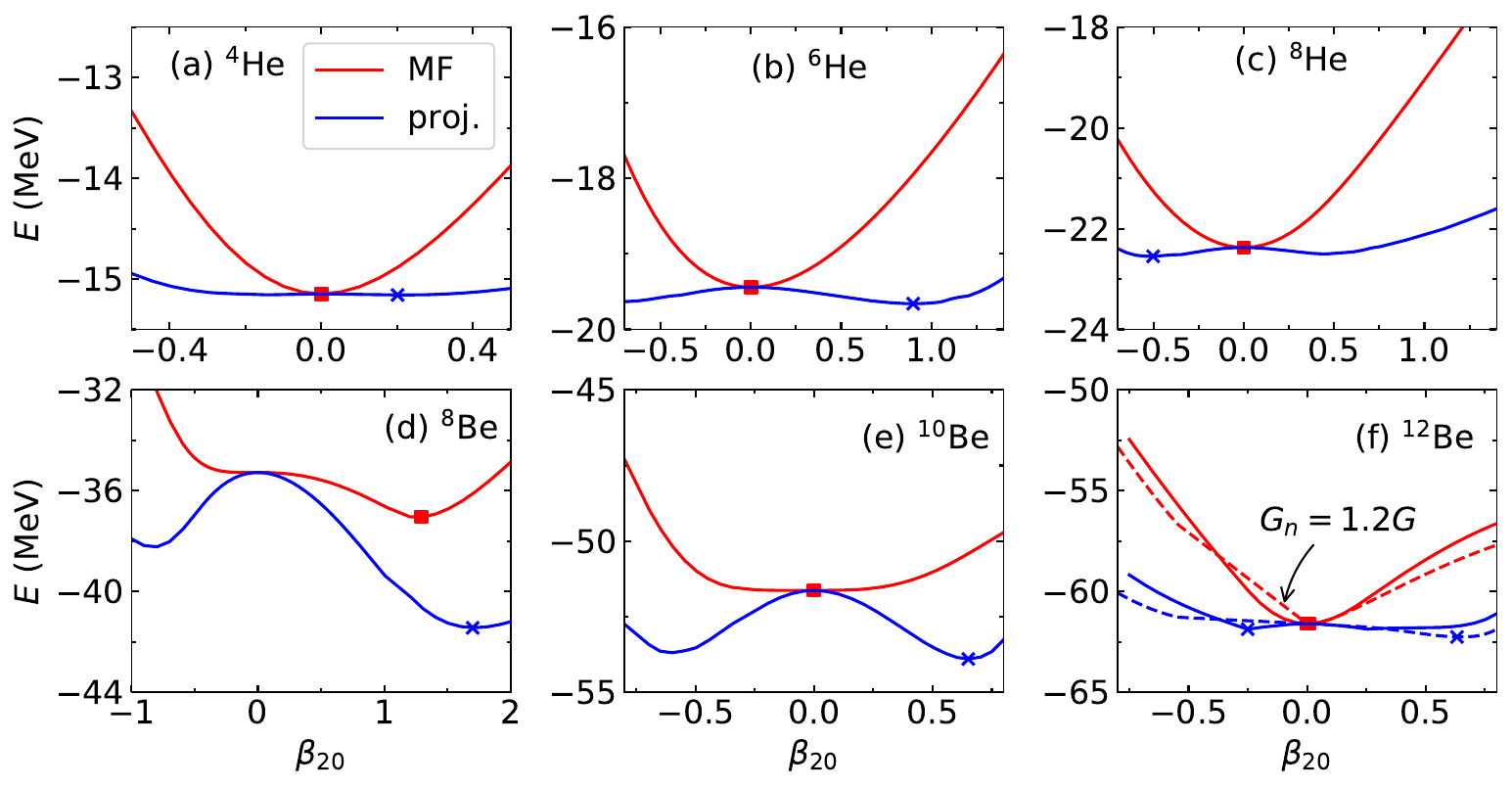}
	\caption{(Color online) The mean-field and the projected potential energy surfaces for $^{4,6,8}$He and $^{8,10,12}$Be calculated with the DD-PC1 effective interaction. The red squares mark the mean-field ground states, and the blue crosses represent the energy mimima after angular momentum projection.}
	\label{fig:He-Be-b20-AMP-PES}
\end{figure*}

Since angular momentum projection usually changes the quadrupole deformation of the mean-field ground state from near zero to a larger value~\cite{Rodriguez-Guzman2000_PRC62-054319,Niksic2006_PRC73-034308},
I attempt to reproduce the experimental deformations for $^6$He and $^8$He by applying angular momentum projection after variation. 
In Fig.~\ref{fig:He-Be-b20-AMP-PES}(a-c), the mean-field and projected potential energy surfaces (PES's) for $^{4,6,8}$He calculated with DD-PC1 effective interaction are shown.
Note that here DD-PC1 is used instead of PK1 since symmetry restoration calculation with the latter effective interaction has not been realized. In Fig.~\ref{fig:He-Be-b20-AMP-PES}(a-c), the quadrupole deformations for these three nuclei are zero in the mean field, consistent with the results in Fig.~\ref{fig:He-PK1-beta-E-radius}(a1). With angular momentum projection, the PES for $^4$He is very soft, while those for $^6$He and $^8$He evidently reach deformed energy minima. Precisely, the locations of the energy minima for
$^{4,6,8}$He are $\beta_{20}= 0.20,0.90$ and $-0.50$. The corresponding
energy differences between the projected and mean-field energy minima are 0.01 MeV, 0.22 MeV, and 0.17 MeV, respectively. One can conclude that the experimental deformations for $^6$He and
$^8$He can be partly explained by doing angular momentum projection,
and the influence of symmetry restoration on $^4$He can be neglected. However, this issue might persist after configuration mixing calculation due to the softness of the PES's, similar to the case of $^{32}$Mg calculated with PC-F1~\cite{Niksic2006_PRC73-034308}. Detailed discussions of configuration mixing are beyond the scope of this article.   

For beryllium isotopes, in Fig.~\ref{fig:He-PK1-beta-E-radius}(a2), the studied nuclei are prolate except for $^{6}$Be and $^{12}$Be. As expected, the deformation of $^8$Be is $\beta_2=1.145$, forming a typical two-alpha cluster structure. With more or fewer neutrons, the alpha cluster structure diminishes or even disappears. 
$^{12}$Be, which consists of four extra neutrons compared to $^8$Be, is a well-known nucleus in discussions about shell evolution. The disappearance of the
magic number $N = 8$ has been suggested in this nucleus through various observables measured in experiments, such as lifetime ~\cite{Imai2009_PLB673-179}, charge radius~\cite{Krieger2012_PRL108-142501}, and single-neutron removal cross sections~\cite{Pain2006_PRL96-032502}. However, as listed in table~\ref{tab:1}, it is predicted to be spherical by the RHB model with all the selected effective interactions. This indicates that the magic number $N=8$ naturally arises in every mean-field description for this nucleus. {\color{black} A natural understanding of this result is that the correlations between nucleons are lost with the mean-field approximation, whereas it is crucial for determining the shell closure in this mass region. 
For example, in the molecular-orbital models, by coupling with the spin-triplet states, the energy of the $(3/2^-)^2 (1/2^+)^2$ configuration for the four valence neutrons is almost the same as, or even lower than the $(3/2^-)^2 (1/2^-)^2$ corresponding to the closed $p$-shell configuration. This results in the breaking of the neutron magic number $N=8$~\cite{Itagaki2000_PRC62-034301,KanadaEnyo2012_PTEP2012-01A202}. However, the inclusion of this effect is difficult within a single-reference configuration.
}

Similar to the discussion in helium isotopes, the radius for $^{12}$Be calculated using Eq.~(\ref{eq:Radius-deformed-nuclei}) with the spherical mean-field radius of 2.312 fm and experimental quadrupole deformation $\beta_2=0.88$ is 2.644 fm, which is slightly larger than the measured charge radius of 2.503(15) fm~\cite{Krieger2012_PRL108-142501}.
The projected PES's calculated with the DD-PC1 effective interaction are presented in Fig.~\ref{fig:He-Be-b20-AMP-PES}(d-f). In this figure, one could observe the distance between the two alpha particles in $^8$Be enlarges after projection, as the corresponding $\beta_{20}$ in the projected energy minimum is larger than that in the mean field. The same phenomenon occurs for $^{10}$Be. However, it is hard to conclude that $^{12}$Be is well-deformed after projection, because the projected PES appears very soft in the figure. I increase the pairing strength for neutrons by $20\%$ and obtain a stiffer projected PES with a prolate energy minimum. This suggests that
an enhancement of pairing interactions may contribute to breaking up the shell closure in this nucleus.

Finally, I discuss the carbon isotopes.
Calculated with PK1, in Fig.~\ref{fig:He-PK1-beta-E-radius}(a3), $^{10,16}$C are prolate, $^{14}$C is spherical, and $^{12,18,20}$C are oblate. These shapes align with those calculated with the MDC-RMF model, except for $^{18}$C~\cite{Lu2011_PRC84-014328}, which is predicted to be triaxially deformed its ground state. In comparison with the DRHBc model, my results are consistent with theirs, except for $^{16}$C. The discrepancy arises from the softness of the PES in the $\gamma$ direction with triaxial degree of freedom. It's worth noting that in Ref.~\cite{Pritychenko2016_ADNDT107-1}, the deformation parameter is obtained from electric quadrupole transition of the nucleus, and distinguishing between prolate ($\beta_{20}>0$) and oblate ($\beta_{20}<0$) shapes is not feasible. Consequently, when compared to the results from Ref.~\cite{Pritychenko2016_ADNDT107-1}, the calculated deformations for $^{16,18,20}$C are quite accurate. 
For $^{10}$C, prediction from the MDCRHB model indicates a smaller quadrupole deformation parameter compared to the experimental value of 0.701~\cite{Pritychenko2016_ADNDT107-1}.  
As for $^{12}$C, this work supports an oblate shape with $\beta_{20}= -0.40(2)$~\cite{Yasue1983_NPA394-29}. 

$^{14}$C, with six protons and eight neutrons, provides an important platform to study the possible existence of the magic number 6 in certain semimagic unstable nuclei. 
The fact that the systematics of proton radii, $B(E2)$ values and the empirical proton-subshell gaps for most carbon isotopes are comparable to those for proton-closed shell Oxygen isotopes manifests $^{14}$C is a doubly-magic nucleus~\cite{Tran2018_NutureCommu9-1594}. 
In the MDCRHB model,
$^{14}$C is predicted to be
spherical in its ground state using all the effective interactions I employed. When compare with the cases of $^8$He and $^{12}$Be, the mean-field calculations consistently yield strong spin-orbit coupling for the $1p$ state, leading to the magic number 6. As a result, shell closures are achieved in $^8$He and $^{14}$C. However, the poor description of deformations in helium and beryllium isotopes shows the absence of the many-body correlations in the mean-field approximation. 

\section{SUMMARY}\label{sec4}

This study delves into the efficacy of the mean-field approach in describing light nuclei. To this end, the ground-state properties of helium, beryllium, and carbon isotopes are examined using the MDCRHB model and its associated corrections. These properties include binding energy, quadrupole deformation, root-mean-square (rms) charge radius, rms matter radius, and neutron skin. Eight effective interactions are employed to assess the theoretical uncertainty arising from effective interactions. Notably, the full microscopic center-of-mass correction for radius, which has been neglected in the descriptions of medium and heavy nuclei, is incorporated for the first time. Additionally, angular momentum projection is performed on the potential energy surfaces after mean-field calculations.


The binding energies of the investigated nuclei are accurately described by the MDCRHB model incorporating microscopic center-of-mass correction and rotational correction. The average mean-field energies calculated with these corrections exhibit an exponential relation to the calculated binding energies, with a coefficient of mass correlation.
Regarding the radius, the exchange terms in the center-of-mass correction cannot be neglected, which contrasts with the situation in heavier mass regions. With the PK1 effective interaction, most of the charge and matter radii closely match the experimental values. The neutron skin ratio can be used to distinguish the type of center-of-mass correction, and the neutron skin becomes smaller with microscopic center-of-mass correction.
Shell closures are achieved in $^8$He and $^{14}$C due to the consistently strong spin-orbit coupling for the $1p$ state predicted by mean-field calculations. However, the poor description of deformations in helium and beryllium isotopes indicates the absence of many-body correlations in the mean-field approximation. Deformation is a key property to test the ability of mean-field models in describing light nuclei, and angular momentum projection after variation can partially aid in reproducing the deformations in helium and beryllium isotopes.
Further utilization of realistic nucleon-nucleon interactions to calculate ground states within this framework would be of interest.

	\begin{acknowledgments}
		Helpful discussions with Xiang-Xiang Sun, Shan-Gui Zhou and Bing-Nan Lu are
		gratefully acknowledged. This work has been supported by the National Natural Science Foundation of China (Grants No. 12205057), the Central Government Guides Local Scientific and Technological Development Fund Projects (Grants No. Guike ZY22096024), the Science and Technology Plan Project of Guangxi (Grants No. Guike AD23026250), the promotion project of scientific research ability of young teachers in universities of Guangxi (Grant No. 2022KY0055), and the the Natural
		Science Foundation of Guangxi (Grant No. 2023GXNSFAA026016). The results in this paper are obtained on the High-performance
		Computing Cluster of ITP-CAS. 
	\end{acknowledgments}
	
	\bibliographystyle{apsrev4-2}
	\bibliography{../../../../Notes/bib/ref}

\begin{thebibliography}{88}%
\makeatletter
\providecommand \@ifxundefined [1]{%
 \@ifx{#1\undefined}
}%
\providecommand \@ifnum [1]{%
 \ifnum #1\expandafter \@firstoftwo
 \else \expandafter \@secondoftwo
 \fi
}%
\providecommand \@ifx [1]{%
 \ifx #1\expandafter \@firstoftwo
 \else \expandafter \@secondoftwo
 \fi
}%
\providecommand \natexlab [1]{#1}%
\providecommand \enquote  [1]{``#1''}%
\providecommand \bibnamefont  [1]{#1}%
\providecommand \bibfnamefont [1]{#1}%
\providecommand \citenamefont [1]{#1}%
\providecommand \href@noop [0]{\@secondoftwo}%
\providecommand \href [0]{\begingroup \@sanitize@url \@href}%
\providecommand \@href[1]{\@@startlink{#1}\@@href}%
\providecommand \@@href[1]{\endgroup#1\@@endlink}%
\providecommand \@sanitize@url [0]{\catcode `\\12\catcode `\$12\catcode
  `\&12\catcode `\#12\catcode `\^12\catcode `\_12\catcode `\%12\relax}%
\providecommand \@@startlink[1]{}%
\providecommand \@@endlink[0]{}%
\providecommand \url  [0]{\begingroup\@sanitize@url \@url }%
\providecommand \@url [1]{\endgroup\@href {#1}{\urlprefix }}%
\providecommand \urlprefix  [0]{URL }%
\providecommand \Eprint [0]{\href }%
\providecommand \doibase [0]{https://doi.org/}%
\providecommand \selectlanguage [0]{\@gobble}%
\providecommand \bibinfo  [0]{\@secondoftwo}%
\providecommand \bibfield  [0]{\@secondoftwo}%
\providecommand \translation [1]{[#1]}%
\providecommand \BibitemOpen [0]{}%
\providecommand \bibitemStop [0]{}%
\providecommand \bibitemNoStop [0]{.\EOS\space}%
\providecommand \EOS [0]{\spacefactor3000\relax}%
\providecommand \BibitemShut  [1]{\csname bibitem#1\endcsname}%
\let\auto@bib@innerbib\@empty
\bibitem [{\citenamefont {Freer}\ \emph {et~al.}(2018)\citenamefont {Freer},
  \citenamefont {Horiuchi}, \citenamefont {Kanada-En'yo}, \citenamefont {Lee},\
  and\ \citenamefont {{Ulf-G. Mei\ss{}ner}}}]{Freer2018_RMP90-035004}%
  \BibitemOpen
  \bibfield  {author} {\bibinfo {author} {\bibfnamefont {M.}~\bibnamefont
  {Freer}}, \bibinfo {author} {\bibfnamefont {H.}~\bibnamefont {Horiuchi}},
  \bibinfo {author} {\bibfnamefont {Y.}~\bibnamefont {Kanada-En'yo}}, \bibinfo
  {author} {\bibfnamefont {D.}~\bibnamefont {Lee}},\ and\ \bibinfo {author}
  {\bibnamefont {{Ulf-G. Mei\ss{}ner}}},\ }\href
  {https://doi.org/10.1103/RevModPhys.90.035004} {\bibfield  {journal}
  {\bibinfo  {journal} {Rev. Mod. Phys.}\ }\textbf {\bibinfo {volume} {90}},\
  \bibinfo {pages} {035004} (\bibinfo {year} {2018})}\BibitemShut {NoStop}%
\bibitem [{\citenamefont {Bijker}\ and\ \citenamefont
  {Iachello}(2020)}]{Bijker2020_PPNP110-103735}%
  \BibitemOpen
  \bibfield  {author} {\bibinfo {author} {\bibfnamefont {R.}~\bibnamefont
  {Bijker}}\ and\ \bibinfo {author} {\bibfnamefont {F.}~\bibnamefont
  {Iachello}},\ }\href {https://doi.org/10.1016/j.ppnp.2019.103735} {\bibfield
  {journal} {\bibinfo  {journal} {Prog. Part. Nucl. Phys.}\ }\textbf {\bibinfo
  {volume} {110}},\ \bibinfo {pages} {103735} (\bibinfo {year}
  {2020})}\BibitemShut {NoStop}%
\bibitem [{\citenamefont {Bishop}\ \emph {et~al.}(2019)\citenamefont {Bishop},
  \citenamefont {Kokalova}, \citenamefont {Freer}, \citenamefont {Acosta},
  \citenamefont {Assi\'e}, \citenamefont {Bailey}, \citenamefont {Cardella},
  \citenamefont {Curtis}, \citenamefont {De~Filippo}, \citenamefont
  {Dell'Aquila}, \citenamefont {De~Luca}, \citenamefont {Francalanza},
  \citenamefont {Gnoffo}, \citenamefont {Lanzalone}, \citenamefont {Lombardo},
  \citenamefont {Martorana}, \citenamefont {Norella}, \citenamefont {Pagano},
  \citenamefont {Pagano}, \citenamefont {Papa}, \citenamefont {Pirrone},
  \citenamefont {Politi}, \citenamefont {Rizzo}, \citenamefont {Russotto},
  \citenamefont {Quattrocchi}, \citenamefont {Smith}, \citenamefont {Stefan},
  \citenamefont {Trifir\`o}, \citenamefont {Trimarch\`{\i}}, \citenamefont
  {Verde}, \citenamefont {Vigilante},\ and\ \citenamefont
  {Wheldon}}]{Bishop2019_PRC100-034320}%
  \BibitemOpen
  \bibfield  {author} {\bibinfo {author} {\bibfnamefont {J.}~\bibnamefont
  {Bishop}}, \bibinfo {author} {\bibfnamefont {T.}~\bibnamefont {Kokalova}},
  \bibinfo {author} {\bibfnamefont {M.}~\bibnamefont {Freer}}, \bibinfo
  {author} {\bibfnamefont {L.}~\bibnamefont {Acosta}}, \bibinfo {author}
  {\bibfnamefont {M.}~\bibnamefont {Assi\'e}}, \bibinfo {author} {\bibfnamefont
  {S.}~\bibnamefont {Bailey}}, \bibinfo {author} {\bibfnamefont
  {G.}~\bibnamefont {Cardella}}, \bibinfo {author} {\bibfnamefont
  {N.}~\bibnamefont {Curtis}}, \bibinfo {author} {\bibfnamefont
  {E.}~\bibnamefont {De~Filippo}}, \bibinfo {author} {\bibfnamefont
  {D.}~\bibnamefont {Dell'Aquila}}, \bibinfo {author} {\bibfnamefont
  {S.}~\bibnamefont {De~Luca}}, \bibinfo {author} {\bibfnamefont
  {L.}~\bibnamefont {Francalanza}}, \bibinfo {author} {\bibfnamefont
  {B.}~\bibnamefont {Gnoffo}}, \bibinfo {author} {\bibfnamefont
  {G.}~\bibnamefont {Lanzalone}}, \bibinfo {author} {\bibfnamefont
  {I.}~\bibnamefont {Lombardo}}, \bibinfo {author} {\bibfnamefont {N.~S.}\
  \bibnamefont {Martorana}}, \bibinfo {author} {\bibfnamefont {S.}~\bibnamefont
  {Norella}}, \bibinfo {author} {\bibfnamefont {A.}~\bibnamefont {Pagano}},
  \bibinfo {author} {\bibfnamefont {E.~V.}\ \bibnamefont {Pagano}}, \bibinfo
  {author} {\bibfnamefont {M.}~\bibnamefont {Papa}}, \bibinfo {author}
  {\bibfnamefont {S.}~\bibnamefont {Pirrone}}, \bibinfo {author} {\bibfnamefont
  {G.}~\bibnamefont {Politi}}, \bibinfo {author} {\bibfnamefont
  {F.}~\bibnamefont {Rizzo}}, \bibinfo {author} {\bibfnamefont
  {P.}~\bibnamefont {Russotto}}, \bibinfo {author} {\bibfnamefont
  {L.}~\bibnamefont {Quattrocchi}}, \bibinfo {author} {\bibfnamefont
  {R.}~\bibnamefont {Smith}}, \bibinfo {author} {\bibfnamefont
  {I.}~\bibnamefont {Stefan}}, \bibinfo {author} {\bibfnamefont
  {A.}~\bibnamefont {Trifir\`o}}, \bibinfo {author} {\bibfnamefont
  {M.}~\bibnamefont {Trimarch\`{\i}}}, \bibinfo {author} {\bibfnamefont
  {G.}~\bibnamefont {Verde}}, \bibinfo {author} {\bibfnamefont
  {M.}~\bibnamefont {Vigilante}},\ and\ \bibinfo {author} {\bibfnamefont
  {C.}~\bibnamefont {Wheldon}},\ }\href
  {https://doi.org/10.1103/PhysRevC.100.034320} {\bibfield  {journal} {\bibinfo
   {journal} {Phys. Rev. C}\ }\textbf {\bibinfo {volume} {100}},\ \bibinfo
  {pages} {034320} (\bibinfo {year} {2019})}\BibitemShut {NoStop}%
\bibitem [{\citenamefont {Epelbaum}\ \emph {et~al.}(2012)\citenamefont
  {Epelbaum}, \citenamefont {Krebs}, \citenamefont {L\"ahde}, \citenamefont
  {Lee},\ and\ \citenamefont {{Ulf-G.
  Mei\ss{}ner}}}]{Epelbaum2012_PRL109-252501}%
  \BibitemOpen
  \bibfield  {author} {\bibinfo {author} {\bibfnamefont {E.}~\bibnamefont
  {Epelbaum}}, \bibinfo {author} {\bibfnamefont {H.}~\bibnamefont {Krebs}},
  \bibinfo {author} {\bibfnamefont {T.~A.}\ \bibnamefont {L\"ahde}}, \bibinfo
  {author} {\bibfnamefont {D.}~\bibnamefont {Lee}},\ and\ \bibinfo {author}
  {\bibnamefont {{Ulf-G. Mei\ss{}ner}}},\ }\href
  {https://doi.org/10.1103/PhysRevLett.109.252501} {\bibfield  {journal}
  {\bibinfo  {journal} {Phys. Rev. Lett.}\ }\textbf {\bibinfo {volume} {109}},\
  \bibinfo {pages} {252501} (\bibinfo {year} {2012})}\BibitemShut {NoStop}%
\bibitem [{\citenamefont {Dell'Aquila}\ \emph {et~al.}(2017)\citenamefont
  {Dell'Aquila}, \citenamefont {Lombardo}, \citenamefont {Verde}, \citenamefont
  {Vigilante}, \citenamefont {Acosta}, \citenamefont {Agodi}, \citenamefont
  {Cappuzzello}, \citenamefont {Carbone}, \citenamefont {Cavallaro},
  \citenamefont {Cherubini}, \citenamefont {Cvetinovic}, \citenamefont
  {D'Agata}, \citenamefont {Francalanza}, \citenamefont {Guardo}, \citenamefont
  {Gulino}, \citenamefont {Indelicato}, \citenamefont {La~Cognata},
  \citenamefont {Lamia}, \citenamefont {Ordine}, \citenamefont {Pizzone},
  \citenamefont {Puglia}, \citenamefont {Rapisarda}, \citenamefont {Romano},
  \citenamefont {Santagati}, \citenamefont {Spart\`a}, \citenamefont
  {Spadaccini}, \citenamefont {Spitaleri},\ and\ \citenamefont
  {Tumino}}]{DellAquila2017_PRL119-132501}%
  \BibitemOpen
  \bibfield  {author} {\bibinfo {author} {\bibfnamefont {D.}~\bibnamefont
  {Dell'Aquila}}, \bibinfo {author} {\bibfnamefont {I.}~\bibnamefont
  {Lombardo}}, \bibinfo {author} {\bibfnamefont {G.}~\bibnamefont {Verde}},
  \bibinfo {author} {\bibfnamefont {M.}~\bibnamefont {Vigilante}}, \bibinfo
  {author} {\bibfnamefont {L.}~\bibnamefont {Acosta}}, \bibinfo {author}
  {\bibfnamefont {C.}~\bibnamefont {Agodi}}, \bibinfo {author} {\bibfnamefont
  {F.}~\bibnamefont {Cappuzzello}}, \bibinfo {author} {\bibfnamefont
  {D.}~\bibnamefont {Carbone}}, \bibinfo {author} {\bibfnamefont
  {M.}~\bibnamefont {Cavallaro}}, \bibinfo {author} {\bibfnamefont
  {S.}~\bibnamefont {Cherubini}}, \bibinfo {author} {\bibfnamefont
  {A.}~\bibnamefont {Cvetinovic}}, \bibinfo {author} {\bibfnamefont
  {G.}~\bibnamefont {D'Agata}}, \bibinfo {author} {\bibfnamefont
  {L.}~\bibnamefont {Francalanza}}, \bibinfo {author} {\bibfnamefont {G.~L.}\
  \bibnamefont {Guardo}}, \bibinfo {author} {\bibfnamefont {M.}~\bibnamefont
  {Gulino}}, \bibinfo {author} {\bibfnamefont {I.}~\bibnamefont {Indelicato}},
  \bibinfo {author} {\bibfnamefont {M.}~\bibnamefont {La~Cognata}}, \bibinfo
  {author} {\bibfnamefont {L.}~\bibnamefont {Lamia}}, \bibinfo {author}
  {\bibfnamefont {A.}~\bibnamefont {Ordine}}, \bibinfo {author} {\bibfnamefont
  {R.~G.}\ \bibnamefont {Pizzone}}, \bibinfo {author} {\bibfnamefont
  {S.~M.~R.}\ \bibnamefont {Puglia}}, \bibinfo {author} {\bibfnamefont {G.~G.}\
  \bibnamefont {Rapisarda}}, \bibinfo {author} {\bibfnamefont {S.}~\bibnamefont
  {Romano}}, \bibinfo {author} {\bibfnamefont {G.}~\bibnamefont {Santagati}},
  \bibinfo {author} {\bibfnamefont {R.}~\bibnamefont {Spart\`a}}, \bibinfo
  {author} {\bibfnamefont {G.}~\bibnamefont {Spadaccini}}, \bibinfo {author}
  {\bibfnamefont {C.}~\bibnamefont {Spitaleri}},\ and\ \bibinfo {author}
  {\bibfnamefont {A.}~\bibnamefont {Tumino}},\ }\href
  {https://doi.org/10.1103/PhysRevLett.119.132501} {\bibfield  {journal}
  {\bibinfo  {journal} {Phys. Rev. Lett.}\ }\textbf {\bibinfo {volume} {119}},\
  \bibinfo {pages} {132501} (\bibinfo {year} {2017})}\BibitemShut {NoStop}%
\bibitem [{\citenamefont {Shen}\ \emph {et~al.}(2021)\citenamefont {Shen},
  \citenamefont {L$\ddot{\rm a}$hde}, \citenamefont {Lee},\ and\ \citenamefont
  {Mei{\ss}ner}}]{Shen2021_EPJA57-276}%
  \BibitemOpen
  \bibfield  {author} {\bibinfo {author} {\bibfnamefont {S.}~\bibnamefont
  {Shen}}, \bibinfo {author} {\bibfnamefont {T.~A.}\ \bibnamefont {L$\ddot{\rm
  a}$hde}}, \bibinfo {author} {\bibfnamefont {D.}~\bibnamefont {Lee}},\ and\
  \bibinfo {author} {\bibfnamefont {U.-G.}\ \bibnamefont {Mei{\ss}ner}},\
  }\href {https://doi.org/10.1140/epja/s10050-021-00586-6} {\bibfield
  {journal} {\bibinfo  {journal} {Eur. Phys. J. A}\ }\textbf {\bibinfo {volume}
  {57}},\ \bibinfo {pages} {276} (\bibinfo {year} {2021})}\BibitemShut
  {NoStop}%
\bibitem [{\citenamefont {Robson}(1982)}]{Robson1982_PRC25-1108}%
  \BibitemOpen
  \bibfield  {author} {\bibinfo {author} {\bibfnamefont {D.}~\bibnamefont
  {Robson}},\ }\href {https://doi.org/10.1103/PhysRevC.25.1108} {\bibfield
  {journal} {\bibinfo  {journal} {Phys. Rev. C}\ }\textbf {\bibinfo {volume}
  {25}},\ \bibinfo {pages} {1108} (\bibinfo {year} {1982})}\BibitemShut
  {NoStop}%
\bibitem [{\citenamefont {Bijker}\ and\ \citenamefont
  {Iachello}(2014)}]{Bijker2014_PRL112-152501}%
  \BibitemOpen
  \bibfield  {author} {\bibinfo {author} {\bibfnamefont {R.}~\bibnamefont
  {Bijker}}\ and\ \bibinfo {author} {\bibfnamefont {F.}~\bibnamefont
  {Iachello}},\ }\href {https://doi.org/10.1103/PhysRevLett.112.152501}
  {\bibfield  {journal} {\bibinfo  {journal} {Phys. Rev. Lett.}\ }\textbf
  {\bibinfo {volume} {112}},\ \bibinfo {pages} {152501} (\bibinfo {year}
  {2014})}\BibitemShut {NoStop}%
\bibitem [{\citenamefont {Halcrow}\ and\ \citenamefont
  {Manton}(2020)}]{Halcrow2020_JPCS1643-012136}%
  \BibitemOpen
  \bibfield  {author} {\bibinfo {author} {\bibfnamefont {C.~J.}\ \bibnamefont
  {Halcrow}}\ and\ \bibinfo {author} {\bibfnamefont {N.~S.}\ \bibnamefont
  {Manton}},\ }\href {https://doi.org/10.1088/1742-6596/1643/1/012136}
  {\bibfield  {journal} {\bibinfo  {journal} {J. Phys.: Conf. Ser.}\ }\textbf
  {\bibinfo {volume} {1643}},\ \bibinfo {pages} {012136} (\bibinfo {year}
  {2020})}\BibitemShut {NoStop}%
\bibitem [{\citenamefont {Pf$\ddot{\rm u}$tzner}\ \emph
  {et~al.}(2012)\citenamefont {Pf$\ddot{\rm u}$tzner}, \citenamefont {Karny},
  \citenamefont {Grigorenko},\ and\ \citenamefont
  {Riisager}}]{Pfuetzner2012_RMP84-567}%
  \BibitemOpen
  \bibfield  {author} {\bibinfo {author} {\bibfnamefont {M.}~\bibnamefont
  {Pf$\ddot{\rm u}$tzner}}, \bibinfo {author} {\bibfnamefont {M.}~\bibnamefont
  {Karny}}, \bibinfo {author} {\bibfnamefont {L.~V.}\ \bibnamefont
  {Grigorenko}},\ and\ \bibinfo {author} {\bibfnamefont {K.}~\bibnamefont
  {Riisager}},\ }\href {https://doi.org/10.1103/RevModPhys.84.567} {\bibfield
  {journal} {\bibinfo  {journal} {Rev. Mod. Phys.}\ }\textbf {\bibinfo {volume}
  {84}},\ \bibinfo {pages} {567} (\bibinfo {year} {2012})}\BibitemShut
  {NoStop}%
\bibitem [{\citenamefont {Ahn}\ \emph {et~al.}(2019)\citenamefont {Ahn},
  \citenamefont {Fukuda}, \citenamefont {Geissel}, \citenamefont {Inabe},
  \citenamefont {Iwasa}, \citenamefont {Kubo}, \citenamefont {Kusaka},
  \citenamefont {Morrissey}, \citenamefont {Murai}, \citenamefont {Nakamura},
  \citenamefont {Ohtake}, \citenamefont {Otsu}, \citenamefont {Sato},
  \citenamefont {Sherrill}, \citenamefont {Shimizu}, \citenamefont {Suzuki},
  \citenamefont {Takeda}, \citenamefont {Tarasov}, \citenamefont {Ueno},
  \citenamefont {Yanagisawa},\ and\ \citenamefont
  {Yoshida}}]{Ahn2019_PRL123-212501}%
  \BibitemOpen
  \bibfield  {author} {\bibinfo {author} {\bibfnamefont {D.~S.}\ \bibnamefont
  {Ahn}}, \bibinfo {author} {\bibfnamefont {N.}~\bibnamefont {Fukuda}},
  \bibinfo {author} {\bibfnamefont {H.}~\bibnamefont {Geissel}}, \bibinfo
  {author} {\bibfnamefont {N.}~\bibnamefont {Inabe}}, \bibinfo {author}
  {\bibfnamefont {N.}~\bibnamefont {Iwasa}}, \bibinfo {author} {\bibfnamefont
  {T.}~\bibnamefont {Kubo}}, \bibinfo {author} {\bibfnamefont {K.}~\bibnamefont
  {Kusaka}}, \bibinfo {author} {\bibfnamefont {D.~J.}\ \bibnamefont
  {Morrissey}}, \bibinfo {author} {\bibfnamefont {D.}~\bibnamefont {Murai}},
  \bibinfo {author} {\bibfnamefont {T.}~\bibnamefont {Nakamura}}, \bibinfo
  {author} {\bibfnamefont {M.}~\bibnamefont {Ohtake}}, \bibinfo {author}
  {\bibfnamefont {H.}~\bibnamefont {Otsu}}, \bibinfo {author} {\bibfnamefont
  {H.}~\bibnamefont {Sato}}, \bibinfo {author} {\bibfnamefont {B.~M.}\
  \bibnamefont {Sherrill}}, \bibinfo {author} {\bibfnamefont {Y.}~\bibnamefont
  {Shimizu}}, \bibinfo {author} {\bibfnamefont {H.}~\bibnamefont {Suzuki}},
  \bibinfo {author} {\bibfnamefont {H.}~\bibnamefont {Takeda}}, \bibinfo
  {author} {\bibfnamefont {O.~B.}\ \bibnamefont {Tarasov}}, \bibinfo {author}
  {\bibfnamefont {H.}~\bibnamefont {Ueno}}, \bibinfo {author} {\bibfnamefont
  {Y.}~\bibnamefont {Yanagisawa}},\ and\ \bibinfo {author} {\bibfnamefont
  {K.}~\bibnamefont {Yoshida}},\ }\href
  {https://doi.org/10.1103/PhysRevLett.123.212501} {\bibfield  {journal}
  {\bibinfo  {journal} {Phys. Rev. Lett.}\ }\textbf {\bibinfo {volume} {123}},\
  \bibinfo {pages} {212501} (\bibinfo {year} {2019})}\BibitemShut {NoStop}%
\bibitem [{\citenamefont {Tanihata}\ \emph {et~al.}(2013)\citenamefont
  {Tanihata}, \citenamefont {Savajols},\ and\ \citenamefont
  {Kanungo}}]{Tanihata2013_PPNP68-215}%
  \BibitemOpen
  \bibfield  {author} {\bibinfo {author} {\bibfnamefont {I.}~\bibnamefont
  {Tanihata}}, \bibinfo {author} {\bibfnamefont {H.}~\bibnamefont {Savajols}},\
  and\ \bibinfo {author} {\bibfnamefont {R.}~\bibnamefont {Kanungo}},\ }\href
  {https://doi.org/10.1016/j.ppnp.2012.07.001} {\bibfield  {journal} {\bibinfo
  {journal} {Prog. Part. Nucl. Phys.}\ }\textbf {\bibinfo {volume} {68}},\
  \bibinfo {pages} {215} (\bibinfo {year} {2013})}\BibitemShut {NoStop}%
\bibitem [{\citenamefont {Skaza}\ \emph {et~al.}(2006)\citenamefont {Skaza},
  \citenamefont {Lapoux}, \citenamefont {Keeley}, \citenamefont {Alamanos},
  \citenamefont {Pollacco}, \citenamefont {Auger}, \citenamefont {Drouart},
  \citenamefont {Gillibert}, \citenamefont {Beaumel}, \citenamefont {Becheva},
  \citenamefont {Blumenfeld}, \citenamefont {Delaunay}, \citenamefont {Giot},
  \citenamefont {Kemper}, \citenamefont {Nalpas}, \citenamefont {Obertelli},
  \citenamefont {Pakou}, \citenamefont {Raabe}, \citenamefont {Roussel-Chomaz},
  \citenamefont {Sida}, \citenamefont {Scarpaci}, \citenamefont {Stepantsov},\
  and\ \citenamefont {Wolski}}]{Skaza2006_PRC73-044301}%
  \BibitemOpen
  \bibfield  {author} {\bibinfo {author} {\bibfnamefont {F.}~\bibnamefont
  {Skaza}}, \bibinfo {author} {\bibfnamefont {V.}~\bibnamefont {Lapoux}},
  \bibinfo {author} {\bibfnamefont {N.}~\bibnamefont {Keeley}}, \bibinfo
  {author} {\bibfnamefont {N.}~\bibnamefont {Alamanos}}, \bibinfo {author}
  {\bibfnamefont {E.~C.}\ \bibnamefont {Pollacco}}, \bibinfo {author}
  {\bibfnamefont {F.}~\bibnamefont {Auger}}, \bibinfo {author} {\bibfnamefont
  {A.}~\bibnamefont {Drouart}}, \bibinfo {author} {\bibfnamefont
  {A.}~\bibnamefont {Gillibert}}, \bibinfo {author} {\bibfnamefont
  {D.}~\bibnamefont {Beaumel}}, \bibinfo {author} {\bibfnamefont
  {E.}~\bibnamefont {Becheva}}, \bibinfo {author} {\bibfnamefont
  {Y.}~\bibnamefont {Blumenfeld}}, \bibinfo {author} {\bibfnamefont
  {F.}~\bibnamefont {Delaunay}}, \bibinfo {author} {\bibfnamefont
  {L.}~\bibnamefont {Giot}}, \bibinfo {author} {\bibfnamefont {K.~W.}\
  \bibnamefont {Kemper}}, \bibinfo {author} {\bibfnamefont {L.}~\bibnamefont
  {Nalpas}}, \bibinfo {author} {\bibfnamefont {A.}~\bibnamefont {Obertelli}},
  \bibinfo {author} {\bibfnamefont {A.}~\bibnamefont {Pakou}}, \bibinfo
  {author} {\bibfnamefont {R.}~\bibnamefont {Raabe}}, \bibinfo {author}
  {\bibfnamefont {P.}~\bibnamefont {Roussel-Chomaz}}, \bibinfo {author}
  {\bibfnamefont {J.-L.}\ \bibnamefont {Sida}}, \bibinfo {author}
  {\bibfnamefont {J.-A.}\ \bibnamefont {Scarpaci}}, \bibinfo {author}
  {\bibfnamefont {S.}~\bibnamefont {Stepantsov}},\ and\ \bibinfo {author}
  {\bibfnamefont {R.}~\bibnamefont {Wolski}},\ }\href
  {https://doi.org/10.1103/PhysRevC.73.044301} {\bibfield  {journal} {\bibinfo
  {journal} {Phys. Rev. C}\ }\textbf {\bibinfo {volume} {73}},\ \bibinfo
  {pages} {044301} (\bibinfo {year} {2006})}\BibitemShut {NoStop}%
\bibitem [{\citenamefont {Otsuka}\ \emph {et~al.}(2001)\citenamefont {Otsuka},
  \citenamefont {Fujimoto}, \citenamefont {Utsuno}, \citenamefont {Brown},
  \citenamefont {Honma},\ and\ \citenamefont
  {Mizusaki}}]{Otsuka2001_PRL87-082502}%
  \BibitemOpen
  \bibfield  {author} {\bibinfo {author} {\bibfnamefont {T.}~\bibnamefont
  {Otsuka}}, \bibinfo {author} {\bibfnamefont {R.}~\bibnamefont {Fujimoto}},
  \bibinfo {author} {\bibfnamefont {Y.}~\bibnamefont {Utsuno}}, \bibinfo
  {author} {\bibfnamefont {B.~A.}\ \bibnamefont {Brown}}, \bibinfo {author}
  {\bibfnamefont {M.}~\bibnamefont {Honma}},\ and\ \bibinfo {author}
  {\bibfnamefont {T.}~\bibnamefont {Mizusaki}},\ }\href
  {https://doi.org/10.1103/PhysRevLett.87.082502} {\bibfield  {journal}
  {\bibinfo  {journal} {Phys. Rev. Lett.}\ }\textbf {\bibinfo {volume} {87}},\
  \bibinfo {pages} {082502} (\bibinfo {year} {2001})}\BibitemShut {NoStop}%
\bibitem [{\citenamefont {Sorlin}\ and\ \citenamefont
  {Porquet}(2008)}]{Sorlin2008_PPNP61-602}%
  \BibitemOpen
  \bibfield  {author} {\bibinfo {author} {\bibfnamefont {O.}~\bibnamefont
  {Sorlin}}\ and\ \bibinfo {author} {\bibfnamefont {M.-G.}\ \bibnamefont
  {Porquet}},\ }\href {https://doi.org/10.1016/j.ppnp.2008.05.001} {\bibfield
  {journal} {\bibinfo  {journal} {Prog. Part. Nucl. Phys.}\ }\textbf {\bibinfo
  {volume} {61}},\ \bibinfo {pages} {602} (\bibinfo {year} {2008})}\BibitemShut
  {NoStop}%
\bibitem [{\citenamefont {{D.T. Tran, H.J. Ong, G. Hagen \emph{et
  al.}}}(2018)}]{Tran2018_NutureCommu9-1594}%
  \BibitemOpen
  \bibfield  {author} {\bibinfo {author} {\bibnamefont {{D.T. Tran, H.J. Ong,
  G. Hagen \emph{et al.}}}},\ }\href
  {https://doi.org/10.1038/s41467-018-04024-y} {\bibfield  {journal} {\bibinfo
  {journal} {Nature Commu.}\ }\textbf {\bibinfo {volume} {9}},\ \bibinfo
  {pages} {1594} (\bibinfo {year} {2018})}\BibitemShut {NoStop}%
\bibitem [{\citenamefont {{N. Imai, N. Aoi, H.J. Ong \emph{et
  al.}}}(2009)}]{Imai2009_PLB673-179}%
  \BibitemOpen
  \bibfield  {author} {\bibinfo {author} {\bibnamefont {{N. Imai, N. Aoi, H.J.
  Ong \emph{et al.}}}},\ }\href
  {https://doi.org/https://doi.org/10.1016/j.physletb.2009.02.039} {\bibfield
  {journal} {\bibinfo  {journal} {Phys. Lett. B}\ }\textbf {\bibinfo {volume}
  {673}},\ \bibinfo {pages} {179} (\bibinfo {year} {2009})}\BibitemShut
  {NoStop}%
\bibitem [{\citenamefont {Krieger}\ \emph {et~al.}(2012)\citenamefont
  {Krieger}, \citenamefont {Blaum}, \citenamefont {Bissell}, \citenamefont
  {Fr$\ddot{\rm o}$mmgen}, \citenamefont {Geppert}, \citenamefont {Hammen},
  \citenamefont {Kreim}, \citenamefont {Kowalska}, \citenamefont {Kr$\ddot{\rm
  a}$mer}, \citenamefont {Neff}, \citenamefont {Neugart}, \citenamefont
  {Neyens}, \citenamefont {N\"ortersh$\ddot{\rm a}$user}, \citenamefont
  {Novotny}, \citenamefont {S\'anchez},\ and\ \citenamefont
  {Yordanov}}]{Krieger2012_PRL108-142501}%
  \BibitemOpen
  \bibfield  {author} {\bibinfo {author} {\bibfnamefont {A.}~\bibnamefont
  {Krieger}}, \bibinfo {author} {\bibfnamefont {K.}~\bibnamefont {Blaum}},
  \bibinfo {author} {\bibfnamefont {M.~L.}\ \bibnamefont {Bissell}}, \bibinfo
  {author} {\bibfnamefont {N.}~\bibnamefont {Fr$\ddot{\rm o}$mmgen}}, \bibinfo
  {author} {\bibfnamefont {C.}~\bibnamefont {Geppert}}, \bibinfo {author}
  {\bibfnamefont {M.}~\bibnamefont {Hammen}}, \bibinfo {author} {\bibfnamefont
  {K.}~\bibnamefont {Kreim}}, \bibinfo {author} {\bibfnamefont
  {M.}~\bibnamefont {Kowalska}}, \bibinfo {author} {\bibfnamefont
  {J.}~\bibnamefont {Kr$\ddot{\rm a}$mer}}, \bibinfo {author} {\bibfnamefont
  {T.}~\bibnamefont {Neff}}, \bibinfo {author} {\bibfnamefont {R.}~\bibnamefont
  {Neugart}}, \bibinfo {author} {\bibfnamefont {G.}~\bibnamefont {Neyens}},
  \bibinfo {author} {\bibfnamefont {W.}~\bibnamefont {N\"ortersh$\ddot{\rm
  a}$user}}, \bibinfo {author} {\bibfnamefont {C.}~\bibnamefont {Novotny}},
  \bibinfo {author} {\bibfnamefont {R.}~\bibnamefont {S\'anchez}},\ and\
  \bibinfo {author} {\bibfnamefont {D.~T.}\ \bibnamefont {Yordanov}},\ }\href
  {https://doi.org/10.1103/PhysRevLett.108.142501} {\bibfield  {journal}
  {\bibinfo  {journal} {Phys. Rev. Lett.}\ }\textbf {\bibinfo {volume} {108}},\
  \bibinfo {pages} {142501} (\bibinfo {year} {2012})}\BibitemShut {NoStop}%
\bibitem [{\citenamefont {Pain}\ \emph {et~al.}(2006)\citenamefont {Pain},
  \citenamefont {Catford}, \citenamefont {Orr}, \citenamefont {Ang\'elique},
  \citenamefont {Ashwood}, \citenamefont {Bouchat}, \citenamefont {Clarke},
  \citenamefont {Curtis}, \citenamefont {Freer}, \citenamefont {Fulton},
  \citenamefont {Hanappe}, \citenamefont {Labiche}, \citenamefont {Lecouey},
  \citenamefont {Lemmon}, \citenamefont {Mahboub}, \citenamefont {Ninane},
  \citenamefont {Normand}, \citenamefont {Soi\ifmmode~\acute{c}\else
  \'{c}\fi{}}, \citenamefont {Stuttge}, \citenamefont {Timis}, \citenamefont
  {Tostevin}, \citenamefont {Winfield},\ and\ \citenamefont
  {Ziman}}]{Pain2006_PRL96-032502}%
  \BibitemOpen
  \bibfield  {author} {\bibinfo {author} {\bibfnamefont {S.~D.}\ \bibnamefont
  {Pain}}, \bibinfo {author} {\bibfnamefont {W.~N.}\ \bibnamefont {Catford}},
  \bibinfo {author} {\bibfnamefont {N.~A.}\ \bibnamefont {Orr}}, \bibinfo
  {author} {\bibfnamefont {J.~C.}\ \bibnamefont {Ang\'elique}}, \bibinfo
  {author} {\bibfnamefont {N.~I.}\ \bibnamefont {Ashwood}}, \bibinfo {author}
  {\bibfnamefont {V.}~\bibnamefont {Bouchat}}, \bibinfo {author} {\bibfnamefont
  {N.~M.}\ \bibnamefont {Clarke}}, \bibinfo {author} {\bibfnamefont
  {N.}~\bibnamefont {Curtis}}, \bibinfo {author} {\bibfnamefont
  {M.}~\bibnamefont {Freer}}, \bibinfo {author} {\bibfnamefont {B.~R.}\
  \bibnamefont {Fulton}}, \bibinfo {author} {\bibfnamefont {F.}~\bibnamefont
  {Hanappe}}, \bibinfo {author} {\bibfnamefont {M.}~\bibnamefont {Labiche}},
  \bibinfo {author} {\bibfnamefont {J.~L.}\ \bibnamefont {Lecouey}}, \bibinfo
  {author} {\bibfnamefont {R.~C.}\ \bibnamefont {Lemmon}}, \bibinfo {author}
  {\bibfnamefont {D.}~\bibnamefont {Mahboub}}, \bibinfo {author} {\bibfnamefont
  {A.}~\bibnamefont {Ninane}}, \bibinfo {author} {\bibfnamefont
  {G.}~\bibnamefont {Normand}}, \bibinfo {author} {\bibfnamefont
  {N.}~\bibnamefont {Soi\ifmmode~\acute{c}\else \'{c}\fi{}}}, \bibinfo {author}
  {\bibfnamefont {L.}~\bibnamefont {Stuttge}}, \bibinfo {author} {\bibfnamefont
  {C.~N.}\ \bibnamefont {Timis}}, \bibinfo {author} {\bibfnamefont {J.~A.}\
  \bibnamefont {Tostevin}}, \bibinfo {author} {\bibfnamefont {J.~S.}\
  \bibnamefont {Winfield}},\ and\ \bibinfo {author} {\bibfnamefont
  {V.}~\bibnamefont {Ziman}},\ }\href
  {https://doi.org/10.1103/PhysRevLett.96.032502} {\bibfield  {journal}
  {\bibinfo  {journal} {Phys. Rev. Lett.}\ }\textbf {\bibinfo {volume} {96}},\
  \bibinfo {pages} {032502} (\bibinfo {year} {2006})}\BibitemShut {NoStop}%
\bibitem [{\citenamefont {Schunck}(2019)}]{Schunck2019_EDF-Nuclei}%
  \BibitemOpen
  \bibinfo {editor} {\bibfnamefont {N.}~\bibnamefont {Schunck}},\ ed.,\ \href
  {https://doi.org/10.1088/2053-2563/aae0ed} {\emph {\bibinfo {title} {Energy
  Density Functional Methods for Atomic Nuclei}}},\ {IOP} Expanding Physics\
  (\bibinfo  {publisher} {{IOP} Publishing},\ \bibinfo {year}
  {2019})\BibitemShut {NoStop}%
\bibitem [{\citenamefont {Meng}\ \emph {et~al.}(2006)\citenamefont {Meng},
  \citenamefont {Toki}, \citenamefont {Zhou}, \citenamefont {Zhang},
  \citenamefont {Long},\ and\ \citenamefont {Geng}}]{Meng2006_PPNP57-470}%
  \BibitemOpen
  \bibfield  {author} {\bibinfo {author} {\bibfnamefont {J.}~\bibnamefont
  {Meng}}, \bibinfo {author} {\bibfnamefont {H.}~\bibnamefont {Toki}}, \bibinfo
  {author} {\bibfnamefont {S.~G.}\ \bibnamefont {Zhou}}, \bibinfo {author}
  {\bibfnamefont {S.~Q.}\ \bibnamefont {Zhang}}, \bibinfo {author}
  {\bibfnamefont {W.~H.}\ \bibnamefont {Long}},\ and\ \bibinfo {author}
  {\bibfnamefont {L.~S.}\ \bibnamefont {Geng}},\ }\href
  {https://doi.org/10.1016/j.ppnp.2005.06.001} {\bibfield  {journal} {\bibinfo
  {journal} {Prog. Part. Nucl. Phys.}\ }\textbf {\bibinfo {volume} {57}},\
  \bibinfo {pages} {470} (\bibinfo {year} {2006})}\BibitemShut {NoStop}%
\bibitem [{\citenamefont {Meng}(2016)}]{Meng2016_RDFNS}%
  \BibitemOpen
  \bibinfo {editor} {\bibfnamefont {J.}~\bibnamefont {Meng}},\ ed.,\ \href
  {https://doi.org/10.1142/9872} {\emph {\bibinfo {title} {Relativistic
  {D}ensity {F}unctional for {N}uclear {S}tructure}}},\ \bibinfo {edition}
  {{V}ol. 10 of {I}nternational {R}eview of {N}uclear {P}hysics}\ ed.\
  (\bibinfo  {publisher} {World Scientific Pub Co Pte Lt},\ \bibinfo {year}
  {2016})\BibitemShut {NoStop}%
\bibitem [{\citenamefont {Meng}\ and\ \citenamefont
  {Ring}(1996)}]{Meng1996_PRL77-3963}%
  \BibitemOpen
  \bibfield  {author} {\bibinfo {author} {\bibfnamefont {J.}~\bibnamefont
  {Meng}}\ and\ \bibinfo {author} {\bibfnamefont {P.}~\bibnamefont {Ring}},\
  }\href {https://doi.org/10.1103/PhysRevLett.77.3963} {\bibfield  {journal}
  {\bibinfo  {journal} {Phys. Rev. Lett.}\ }\textbf {\bibinfo {volume} {77}},\
  \bibinfo {pages} {3963} (\bibinfo {year} {1996})}\BibitemShut {NoStop}%
\bibitem [{\citenamefont {Zhu}\ \emph {et~al.}(1994)\citenamefont {Zhu},
  \citenamefont {Shen}, \citenamefont {Cai},\ and\ \citenamefont
  {Ma}}]{Zhu1994_PLB328-1}%
  \BibitemOpen
  \bibfield  {author} {\bibinfo {author} {\bibfnamefont {Z.}~\bibnamefont
  {Zhu}}, \bibinfo {author} {\bibfnamefont {W.}~\bibnamefont {Shen}}, \bibinfo
  {author} {\bibfnamefont {Y.}~\bibnamefont {Cai}},\ and\ \bibinfo {author}
  {\bibfnamefont {Y.}~\bibnamefont {Ma}},\ }\href
  {https://doi.org/10.1016/0370-2693(94)90418-9} {\bibfield  {journal}
  {\bibinfo  {journal} {Phys, Lett. B}\ }\textbf {\bibinfo {volume} {328}},\
  \bibinfo {pages} {1} (\bibinfo {year} {1994})}\BibitemShut {NoStop}%
\bibitem [{\citenamefont {Arumugam}\ \emph {et~al.}(2005)\citenamefont
  {Arumugam}, \citenamefont {Sharma}, \citenamefont {Patra},\ and\
  \citenamefont {Gupta}}]{Arumugam2005_PRC71-064308}%
  \BibitemOpen
  \bibfield  {author} {\bibinfo {author} {\bibfnamefont {P.}~\bibnamefont
  {Arumugam}}, \bibinfo {author} {\bibfnamefont {B.~K.}\ \bibnamefont
  {Sharma}}, \bibinfo {author} {\bibfnamefont {S.~K.}\ \bibnamefont {Patra}},\
  and\ \bibinfo {author} {\bibfnamefont {R.~K.}\ \bibnamefont {Gupta}},\ }\href
  {https://doi.org/10.1103/PhysRevC.71.064308} {\bibfield  {journal} {\bibinfo
  {journal} {Phys. Rev. C}\ }\textbf {\bibinfo {volume} {71}},\ \bibinfo
  {pages} {064308} (\bibinfo {year} {2005})}\BibitemShut {NoStop}%
\bibitem [{\citenamefont {Lu}\ \emph {et~al.}(2011)\citenamefont {Lu},
  \citenamefont {Zhao},\ and\ \citenamefont {Zhou}}]{Lu2011_PRC84-014328}%
  \BibitemOpen
  \bibfield  {author} {\bibinfo {author} {\bibfnamefont {B.-N.}\ \bibnamefont
  {Lu}}, \bibinfo {author} {\bibfnamefont {E.-G.}\ \bibnamefont {Zhao}},\ and\
  \bibinfo {author} {\bibfnamefont {S.-G.}\ \bibnamefont {Zhou}},\ }\href
  {https://doi.org/10.1103/PhysRevC.84.014328} {\bibfield  {journal} {\bibinfo
  {journal} {Phys. Rev. C}\ }\textbf {\bibinfo {volume} {84}},\ \bibinfo
  {pages} {014328} (\bibinfo {year} {2011})}\BibitemShut {NoStop}%
\bibitem [{\citenamefont {Tang}\ \emph {et~al.}(2013)\citenamefont {Tang},
  \citenamefont {Li}, \citenamefont {Ji},\ and\ \citenamefont
  {Zhou}}]{Tang2013_CPL30-012101}%
  \BibitemOpen
  \bibfield  {author} {\bibinfo {author} {\bibfnamefont {Z.-H.}\ \bibnamefont
  {Tang}}, \bibinfo {author} {\bibfnamefont {J.-X.}\ \bibnamefont {Li}},
  \bibinfo {author} {\bibfnamefont {J.-X.}\ \bibnamefont {Ji}},\ and\ \bibinfo
  {author} {\bibfnamefont {T.}~\bibnamefont {Zhou}},\ }\href
  {https://doi.org/10.1088/0256-307x/30/1/012101} {\bibfield  {journal}
  {\bibinfo  {journal} {Chin. Phys. Lett.}\ }\textbf {\bibinfo {volume} {30}},\
  \bibinfo {pages} {012101} (\bibinfo {year} {2013})}\BibitemShut {NoStop}%
\bibitem [{\citenamefont {Sun}\ \emph {et~al.}(2018)\citenamefont {Sun},
  \citenamefont {Zhao},\ and\ \citenamefont {Zhou}}]{Sun2018_PLB785-530}%
  \BibitemOpen
  \bibfield  {author} {\bibinfo {author} {\bibfnamefont {X.-X.}\ \bibnamefont
  {Sun}}, \bibinfo {author} {\bibfnamefont {J.}~\bibnamefont {Zhao}},\ and\
  \bibinfo {author} {\bibfnamefont {S.-G.}\ \bibnamefont {Zhou}},\ }\href
  {https://doi.org/https://doi.org/10.1016/j.physletb.2018.08.071} {\bibfield
  {journal} {\bibinfo  {journal} {Phys. Lett. B}\ }\textbf {\bibinfo {volume}
  {785}},\ \bibinfo {pages} {530 } (\bibinfo {year} {2018})}\BibitemShut
  {NoStop}%
\bibitem [{\citenamefont {{Z. H. Yang, Y. Kubota, A. Corsi \emph{et
  al.}}}(2021)}]{Yang2021_PRL126-082501}%
  \BibitemOpen
  \bibfield  {author} {\bibinfo {author} {\bibnamefont {{Z. H. Yang, Y. Kubota,
  A. Corsi \emph{et al.}}}},\ }\href
  {https://doi.org/10.1103/PhysRevLett.126.082501} {\bibfield  {journal}
  {\bibinfo  {journal} {Phys. Rev. Lett.}\ }\textbf {\bibinfo {volume} {126}},\
  \bibinfo {pages} {082501} (\bibinfo {year} {2021})}\BibitemShut {NoStop}%
\bibitem [{\citenamefont {Wang}\ and\ \citenamefont
  {Lu}(2022)}]{Wang2022_CTP74-015303}%
  \BibitemOpen
  \bibfield  {author} {\bibinfo {author} {\bibfnamefont {K.}~\bibnamefont
  {Wang}}\ and\ \bibinfo {author} {\bibfnamefont {B.}~\bibnamefont {Lu}},\
  }\href {https://doi.org/10.1088/1572-9494/ac3999} {\bibfield  {journal}
  {\bibinfo  {journal} {Commun. Theor. Phys.}\ }\textbf {\bibinfo {volume}
  {74}},\ \bibinfo {pages} {015303} (\bibinfo {year} {2022})}\BibitemShut
  {NoStop}%
\bibitem [{\citenamefont {Lu}\ \emph {et~al.}(2022)\citenamefont {Lu},
  \citenamefont {Wang}, \citenamefont {Xiao}, \citenamefont {Geng},
  \citenamefont {Meng},\ and\ \citenamefont {Ring}}]{Lu2022_PRL128-142002}%
  \BibitemOpen
  \bibfield  {author} {\bibinfo {author} {\bibfnamefont {J.-X.}\ \bibnamefont
  {Lu}}, \bibinfo {author} {\bibfnamefont {C.-X.}\ \bibnamefont {Wang}},
  \bibinfo {author} {\bibfnamefont {Y.}~\bibnamefont {Xiao}}, \bibinfo {author}
  {\bibfnamefont {L.-S.}\ \bibnamefont {Geng}}, \bibinfo {author}
  {\bibfnamefont {J.}~\bibnamefont {Meng}},\ and\ \bibinfo {author}
  {\bibfnamefont {P.}~\bibnamefont {Ring}},\ }\href
  {https://doi.org/10.1103/PhysRevLett.128.142002} {\bibfield  {journal}
  {\bibinfo  {journal} {Phys. Rev. Lett.}\ }\textbf {\bibinfo {volume} {128}},\
  \bibinfo {pages} {142002} (\bibinfo {year} {2022})}\BibitemShut {NoStop}%
\bibitem [{\citenamefont {Ren}\ \emph {et~al.}(2018)\citenamefont {Ren},
  \citenamefont {Li}, \citenamefont {Geng}, \citenamefont {Long}, \citenamefont
  {Ring},\ and\ \citenamefont {Meng}}]{Ren2018_ChinPhysC42-14103}%
  \BibitemOpen
  \bibfield  {author} {\bibinfo {author} {\bibfnamefont {X.-L.}\ \bibnamefont
  {Ren}}, \bibinfo {author} {\bibfnamefont {K.-W.}\ \bibnamefont {Li}},
  \bibinfo {author} {\bibfnamefont {L.-S.}\ \bibnamefont {Geng}}, \bibinfo
  {author} {\bibfnamefont {B.}~\bibnamefont {Long}}, \bibinfo {author}
  {\bibfnamefont {P.}~\bibnamefont {Ring}},\ and\ \bibinfo {author}
  {\bibfnamefont {J.}~\bibnamefont {Meng}},\ }\href
  {https://doi.org/10.1088/1674-1137/42/1/014103} {\bibfield  {journal}
  {\bibinfo  {journal} {Chin. Phys. C}\ }\textbf {\bibinfo {volume} {42}},\
  \bibinfo {pages} {014103} (\bibinfo {year} {2018})}\BibitemShut {NoStop}%
\bibitem [{\citenamefont {Zhao}\ \emph {et~al.}(2017)\citenamefont {Zhao},
  \citenamefont {Lu}, \citenamefont {Zhao},\ and\ \citenamefont
  {Zhou}}]{Zhao2017_PRC95-014320}%
  \BibitemOpen
  \bibfield  {author} {\bibinfo {author} {\bibfnamefont {J.}~\bibnamefont
  {Zhao}}, \bibinfo {author} {\bibfnamefont {B.-N.}\ \bibnamefont {Lu}},
  \bibinfo {author} {\bibfnamefont {E.-G.}\ \bibnamefont {Zhao}},\ and\
  \bibinfo {author} {\bibfnamefont {S.-G.}\ \bibnamefont {Zhou}},\ }\href
  {https://doi.org/10.1103/PhysRevC.95.014320} {\bibfield  {journal} {\bibinfo
  {journal} {Phys. Rev. C}\ }\textbf {\bibinfo {volume} {95}},\ \bibinfo
  {pages} {014320} (\bibinfo {year} {2017})}\BibitemShut {NoStop}%
\bibitem [{\citenamefont {Zhou}(2016)}]{Zhou2016_PS91-063008}%
  \BibitemOpen
  \bibfield  {author} {\bibinfo {author} {\bibfnamefont {S.-G.}\ \bibnamefont
  {Zhou}},\ }\href {https://doi.org/10.1088/0031-8949/91/6/063008} {\bibfield
  {journal} {\bibinfo  {journal} {Phys. Scr.}\ }\textbf {\bibinfo {volume}
  {91}},\ \bibinfo {pages} {063008} (\bibinfo {year} {2016})}\BibitemShut
  {NoStop}%
\bibitem [{\citenamefont {Long}\ \emph {et~al.}(2004)\citenamefont {Long},
  \citenamefont {Meng}, \citenamefont {{N. Van Giai}},\ and\ \citenamefont
  {Zhou}}]{Long2004_PRC69-034319}%
  \BibitemOpen
  \bibfield  {author} {\bibinfo {author} {\bibfnamefont {W.}~\bibnamefont
  {Long}}, \bibinfo {author} {\bibfnamefont {J.}~\bibnamefont {Meng}}, \bibinfo
  {author} {\bibnamefont {{N. Van Giai}}},\ and\ \bibinfo {author}
  {\bibfnamefont {S.~G.}\ \bibnamefont {Zhou}},\ }\href
  {https://doi.org/10.1103/PhysRevC.69.034319} {\bibfield  {journal} {\bibinfo
  {journal} {Phys. Rev. C}\ }\textbf {\bibinfo {volume} {69}},\ \bibinfo
  {pages} {034319} (\bibinfo {year} {2004})}\BibitemShut {NoStop}%
\bibitem [{\citenamefont {Ring}\ and\ \citenamefont {Schuck}(1980)}]{Ring1980}%
  \BibitemOpen
  \bibfield  {author} {\bibinfo {author} {\bibfnamefont {P.}~\bibnamefont
  {Ring}}\ and\ \bibinfo {author} {\bibfnamefont {P.}~\bibnamefont {Schuck}},\
  }\href@noop {} {\emph {\bibinfo {title} {The {N}uclear {M}any-{B}ody
  {P}roblem}}}\ (\bibinfo  {publisher} {Springer-{V}erlag {B}erlin
  {H}eidelberg},\ \bibinfo {year} {1980})\BibitemShut {NoStop}%
\bibitem [{\citenamefont {Tian}\ \emph {et~al.}(2009)\citenamefont {Tian},
  \citenamefont {Ma},\ and\ \citenamefont {Ring}}]{Tian2009_PLB676-44}%
  \BibitemOpen
  \bibfield  {author} {\bibinfo {author} {\bibfnamefont {Y.}~\bibnamefont
  {Tian}}, \bibinfo {author} {\bibfnamefont {Z.~Y.}\ \bibnamefont {Ma}},\ and\
  \bibinfo {author} {\bibfnamefont {P.}~\bibnamefont {Ring}},\ }\href
  {https://doi.org/10.1016/j.physletb.2009.04.067} {\bibfield  {journal}
  {\bibinfo  {journal} {Phys. Lett. B}\ }\textbf {\bibinfo {volume} {676}},\
  \bibinfo {pages} {44} (\bibinfo {year} {2009})}\BibitemShut {NoStop}%
\bibitem [{\citenamefont {Bender}\ \emph {et~al.}(2000)\citenamefont {Bender},
  \citenamefont {Rutz}, \citenamefont {Reinhard},\ and\ \citenamefont
  {Maruhn}}]{Bender2000_EPJA7-467}%
  \BibitemOpen
  \bibfield  {author} {\bibinfo {author} {\bibfnamefont {M.}~\bibnamefont
  {Bender}}, \bibinfo {author} {\bibfnamefont {K.}~\bibnamefont {Rutz}},
  \bibinfo {author} {\bibfnamefont {P.-G.}\ \bibnamefont {Reinhard}},\ and\
  \bibinfo {author} {\bibfnamefont {J.~A.}\ \bibnamefont {Maruhn}},\ }\href
  {https://doi.org/10.1007/PL00013645} {\bibfield  {journal} {\bibinfo
  {journal} {Eur. Phys. J. A}\ }\textbf {\bibinfo {volume} {7}},\ \bibinfo
  {pages} {467} (\bibinfo {year} {2000})}\BibitemShut {NoStop}%
\bibitem [{\citenamefont {Inglis}(1956)}]{Inglis1956_PR103-1786}%
  \BibitemOpen
  \bibfield  {author} {\bibinfo {author} {\bibfnamefont {D.~R.}\ \bibnamefont
  {Inglis}},\ }\href {https://doi.org/10.1103/PhysRev.103.1786} {\bibfield
  {journal} {\bibinfo  {journal} {Phys. Rev.}\ }\textbf {\bibinfo {volume}
  {103}},\ \bibinfo {pages} {1786} (\bibinfo {year} {1956})}\BibitemShut
  {NoStop}%
\bibitem [{\citenamefont {Beliaev}(1961)}]{Beliaev1961_NP24-322}%
  \BibitemOpen
  \bibfield  {author} {\bibinfo {author} {\bibfnamefont {S.}~\bibnamefont
  {Beliaev}},\ }\href {https://doi.org/10.1016/0029-5582(61)90384-4} {\bibfield
   {journal} {\bibinfo  {journal} {Nucl. Phys.}\ }\textbf {\bibinfo {volume}
  {24}},\ \bibinfo {pages} {322} (\bibinfo {year} {1961})}\BibitemShut
  {NoStop}%
\bibitem [{\citenamefont {Sugahara}\ and\ \citenamefont
  {Toki}(1994)}]{Sugahara1994_NPA579-557}%
  \BibitemOpen
  \bibfield  {author} {\bibinfo {author} {\bibfnamefont {Y.}~\bibnamefont
  {Sugahara}}\ and\ \bibinfo {author} {\bibfnamefont {H.}~\bibnamefont
  {Toki}},\ }\href {https://doi.org/10.1016/0375-9474(94)90923-7} {\bibfield
  {journal} {\bibinfo  {journal} {Nucl. Phys. A}\ }\textbf {\bibinfo {volume}
  {579}},\ \bibinfo {pages} {557} (\bibinfo {year} {1994})}\BibitemShut
  {NoStop}%
\bibitem [{\citenamefont {Horowitz}\ and\ \citenamefont
  {Piekarewicz}(2012)}]{Horowitz2012_PRC86-045503}%
  \BibitemOpen
  \bibfield  {author} {\bibinfo {author} {\bibfnamefont {C.~J.}\ \bibnamefont
  {Horowitz}}\ and\ \bibinfo {author} {\bibfnamefont {J.}~\bibnamefont
  {Piekarewicz}},\ }\href {https://doi.org/10.1103/PhysRevC.86.045503}
  {\bibfield  {journal} {\bibinfo  {journal} {Phys. Rev. C}\ }\textbf {\bibinfo
  {volume} {86}},\ \bibinfo {pages} {045503} (\bibinfo {year}
  {2012})}\BibitemShut {NoStop}%
\bibitem [{\citenamefont {Kurasawa}\ and\ \citenamefont
  {Suzuki}(2019)}]{Kurasawa2019_PTEP2019-113D01}%
  \BibitemOpen
  \bibfield  {author} {\bibinfo {author} {\bibfnamefont {H.}~\bibnamefont
  {Kurasawa}}\ and\ \bibinfo {author} {\bibfnamefont {T.}~\bibnamefont
  {Suzuki}},\ }\href {https://doi.org/10.1093/ptep/ptz121} {\bibfield
  {journal} {\bibinfo  {journal} {Prog. Theo. Exp. Phys.}\ }\textbf {\bibinfo
  {volume} {2019}},\ \bibinfo {pages} {113D01} (\bibinfo {year}
  {2019})}\BibitemShut {NoStop}%
\bibitem [{\citenamefont {Bender}\ \emph {et~al.}(2003)\citenamefont {Bender},
  \citenamefont {Heenen},\ and\ \citenamefont
  {Reinhard}}]{Bender2003_RMP75-121}%
  \BibitemOpen
  \bibfield  {author} {\bibinfo {author} {\bibfnamefont {M.}~\bibnamefont
  {Bender}}, \bibinfo {author} {\bibfnamefont {P.-H.}\ \bibnamefont {Heenen}},\
  and\ \bibinfo {author} {\bibfnamefont {P.-G.}\ \bibnamefont {Reinhard}},\
  }\href {https://doi.org/10.1103/RevModPhys.75.121} {\bibfield  {journal}
  {\bibinfo  {journal} {Rev. Mod. Phys.}\ }\textbf {\bibinfo {volume} {75}},\
  \bibinfo {pages} {121} (\bibinfo {year} {2003})}\BibitemShut {NoStop}%
\bibitem [{\citenamefont {Nik$\check{\text{s}}$i$\acute{\text{c}}$}\ \emph
  {et~al.}(2011)\citenamefont {Nik$\check{\text{s}}$i$\acute{\text{c}}$},
  \citenamefont {Vretenar},\ and\ \citenamefont
  {Ring}}]{Niksic2011_PPNP66-519}%
  \BibitemOpen
  \bibfield  {author} {\bibinfo {author} {\bibfnamefont {T.}~\bibnamefont
  {Nik$\check{\text{s}}$i$\acute{\text{c}}$}}, \bibinfo {author} {\bibfnamefont
  {D.}~\bibnamefont {Vretenar}},\ and\ \bibinfo {author} {\bibfnamefont
  {P.}~\bibnamefont {Ring}},\ }\href
  {https://doi.org/10.1016/j.ppnp.2011.01.055} {\bibfield  {journal} {\bibinfo
  {journal} {Prog. Part. Nucl. Phys.}\ }\textbf {\bibinfo {volume} {66}},\
  \bibinfo {pages} {519} (\bibinfo {year} {2011})}\BibitemShut {NoStop}%
\bibitem [{\citenamefont {Egido}(2016)}]{Egido2016_PS91-073003}%
  \BibitemOpen
  \bibfield  {author} {\bibinfo {author} {\bibfnamefont {J.~L.}\ \bibnamefont
  {Egido}},\ }\href {https://doi.org/10.1088/0031-8949/91/7/073003} {\bibfield
  {journal} {\bibinfo  {journal} {Phys. Scr.}\ }\textbf {\bibinfo {volume}
  {91}},\ \bibinfo {pages} {073003} (\bibinfo {year} {2016})}\BibitemShut
  {NoStop}%
\bibitem [{\citenamefont {Robledo}\ \emph {et~al.}(2018)\citenamefont
  {Robledo}, \citenamefont {Rodr{\'{\i}}guez},\ and\ \citenamefont
  {Rodr{\'{\i}}guez-Guzm{\'{a}}n}}]{Robledo2018_JPG46-013001}%
  \BibitemOpen
  \bibfield  {author} {\bibinfo {author} {\bibfnamefont {L.~M.}\ \bibnamefont
  {Robledo}}, \bibinfo {author} {\bibfnamefont {T.~R.}\ \bibnamefont
  {Rodr{\'{\i}}guez}},\ and\ \bibinfo {author} {\bibfnamefont {R.~R.}\
  \bibnamefont {Rodr{\'{\i}}guez-Guzm{\'{a}}n}},\ }\href
  {https://doi.org/10.1088/1361-6471/aadebd} {\bibfield  {journal} {\bibinfo
  {journal} {J. Phys. G: Nucl. Part. Phys.}\ }\textbf {\bibinfo {volume}
  {46}},\ \bibinfo {pages} {013001} (\bibinfo {year} {2018})}\BibitemShut
  {NoStop}%
\bibitem [{\citenamefont {Sheikh}\ \emph {et~al.}(2021)\citenamefont {Sheikh},
  \citenamefont {Dobaczewski}, \citenamefont {Ring}, \citenamefont {Robledo},\
  and\ \citenamefont {Yannouleas}}]{Sheikh2021_JPG48-123001}%
  \BibitemOpen
  \bibfield  {author} {\bibinfo {author} {\bibfnamefont {J.~A.}\ \bibnamefont
  {Sheikh}}, \bibinfo {author} {\bibfnamefont {J.~J.}\ \bibnamefont
  {Dobaczewski}}, \bibinfo {author} {\bibfnamefont {P.}~\bibnamefont {Ring}},
  \bibinfo {author} {\bibfnamefont {L.~M.}\ \bibnamefont {Robledo}},\ and\
  \bibinfo {author} {\bibfnamefont {C.}~\bibnamefont {Yannouleas}},\ }\href
  {https://doi.org/10.1088/1361-6471/ac288a} {\bibfield  {journal} {\bibinfo
  {journal} {J. Phys. G: Nucl. Part. Phys.}\ }\textbf {\bibinfo {volume}
  {48}},\ \bibinfo {pages} {123001} (\bibinfo {year} {2021})}\BibitemShut
  {NoStop}%
\bibitem [{\citenamefont {Sun}\ and\ \citenamefont
  {Zhou}(2021)}]{Sun2021_SciBulletin66-2072}%
  \BibitemOpen
  \bibfield  {author} {\bibinfo {author} {\bibfnamefont {X.-X.}\ \bibnamefont
  {Sun}}\ and\ \bibinfo {author} {\bibfnamefont {S.-G.}\ \bibnamefont {Zhou}},\
  }\href {https://doi.org/10.1016/j.scib.2021.07.005} {\bibfield  {journal}
  {\bibinfo  {journal} {Sci. Bulletin}\ }\textbf {\bibinfo {volume} {66}},\
  \bibinfo {pages} {2072} (\bibinfo {year} {2021})}\BibitemShut {NoStop}%
\bibitem [{\citenamefont {Yao}\ \emph {et~al.}(2010)\citenamefont {Yao},
  \citenamefont {Meng}, \citenamefont {Ring},\ and\ \citenamefont
  {Vretenar}}]{Yao2010_PRC81-044311}%
  \BibitemOpen
  \bibfield  {author} {\bibinfo {author} {\bibfnamefont {J.~M.}\ \bibnamefont
  {Yao}}, \bibinfo {author} {\bibfnamefont {J.}~\bibnamefont {Meng}}, \bibinfo
  {author} {\bibfnamefont {P.}~\bibnamefont {Ring}},\ and\ \bibinfo {author}
  {\bibfnamefont {D.}~\bibnamefont {Vretenar}},\ }\href
  {https://doi.org/10.1103/PhysRevC.81.044311} {\bibfield  {journal} {\bibinfo
  {journal} {Phys. Rev. C}\ }\textbf {\bibinfo {volume} {81}},\ \bibinfo
  {pages} {044311} (\bibinfo {year} {2010})}\BibitemShut {NoStop}%
\bibitem [{\citenamefont {Yao}\ \emph {et~al.}(2009)\citenamefont {Yao},
  \citenamefont {Meng}, \citenamefont {Ring},\ and\ \citenamefont {{D. Pena
  Arteaga}}}]{Yao2009_PRC79-044312}%
  \BibitemOpen
  \bibfield  {author} {\bibinfo {author} {\bibfnamefont {J.~M.}\ \bibnamefont
  {Yao}}, \bibinfo {author} {\bibfnamefont {J.}~\bibnamefont {Meng}}, \bibinfo
  {author} {\bibfnamefont {P.}~\bibnamefont {Ring}},\ and\ \bibinfo {author}
  {\bibnamefont {{D. Pena Arteaga}}},\ }\href
  {https://doi.org/10.1103/PhysRevC.79.044312} {\bibfield  {journal} {\bibinfo
  {journal} {Phys. Rev. C}\ }\textbf {\bibinfo {volume} {79}},\ \bibinfo
  {pages} {044312} (\bibinfo {year} {2009})}\BibitemShut {NoStop}%
\bibitem [{\citenamefont {Rong}\ \emph {et~al.}(2023)\citenamefont {Rong},
  \citenamefont {Wu}, \citenamefont {Lu},\ and\ \citenamefont
  {Yao}}]{Rong2023_PLB840-137896}%
  \BibitemOpen
  \bibfield  {author} {\bibinfo {author} {\bibfnamefont {Y.-T.}\ \bibnamefont
  {Rong}}, \bibinfo {author} {\bibfnamefont {X.-Y.}\ \bibnamefont {Wu}},
  \bibinfo {author} {\bibfnamefont {B.-N.}\ \bibnamefont {Lu}},\ and\ \bibinfo
  {author} {\bibfnamefont {J.-M.}\ \bibnamefont {Yao}},\ }\href
  {https://doi.org/https://doi.org/10.1016/j.physletb.2023.137896} {\bibfield
  {journal} {\bibinfo  {journal} {Phys. Lett. B}\ }\textbf {\bibinfo {volume}
  {840}},\ \bibinfo {pages} {137896} (\bibinfo {year} {2023})}\BibitemShut
  {NoStop}%
\bibitem [{\citenamefont {Sharma}\ \emph {et~al.}(1993)\citenamefont {Sharma},
  \citenamefont {Nagarajan},\ and\ \citenamefont
  {Ring}}]{Sharma1993_PLB312-377}%
  \BibitemOpen
  \bibfield  {author} {\bibinfo {author} {\bibfnamefont {M.~M.}\ \bibnamefont
  {Sharma}}, \bibinfo {author} {\bibfnamefont {M.~A.}\ \bibnamefont
  {Nagarajan}},\ and\ \bibinfo {author} {\bibfnamefont {P.}~\bibnamefont
  {Ring}},\ }\href {https://doi.org/10.1016/0370-2693(93)90970-S} {\bibfield
  {journal} {\bibinfo  {journal} {Phys. Lett. B}\ }\textbf {\bibinfo {volume}
  {312}},\ \bibinfo {pages} {377} (\bibinfo {year} {1993})}\BibitemShut
  {NoStop}%
\bibitem [{\citenamefont {Agbemava}\ \emph {et~al.}(2019)\citenamefont
  {Agbemava}, \citenamefont {Afanasjev},\ and\ \citenamefont
  {Taninah}}]{Agbemava2019_PRC99-014318}%
  \BibitemOpen
  \bibfield  {author} {\bibinfo {author} {\bibfnamefont {S.~E.}\ \bibnamefont
  {Agbemava}}, \bibinfo {author} {\bibfnamefont {A.~V.}\ \bibnamefont
  {Afanasjev}},\ and\ \bibinfo {author} {\bibfnamefont {A.}~\bibnamefont
  {Taninah}},\ }\href {https://doi.org/10.1103/PhysRevC.99.014318} {\bibfield
  {journal} {\bibinfo  {journal} {Phys. Rev. C}\ }\textbf {\bibinfo {volume}
  {99}},\ \bibinfo {pages} {014318} (\bibinfo {year} {2019})}\BibitemShut
  {NoStop}%
\bibitem [{\citenamefont {Lalazissis}\ \emph {et~al.}(2005)\citenamefont
  {Lalazissis}, \citenamefont {Nik\ifmmode \check{s}\else
  \v{s}\fi{}i\ifmmode~\acute{c}\else \'{c}\fi{}}, \citenamefont {Vretenar},\
  and\ \citenamefont {Ring}}]{Lalazissis2005_PRC71-024312}%
  \BibitemOpen
  \bibfield  {author} {\bibinfo {author} {\bibfnamefont {G.~A.}\ \bibnamefont
  {Lalazissis}}, \bibinfo {author} {\bibfnamefont {T.}~\bibnamefont
  {Nik\ifmmode \check{s}\else \v{s}\fi{}i\ifmmode~\acute{c}\else \'{c}\fi{}}},
  \bibinfo {author} {\bibfnamefont {D.}~\bibnamefont {Vretenar}},\ and\
  \bibinfo {author} {\bibfnamefont {P.}~\bibnamefont {Ring}},\ }\href
  {https://doi.org/10.1103/PhysRevC.71.024312} {\bibfield  {journal} {\bibinfo
  {journal} {Phys. Rev. C}\ }\textbf {\bibinfo {volume} {71}},\ \bibinfo
  {pages} {024312} (\bibinfo {year} {2005})}\BibitemShut {NoStop}%
\bibitem [{\citenamefont {Wei}\ \emph {et~al.}(2020)\citenamefont {Wei},
  \citenamefont {Zhao}, \citenamefont {Wang}, \citenamefont {Geng},
  \citenamefont {Sun}, \citenamefont {Niu},\ and\ \citenamefont
  {Long}}]{Wei2020_ChinPhysC44-074107}%
  \BibitemOpen
  \bibfield  {author} {\bibinfo {author} {\bibfnamefont {B.}~\bibnamefont
  {Wei}}, \bibinfo {author} {\bibfnamefont {Q.}~\bibnamefont {Zhao}}, \bibinfo
  {author} {\bibfnamefont {Z.-H.}\ \bibnamefont {Wang}}, \bibinfo {author}
  {\bibfnamefont {J.}~\bibnamefont {Geng}}, \bibinfo {author} {\bibfnamefont
  {B.-Y.}\ \bibnamefont {Sun}}, \bibinfo {author} {\bibfnamefont {Y.-F.}\
  \bibnamefont {Niu}},\ and\ \bibinfo {author} {\bibfnamefont {W.-H.}\
  \bibnamefont {Long}},\ }\href {https://doi.org/10.1088/1674-1137/44/7/074107}
  {\bibfield  {journal} {\bibinfo  {journal} {Chin. Phys. C}\ }\textbf
  {\bibinfo {volume} {44}},\ \bibinfo {pages} {074107} (\bibinfo {year}
  {2020})}\BibitemShut {NoStop}%
\bibitem [{\citenamefont {Nik\ifmmode \check{s}\else
  \v{s}\fi{}i\ifmmode~\acute{c}\else \'{c}\fi{}}\ \emph
  {et~al.}(2008)\citenamefont {Nik\ifmmode \check{s}\else
  \v{s}\fi{}i\ifmmode~\acute{c}\else \'{c}\fi{}}, \citenamefont {Vretenar},\
  and\ \citenamefont {Ring}}]{Niksic2008_PRC78-034318}%
  \BibitemOpen
  \bibfield  {author} {\bibinfo {author} {\bibfnamefont {T.}~\bibnamefont
  {Nik\ifmmode \check{s}\else \v{s}\fi{}i\ifmmode~\acute{c}\else \'{c}\fi{}}},
  \bibinfo {author} {\bibfnamefont {D.}~\bibnamefont {Vretenar}},\ and\
  \bibinfo {author} {\bibfnamefont {P.}~\bibnamefont {Ring}},\ }\href
  {https://doi.org/10.1103/PhysRevC.78.034318} {\bibfield  {journal} {\bibinfo
  {journal} {Phys. Rev. C}\ }\textbf {\bibinfo {volume} {78}},\ \bibinfo
  {pages} {034318} (\bibinfo {year} {2008})}\BibitemShut {NoStop}%
\bibitem [{\citenamefont {Wang}\ \emph {et~al.}(2021)\citenamefont {Wang},
  \citenamefont {Huang}, \citenamefont {Kondev}, \citenamefont {Audi},\ and\
  \citenamefont {Naimi}}]{Wang2021_ChinPhysC45-030003}%
  \BibitemOpen
  \bibfield  {author} {\bibinfo {author} {\bibfnamefont {M.}~\bibnamefont
  {Wang}}, \bibinfo {author} {\bibfnamefont {W.}~\bibnamefont {Huang}},
  \bibinfo {author} {\bibfnamefont {F.}~\bibnamefont {Kondev}}, \bibinfo
  {author} {\bibfnamefont {G.}~\bibnamefont {Audi}},\ and\ \bibinfo {author}
  {\bibfnamefont {S.}~\bibnamefont {Naimi}},\ }\href
  {https://doi.org/10.1088/1674-1137/abddaf} {\bibfield  {journal} {\bibinfo
  {journal} {Chin. Phys. C}\ }\textbf {\bibinfo {volume} {45}},\ \bibinfo
  {pages} {030003} (\bibinfo {year} {2021})}\BibitemShut {NoStop}%
\bibitem [{\citenamefont {Pritychenko}\ \emph {et~al.}(2016)\citenamefont
  {Pritychenko}, \citenamefont {Birch}, \citenamefont {Singh},\ and\
  \citenamefont {Horoi}}]{Pritychenko2016_ADNDT107-1}%
  \BibitemOpen
  \bibfield  {author} {\bibinfo {author} {\bibfnamefont {B.}~\bibnamefont
  {Pritychenko}}, \bibinfo {author} {\bibfnamefont {M.}~\bibnamefont {Birch}},
  \bibinfo {author} {\bibfnamefont {B.}~\bibnamefont {Singh}},\ and\ \bibinfo
  {author} {\bibfnamefont {M.}~\bibnamefont {Horoi}},\ }\href
  {https://doi.org/https://doi.org/10.1016/j.adt.2015.10.001} {\bibfield
  {journal} {\bibinfo  {journal} {At. Data Nucl. Data Tables}\ }\textbf
  {\bibinfo {volume} {107}},\ \bibinfo {pages} {1 } (\bibinfo {year}
  {2016})}\BibitemShut {NoStop}%
\bibitem [{\citenamefont {Sick}(2008)}]{Sick2008_PRC77-041302(R)}%
  \BibitemOpen
  \bibfield  {author} {\bibinfo {author} {\bibfnamefont {I.}~\bibnamefont
  {Sick}},\ }\href {https://doi.org/10.1103/PhysRevC.77.041302} {\bibfield
  {journal} {\bibinfo  {journal} {Phys. Rev. C}\ }\textbf {\bibinfo {volume}
  {77}},\ \bibinfo {pages} {041302} (\bibinfo {year} {2008})}\BibitemShut
  {NoStop}%
\bibitem [{\citenamefont {Mueller}\ \emph {et~al.}(2007)\citenamefont
  {Mueller}, \citenamefont {Sulai}, \citenamefont {Villari}, \citenamefont
  {Alc\'antara-N\'u\~nez}, \citenamefont {Alves-Cond\'e}, \citenamefont
  {Bailey}, \citenamefont {Drake}, \citenamefont {Dubois}, \citenamefont
  {El\'eon}, \citenamefont {Gaubert}, \citenamefont {Holt}, \citenamefont
  {Janssens}, \citenamefont {Lecesne}, \citenamefont {Lu}, \citenamefont
  {O'Connor}, \citenamefont {Saint-Laurent}, \citenamefont {Thomas},\ and\
  \citenamefont {Wang}}]{Mueller2007_PRL99-252501}%
  \BibitemOpen
  \bibfield  {author} {\bibinfo {author} {\bibfnamefont {P.}~\bibnamefont
  {Mueller}}, \bibinfo {author} {\bibfnamefont {I.~A.}\ \bibnamefont {Sulai}},
  \bibinfo {author} {\bibfnamefont {A.~C.~C.}\ \bibnamefont {Villari}},
  \bibinfo {author} {\bibfnamefont {J.~A.}\ \bibnamefont
  {Alc\'antara-N\'u\~nez}}, \bibinfo {author} {\bibfnamefont {R.}~\bibnamefont
  {Alves-Cond\'e}}, \bibinfo {author} {\bibfnamefont {K.}~\bibnamefont
  {Bailey}}, \bibinfo {author} {\bibfnamefont {G.~W.~F.}\ \bibnamefont
  {Drake}}, \bibinfo {author} {\bibfnamefont {M.}~\bibnamefont {Dubois}},
  \bibinfo {author} {\bibfnamefont {C.}~\bibnamefont {El\'eon}}, \bibinfo
  {author} {\bibfnamefont {G.}~\bibnamefont {Gaubert}}, \bibinfo {author}
  {\bibfnamefont {R.~J.}\ \bibnamefont {Holt}}, \bibinfo {author}
  {\bibfnamefont {R.~V.~F.}\ \bibnamefont {Janssens}}, \bibinfo {author}
  {\bibfnamefont {N.}~\bibnamefont {Lecesne}}, \bibinfo {author} {\bibfnamefont
  {Z.-T.}\ \bibnamefont {Lu}}, \bibinfo {author} {\bibfnamefont {T.~P.}\
  \bibnamefont {O'Connor}}, \bibinfo {author} {\bibfnamefont {M.-G.}\
  \bibnamefont {Saint-Laurent}}, \bibinfo {author} {\bibfnamefont {J.-C.}\
  \bibnamefont {Thomas}},\ and\ \bibinfo {author} {\bibfnamefont {L.-B.}\
  \bibnamefont {Wang}},\ }\href {https://doi.org/10.1103/PhysRevLett.99.252501}
  {\bibfield  {journal} {\bibinfo  {journal} {Phys. Rev. Lett.}\ }\textbf
  {\bibinfo {volume} {99}},\ \bibinfo {pages} {252501} (\bibinfo {year}
  {2007})}\BibitemShut {NoStop}%
\bibitem [{\citenamefont {Alkhazov}\ \emph {et~al.}(2002)\citenamefont
  {Alkhazov}, \citenamefont {Dobrovolsky}, \citenamefont {Egelhof},
  \citenamefont {Geissel}, \citenamefont {Irnich}, \citenamefont {Khanzadeev},
  \citenamefont {Korolev}, \citenamefont {Lobodenko}, \citenamefont
  {M$\ddot{\rm u}$nzenberg}, \citenamefont {Mutterer}, \citenamefont
  {Neumaier}, \citenamefont {Schwab}, \citenamefont {Seliverstov},
  \citenamefont {Suzuki},\ and\ \citenamefont
  {Vorobyov}}]{Alkhazov2002_NPA712-269}%
  \BibitemOpen
  \bibfield  {author} {\bibinfo {author} {\bibfnamefont {G.}~\bibnamefont
  {Alkhazov}}, \bibinfo {author} {\bibfnamefont {A.}~\bibnamefont
  {Dobrovolsky}}, \bibinfo {author} {\bibfnamefont {P.}~\bibnamefont
  {Egelhof}}, \bibinfo {author} {\bibfnamefont {H.}~\bibnamefont {Geissel}},
  \bibinfo {author} {\bibfnamefont {H.}~\bibnamefont {Irnich}}, \bibinfo
  {author} {\bibfnamefont {A.}~\bibnamefont {Khanzadeev}}, \bibinfo {author}
  {\bibfnamefont {G.}~\bibnamefont {Korolev}}, \bibinfo {author} {\bibfnamefont
  {A.}~\bibnamefont {Lobodenko}}, \bibinfo {author} {\bibfnamefont
  {G.}~\bibnamefont {M$\ddot{\rm u}$nzenberg}}, \bibinfo {author}
  {\bibfnamefont {M.}~\bibnamefont {Mutterer}}, \bibinfo {author}
  {\bibfnamefont {S.}~\bibnamefont {Neumaier}}, \bibinfo {author}
  {\bibfnamefont {W.}~\bibnamefont {Schwab}}, \bibinfo {author} {\bibfnamefont
  {D.}~\bibnamefont {Seliverstov}}, \bibinfo {author} {\bibfnamefont
  {T.}~\bibnamefont {Suzuki}},\ and\ \bibinfo {author} {\bibfnamefont
  {A.}~\bibnamefont {Vorobyov}},\ }\href
  {https://doi.org/10.1016/S0375-9474(02)01273-3} {\bibfield  {journal}
  {\bibinfo  {journal} {Nucl. Phys. A}\ }\textbf {\bibinfo {volume} {712}},\
  \bibinfo {pages} {269} (\bibinfo {year} {2002})}\BibitemShut {NoStop}%
\bibitem [{\citenamefont {{M. Holl, R. Kanungo, Z.H. Sun \emph{et
  al.}}}(2021)}]{Holl2021_PLB822-136710}%
  \BibitemOpen
  \bibfield  {author} {\bibinfo {author} {\bibnamefont {{M. Holl, R. Kanungo,
  Z.H. Sun \emph{et al.}}}},\ }\href
  {https://doi.org/https://doi.org/10.1016/j.physletb.2021.136710} {\bibfield
  {journal} {\bibinfo  {journal} {Phys. Lett. B}\ }\textbf {\bibinfo {volume}
  {822}},\ \bibinfo {pages} {136710} (\bibinfo {year} {2021})}\BibitemShut
  {NoStop}%
\bibitem [{\citenamefont {Tanihata}\ \emph {et~al.}(1988)\citenamefont
  {Tanihata}, \citenamefont {Kobayashi}, \citenamefont {Yamakawa},
  \citenamefont {Shimoura}, \citenamefont {Ekuni}, \citenamefont {Sugimoto},
  \citenamefont {Takahashi}, \citenamefont {Shimoda},\ and\ \citenamefont
  {Sato}}]{Tanihata1988_PLB206-592}%
  \BibitemOpen
  \bibfield  {author} {\bibinfo {author} {\bibfnamefont {I.}~\bibnamefont
  {Tanihata}}, \bibinfo {author} {\bibfnamefont {T.}~\bibnamefont {Kobayashi}},
  \bibinfo {author} {\bibfnamefont {O.}~\bibnamefont {Yamakawa}}, \bibinfo
  {author} {\bibfnamefont {S.}~\bibnamefont {Shimoura}}, \bibinfo {author}
  {\bibfnamefont {K.}~\bibnamefont {Ekuni}}, \bibinfo {author} {\bibfnamefont
  {K.}~\bibnamefont {Sugimoto}}, \bibinfo {author} {\bibfnamefont
  {N.}~\bibnamefont {Takahashi}}, \bibinfo {author} {\bibfnamefont
  {T.}~\bibnamefont {Shimoda}},\ and\ \bibinfo {author} {\bibfnamefont
  {H.}~\bibnamefont {Sato}},\ }\href
  {https://doi.org/https://doi.org/10.1016/0370-2693(88)90702-2} {\bibfield
  {journal} {\bibinfo  {journal} {Phys. Lett. B}\ }\textbf {\bibinfo {volume}
  {206}},\ \bibinfo {pages} {592} (\bibinfo {year} {1988})}\BibitemShut
  {NoStop}%
\bibitem [{\citenamefont {Angeli}\ and\ \citenamefont
  {Marinova}(2013)}]{Angeli2013_ADNDT99-69}%
  \BibitemOpen
  \bibfield  {author} {\bibinfo {author} {\bibfnamefont {I.}~\bibnamefont
  {Angeli}}\ and\ \bibinfo {author} {\bibfnamefont {K.}~\bibnamefont
  {Marinova}},\ }\href {https://doi.org/10.1016/j.adt.2011.12.006} {\bibfield
  {journal} {\bibinfo  {journal} {At. Data Nucl. Data Tables}\ }\textbf
  {\bibinfo {volume} {99}},\ \bibinfo {pages} {69 } (\bibinfo {year}
  {2013})}\BibitemShut {NoStop}%
\bibitem [{\citenamefont {Kanungo}\ \emph {et~al.}(2016)\citenamefont
  {Kanungo}, \citenamefont {Horiuchi}, \citenamefont {Hagen}, \citenamefont
  {Jansen}, \citenamefont {Navratil}, \citenamefont {Ameil}, \citenamefont
  {Atkinson}, \citenamefont {Ayyad}, \citenamefont {Cortina-Gil}, \citenamefont
  {Dillmann}, \citenamefont {Estrad\'e}, \citenamefont {Evdokimov},
  \citenamefont {Farinon}, \citenamefont {Geissel}, \citenamefont {Guastalla},
  \citenamefont {Janik}, \citenamefont {Kimura}, \citenamefont {Kn\"obel},
  \citenamefont {Kurcewicz}, \citenamefont {Litvinov}, \citenamefont {Marta},
  \citenamefont {Mostazo}, \citenamefont {Mukha}, \citenamefont {Nociforo},
  \citenamefont {Ong}, \citenamefont {Pietri}, \citenamefont {Prochazka},
  \citenamefont {Scheidenberger}, \citenamefont {Sitar}, \citenamefont
  {Strmen}, \citenamefont {Suzuki}, \citenamefont {Takechi}, \citenamefont
  {Tanaka}, \citenamefont {Tanihata}, \citenamefont {Terashima}, \citenamefont
  {Vargas}, \citenamefont {Weick},\ and\ \citenamefont
  {Winfield}}]{Kanungo2016_PRL117-102501}%
  \BibitemOpen
  \bibfield  {author} {\bibinfo {author} {\bibfnamefont {R.}~\bibnamefont
  {Kanungo}}, \bibinfo {author} {\bibfnamefont {W.}~\bibnamefont {Horiuchi}},
  \bibinfo {author} {\bibfnamefont {G.}~\bibnamefont {Hagen}}, \bibinfo
  {author} {\bibfnamefont {G.~R.}\ \bibnamefont {Jansen}}, \bibinfo {author}
  {\bibfnamefont {P.}~\bibnamefont {Navratil}}, \bibinfo {author}
  {\bibfnamefont {F.}~\bibnamefont {Ameil}}, \bibinfo {author} {\bibfnamefont
  {J.}~\bibnamefont {Atkinson}}, \bibinfo {author} {\bibfnamefont
  {Y.}~\bibnamefont {Ayyad}}, \bibinfo {author} {\bibfnamefont
  {D.}~\bibnamefont {Cortina-Gil}}, \bibinfo {author} {\bibfnamefont
  {I.}~\bibnamefont {Dillmann}}, \bibinfo {author} {\bibfnamefont
  {A.}~\bibnamefont {Estrad\'e}}, \bibinfo {author} {\bibfnamefont
  {A.}~\bibnamefont {Evdokimov}}, \bibinfo {author} {\bibfnamefont
  {F.}~\bibnamefont {Farinon}}, \bibinfo {author} {\bibfnamefont
  {H.}~\bibnamefont {Geissel}}, \bibinfo {author} {\bibfnamefont
  {G.}~\bibnamefont {Guastalla}}, \bibinfo {author} {\bibfnamefont
  {R.}~\bibnamefont {Janik}}, \bibinfo {author} {\bibfnamefont
  {M.}~\bibnamefont {Kimura}}, \bibinfo {author} {\bibfnamefont
  {R.}~\bibnamefont {Kn\"obel}}, \bibinfo {author} {\bibfnamefont
  {J.}~\bibnamefont {Kurcewicz}}, \bibinfo {author} {\bibfnamefont {Y.~A.}\
  \bibnamefont {Litvinov}}, \bibinfo {author} {\bibfnamefont {M.}~\bibnamefont
  {Marta}}, \bibinfo {author} {\bibfnamefont {M.}~\bibnamefont {Mostazo}},
  \bibinfo {author} {\bibfnamefont {I.}~\bibnamefont {Mukha}}, \bibinfo
  {author} {\bibfnamefont {C.}~\bibnamefont {Nociforo}}, \bibinfo {author}
  {\bibfnamefont {H.~J.}\ \bibnamefont {Ong}}, \bibinfo {author} {\bibfnamefont
  {S.}~\bibnamefont {Pietri}}, \bibinfo {author} {\bibfnamefont
  {A.}~\bibnamefont {Prochazka}}, \bibinfo {author} {\bibfnamefont
  {C.}~\bibnamefont {Scheidenberger}}, \bibinfo {author} {\bibfnamefont
  {B.}~\bibnamefont {Sitar}}, \bibinfo {author} {\bibfnamefont
  {P.}~\bibnamefont {Strmen}}, \bibinfo {author} {\bibfnamefont
  {Y.}~\bibnamefont {Suzuki}}, \bibinfo {author} {\bibfnamefont
  {M.}~\bibnamefont {Takechi}}, \bibinfo {author} {\bibfnamefont
  {J.}~\bibnamefont {Tanaka}}, \bibinfo {author} {\bibfnamefont
  {I.}~\bibnamefont {Tanihata}}, \bibinfo {author} {\bibfnamefont
  {S.}~\bibnamefont {Terashima}}, \bibinfo {author} {\bibfnamefont
  {J.}~\bibnamefont {Vargas}}, \bibinfo {author} {\bibfnamefont
  {H.}~\bibnamefont {Weick}},\ and\ \bibinfo {author} {\bibfnamefont {J.~S.}\
  \bibnamefont {Winfield}},\ }\href
  {https://doi.org/10.1103/PhysRevLett.117.102501} {\bibfield  {journal}
  {\bibinfo  {journal} {Phys. Rev. Lett.}\ }\textbf {\bibinfo {volume} {117}},\
  \bibinfo {pages} {102501} (\bibinfo {year} {2016})}\BibitemShut {NoStop}%
\bibitem [{\citenamefont {Yasue}\ \emph {et~al.}(1983)\citenamefont {Yasue},
  \citenamefont {Tanabe}, \citenamefont {Soga}, \citenamefont {Kokame},
  \citenamefont {Shimokoshi}, \citenamefont {Kasagi}, \citenamefont {Toba},
  \citenamefont {Kadota}, \citenamefont {Ohsawa},\ and\ \citenamefont
  {Furuno}}]{Yasue1983_NPA394-29}%
  \BibitemOpen
  \bibfield  {author} {\bibinfo {author} {\bibfnamefont {M.}~\bibnamefont
  {Yasue}}, \bibinfo {author} {\bibfnamefont {T.}~\bibnamefont {Tanabe}},
  \bibinfo {author} {\bibfnamefont {F.}~\bibnamefont {Soga}}, \bibinfo {author}
  {\bibfnamefont {J.}~\bibnamefont {Kokame}}, \bibinfo {author} {\bibfnamefont
  {F.}~\bibnamefont {Shimokoshi}}, \bibinfo {author} {\bibfnamefont
  {J.}~\bibnamefont {Kasagi}}, \bibinfo {author} {\bibfnamefont
  {Y.}~\bibnamefont {Toba}}, \bibinfo {author} {\bibfnamefont {Y.}~\bibnamefont
  {Kadota}}, \bibinfo {author} {\bibfnamefont {T.}~\bibnamefont {Ohsawa}},\
  and\ \bibinfo {author} {\bibfnamefont {K.}~\bibnamefont {Furuno}},\ }\href
  {https://doi.org/10.1016/0375-9474(83)90159-8} {\bibfield  {journal}
  {\bibinfo  {journal} {Nucl. Phys. A}\ }\textbf {\bibinfo {volume} {394}},\
  \bibinfo {pages} {29} (\bibinfo {year} {1983})}\BibitemShut {NoStop}%
\bibitem [{\citenamefont {Jiang}\ \emph {et~al.}(2020)\citenamefont {Jiang},
  \citenamefont {Lou}, \citenamefont {Ye}, \citenamefont {Liu}, \citenamefont
  {Tan}, \citenamefont {Liu}, \citenamefont {Yang}, \citenamefont {Tao},
  \citenamefont {Ma}, \citenamefont {Li}, \citenamefont {Li}, \citenamefont
  {Yang}, \citenamefont {Xu}, \citenamefont {Yu}, \citenamefont {Han},
  \citenamefont {Bai}, \citenamefont {Huang}, \citenamefont {Li}, \citenamefont
  {Wu}, \citenamefont {Zang}, \citenamefont {Feng}, \citenamefont {Chen},
  \citenamefont {Chen}, \citenamefont {Yuan}, \citenamefont {Li}, \citenamefont
  {Hu}, \citenamefont {Xu}, \citenamefont {Wang}, \citenamefont {Yang},
  \citenamefont {Ma}, \citenamefont {Hu}, \citenamefont {Bai}, \citenamefont
  {Gao}, \citenamefont {Duan}, \citenamefont {Hu}, \citenamefont {Tan},
  \citenamefont {Sun}, \citenamefont {Song}, \citenamefont {Ong}, \citenamefont
  {Tran}, \citenamefont {Pang},\ and\ \citenamefont
  {Yuan}}]{Jiang2020_PRC101-024601}%
  \BibitemOpen
  \bibfield  {author} {\bibinfo {author} {\bibfnamefont {Y.}~\bibnamefont
  {Jiang}}, \bibinfo {author} {\bibfnamefont {J.~L.}\ \bibnamefont {Lou}},
  \bibinfo {author} {\bibfnamefont {Y.~L.}\ \bibnamefont {Ye}}, \bibinfo
  {author} {\bibfnamefont {Y.}~\bibnamefont {Liu}}, \bibinfo {author}
  {\bibfnamefont {Z.~W.}\ \bibnamefont {Tan}}, \bibinfo {author} {\bibfnamefont
  {W.}~\bibnamefont {Liu}}, \bibinfo {author} {\bibfnamefont {B.}~\bibnamefont
  {Yang}}, \bibinfo {author} {\bibfnamefont {L.~C.}\ \bibnamefont {Tao}},
  \bibinfo {author} {\bibfnamefont {K.}~\bibnamefont {Ma}}, \bibinfo {author}
  {\bibfnamefont {Z.~H.}\ \bibnamefont {Li}}, \bibinfo {author} {\bibfnamefont
  {Q.~T.}\ \bibnamefont {Li}}, \bibinfo {author} {\bibfnamefont {X.~F.}\
  \bibnamefont {Yang}}, \bibinfo {author} {\bibfnamefont {J.~Y.}\ \bibnamefont
  {Xu}}, \bibinfo {author} {\bibfnamefont {H.~Z.}\ \bibnamefont {Yu}}, \bibinfo
  {author} {\bibfnamefont {J.~X.}\ \bibnamefont {Han}}, \bibinfo {author}
  {\bibfnamefont {S.~W.}\ \bibnamefont {Bai}}, \bibinfo {author} {\bibfnamefont
  {S.~W.}\ \bibnamefont {Huang}}, \bibinfo {author} {\bibfnamefont
  {G.}~\bibnamefont {Li}}, \bibinfo {author} {\bibfnamefont {H.~Y.}\
  \bibnamefont {Wu}}, \bibinfo {author} {\bibfnamefont {H.~L.}\ \bibnamefont
  {Zang}}, \bibinfo {author} {\bibfnamefont {J.}~\bibnamefont {Feng}}, \bibinfo
  {author} {\bibfnamefont {Z.~Q.}\ \bibnamefont {Chen}}, \bibinfo {author}
  {\bibfnamefont {Y.~D.}\ \bibnamefont {Chen}}, \bibinfo {author}
  {\bibfnamefont {Q.}~\bibnamefont {Yuan}}, \bibinfo {author} {\bibfnamefont
  {J.~G.}\ \bibnamefont {Li}}, \bibinfo {author} {\bibfnamefont {B.~S.}\
  \bibnamefont {Hu}}, \bibinfo {author} {\bibfnamefont {F.~R.}\ \bibnamefont
  {Xu}}, \bibinfo {author} {\bibfnamefont {J.~S.}\ \bibnamefont {Wang}},
  \bibinfo {author} {\bibfnamefont {Y.~Y.}\ \bibnamefont {Yang}}, \bibinfo
  {author} {\bibfnamefont {P.}~\bibnamefont {Ma}}, \bibinfo {author}
  {\bibfnamefont {Q.}~\bibnamefont {Hu}}, \bibinfo {author} {\bibfnamefont
  {Z.}~\bibnamefont {Bai}}, \bibinfo {author} {\bibfnamefont {Z.~H.}\
  \bibnamefont {Gao}}, \bibinfo {author} {\bibfnamefont {F.~F.}\ \bibnamefont
  {Duan}}, \bibinfo {author} {\bibfnamefont {L.~Y.}\ \bibnamefont {Hu}},
  \bibinfo {author} {\bibfnamefont {J.~H.}\ \bibnamefont {Tan}}, \bibinfo
  {author} {\bibfnamefont {S.~Q.}\ \bibnamefont {Sun}}, \bibinfo {author}
  {\bibfnamefont {Y.~S.}\ \bibnamefont {Song}}, \bibinfo {author}
  {\bibfnamefont {H.~J.}\ \bibnamefont {Ong}}, \bibinfo {author} {\bibfnamefont
  {D.~T.}\ \bibnamefont {Tran}}, \bibinfo {author} {\bibfnamefont {D.~Y.}\
  \bibnamefont {Pang}},\ and\ \bibinfo {author} {\bibfnamefont {C.~X.}\
  \bibnamefont {Yuan}} (\bibinfo {collaboration} {RIBLL Collaboration}),\
  }\href {https://doi.org/10.1103/PhysRevC.101.024601} {\bibfield  {journal}
  {\bibinfo  {journal} {Phys. Rev. C}\ }\textbf {\bibinfo {volume} {101}},\
  \bibinfo {pages} {024601} (\bibinfo {year} {2020})}\BibitemShut {NoStop}%
\bibitem [{\citenamefont {{Y. Togano, T. Nakamura, Y. Kondo \emph{et
  al.}}}(2016)}]{Togano2016_PLB761-412}%
  \BibitemOpen
  \bibfield  {author} {\bibinfo {author} {\bibnamefont {{Y. Togano, T.
  Nakamura, Y. Kondo \emph{et al.}}}},\ }\href
  {https://doi.org/10.1016/j.physletb.2016.08.062} {\bibfield  {journal}
  {\bibinfo  {journal} {Phys. Lett. B}\ }\textbf {\bibinfo {volume} {761}},\
  \bibinfo {pages} {412} (\bibinfo {year} {2016})}\BibitemShut {NoStop}%
\bibitem [{\citenamefont {{S. Terashima, I. Tanihata, R. Kanungo \emph{et
  al.}}}(2014)}]{Terashima2014_PTEP2014-101D02}%
  \BibitemOpen
  \bibfield  {author} {\bibinfo {author} {\bibnamefont {{S. Terashima, I.
  Tanihata, R. Kanungo \emph{et al.}}}},\ }\href
  {https://doi.org/10.1093/ptep/ptu134} {\bibfield  {journal} {\bibinfo
  {journal} {Prog. Theor. Exp. Phys.}\ }\textbf {\bibinfo {volume} {2014}},\
  \bibinfo {pages} {101D02} (\bibinfo {year} {2014})}\BibitemShut {NoStop}%
\bibitem [{\citenamefont {Tran}\ \emph {et~al.}(2016)\citenamefont {Tran},
  \citenamefont {Ong}, \citenamefont {Nguyen}, \citenamefont {Tanihata},
  \citenamefont {Aoi}, \citenamefont {Ayyad}, \citenamefont {Chan},
  \citenamefont {Fukuda}, \citenamefont {Hashimoto}, \citenamefont {Hoang},
  \citenamefont {Ideguchi}, \citenamefont {Inoue}, \citenamefont {Kawabata},
  \citenamefont {Khiem}, \citenamefont {Lin}, \citenamefont {Matsuta},
  \citenamefont {Mihara}, \citenamefont {Momota}, \citenamefont {Nagae},
  \citenamefont {Nguyen}, \citenamefont {Nishimura}, \citenamefont {Ozawa},
  \citenamefont {Ren}, \citenamefont {Sakaguchi}, \citenamefont {Tanaka},
  \citenamefont {Takechi}, \citenamefont {Terashima}, \citenamefont {Wada},\
  and\ \citenamefont {Yamamoto}}]{Tran2016_PRC94-064604}%
  \BibitemOpen
  \bibfield  {author} {\bibinfo {author} {\bibfnamefont {D.~T.}\ \bibnamefont
  {Tran}}, \bibinfo {author} {\bibfnamefont {H.~J.}\ \bibnamefont {Ong}},
  \bibinfo {author} {\bibfnamefont {T.~T.}\ \bibnamefont {Nguyen}}, \bibinfo
  {author} {\bibfnamefont {I.}~\bibnamefont {Tanihata}}, \bibinfo {author}
  {\bibfnamefont {N.}~\bibnamefont {Aoi}}, \bibinfo {author} {\bibfnamefont
  {Y.}~\bibnamefont {Ayyad}}, \bibinfo {author} {\bibfnamefont {P.~Y.}\
  \bibnamefont {Chan}}, \bibinfo {author} {\bibfnamefont {M.}~\bibnamefont
  {Fukuda}}, \bibinfo {author} {\bibfnamefont {T.}~\bibnamefont {Hashimoto}},
  \bibinfo {author} {\bibfnamefont {T.~H.}\ \bibnamefont {Hoang}}, \bibinfo
  {author} {\bibfnamefont {E.}~\bibnamefont {Ideguchi}}, \bibinfo {author}
  {\bibfnamefont {A.}~\bibnamefont {Inoue}}, \bibinfo {author} {\bibfnamefont
  {T.}~\bibnamefont {Kawabata}}, \bibinfo {author} {\bibfnamefont {L.~H.}\
  \bibnamefont {Khiem}}, \bibinfo {author} {\bibfnamefont {W.~P.}\ \bibnamefont
  {Lin}}, \bibinfo {author} {\bibfnamefont {K.}~\bibnamefont {Matsuta}},
  \bibinfo {author} {\bibfnamefont {M.}~\bibnamefont {Mihara}}, \bibinfo
  {author} {\bibfnamefont {S.}~\bibnamefont {Momota}}, \bibinfo {author}
  {\bibfnamefont {D.}~\bibnamefont {Nagae}}, \bibinfo {author} {\bibfnamefont
  {N.~D.}\ \bibnamefont {Nguyen}}, \bibinfo {author} {\bibfnamefont
  {D.}~\bibnamefont {Nishimura}}, \bibinfo {author} {\bibfnamefont
  {A.}~\bibnamefont {Ozawa}}, \bibinfo {author} {\bibfnamefont {P.~P.}\
  \bibnamefont {Ren}}, \bibinfo {author} {\bibfnamefont {H.}~\bibnamefont
  {Sakaguchi}}, \bibinfo {author} {\bibfnamefont {J.}~\bibnamefont {Tanaka}},
  \bibinfo {author} {\bibfnamefont {M.}~\bibnamefont {Takechi}}, \bibinfo
  {author} {\bibfnamefont {S.}~\bibnamefont {Terashima}}, \bibinfo {author}
  {\bibfnamefont {R.}~\bibnamefont {Wada}},\ and\ \bibinfo {author}
  {\bibfnamefont {T.}~\bibnamefont {Yamamoto}} (\bibinfo {collaboration}
  {RCNP-E372 Collaboration}),\ }\href
  {https://doi.org/10.1103/PhysRevC.94.064604} {\bibfield  {journal} {\bibinfo
  {journal} {Phys. Rev. C}\ }\textbf {\bibinfo {volume} {94}},\ \bibinfo
  {pages} {064604} (\bibinfo {year} {2016})}\BibitemShut {NoStop}%
\bibitem [{\citenamefont {Charlwood}\ \emph {et~al.}(2009)\citenamefont
  {Charlwood}, \citenamefont {Baczynska}, \citenamefont {Billowes},
  \citenamefont {Campbell}, \citenamefont {Cheal}, \citenamefont {Eronen},
  \citenamefont {Forest}, \citenamefont {Jokinen}, \citenamefont {Kessler},
  \citenamefont {Moore}, \citenamefont {Penttil$\ddot{\rm a}$}, \citenamefont
  {Powis}, \citenamefont {R$\ddot{\rm u}$ffer}, \citenamefont {Saastamoinen},
  \citenamefont {Tungate},\ and\ \citenamefont {$\ddot{\rm A}$yst$\ddot{\rm
  o}$}}]{Charlwood2009_PLB674-23}%
  \BibitemOpen
  \bibfield  {author} {\bibinfo {author} {\bibfnamefont {F.}~\bibnamefont
  {Charlwood}}, \bibinfo {author} {\bibfnamefont {K.}~\bibnamefont
  {Baczynska}}, \bibinfo {author} {\bibfnamefont {J.}~\bibnamefont {Billowes}},
  \bibinfo {author} {\bibfnamefont {P.}~\bibnamefont {Campbell}}, \bibinfo
  {author} {\bibfnamefont {B.}~\bibnamefont {Cheal}}, \bibinfo {author}
  {\bibfnamefont {T.}~\bibnamefont {Eronen}}, \bibinfo {author} {\bibfnamefont
  {D.}~\bibnamefont {Forest}}, \bibinfo {author} {\bibfnamefont
  {A.}~\bibnamefont {Jokinen}}, \bibinfo {author} {\bibfnamefont
  {T.}~\bibnamefont {Kessler}}, \bibinfo {author} {\bibfnamefont
  {I.}~\bibnamefont {Moore}}, \bibinfo {author} {\bibfnamefont
  {H.}~\bibnamefont {Penttil$\ddot{\rm a}$}}, \bibinfo {author} {\bibfnamefont
  {R.}~\bibnamefont {Powis}}, \bibinfo {author} {\bibfnamefont
  {M.}~\bibnamefont {R$\ddot{\rm u}$ffer}}, \bibinfo {author} {\bibfnamefont
  {A.}~\bibnamefont {Saastamoinen}}, \bibinfo {author} {\bibfnamefont
  {G.}~\bibnamefont {Tungate}},\ and\ \bibinfo {author} {\bibfnamefont
  {J.}~\bibnamefont {$\ddot{\rm A}$yst$\ddot{\rm o}$}},\ }\href
  {https://doi.org/https://doi.org/10.1016/j.physletb.2009.02.050} {\bibfield
  {journal} {\bibinfo  {journal} {Phys. Lett. B}\ }\textbf {\bibinfo {volume}
  {674}},\ \bibinfo {pages} {23} (\bibinfo {year} {2009})}\BibitemShut
  {NoStop}%
\bibitem [{\citenamefont {{S. Abrahamyan \emph{et
  al.}}}(2012)}]{Abrahamyan2012_PRL108-112502}%
  \BibitemOpen
  \bibfield  {author} {\bibinfo {author} {\bibnamefont {{S. Abrahamyan \emph{et
  al.}}}} (\bibinfo {collaboration} {PREX Collaboration}),\ }\href
  {https://doi.org/10.1103/PhysRevLett.108.112502} {\bibfield  {journal}
  {\bibinfo  {journal} {Phys. Rev. Lett.}\ }\textbf {\bibinfo {volume} {108}},\
  \bibinfo {pages} {112502} (\bibinfo {year} {2012})}\BibitemShut {NoStop}%
\bibitem [{\citenamefont {Weiss}\ \emph {et~al.}(2019)\citenamefont {Weiss},
  \citenamefont {Schmidt}, \citenamefont {Miller},\ and\ \citenamefont
  {Barnea}}]{Weiss2019_PLB790-484}%
  \BibitemOpen
  \bibfield  {author} {\bibinfo {author} {\bibfnamefont {R.}~\bibnamefont
  {Weiss}}, \bibinfo {author} {\bibfnamefont {A.}~\bibnamefont {Schmidt}},
  \bibinfo {author} {\bibfnamefont {G.~A.}\ \bibnamefont {Miller}},\ and\
  \bibinfo {author} {\bibfnamefont {N.}~\bibnamefont {Barnea}},\ }\href
  {https://doi.org/https://doi.org/10.1016/j.physletb.2019.01.053} {\bibfield
  {journal} {\bibinfo  {journal} {Phys. Lett. B}\ }\textbf {\bibinfo {volume}
  {790}},\ \bibinfo {pages} {484 } (\bibinfo {year} {2019})}\BibitemShut
  {NoStop}%
\bibitem [{\citenamefont {Reinhard}\ and\ \citenamefont
  {Nazarewicz}(2021)}]{Reinhard2021_PRC103-054310}%
  \BibitemOpen
  \bibfield  {author} {\bibinfo {author} {\bibfnamefont {P.-G.}\ \bibnamefont
  {Reinhard}}\ and\ \bibinfo {author} {\bibfnamefont {W.}~\bibnamefont
  {Nazarewicz}},\ }\href {https://doi.org/10.1103/PhysRevC.103.054310}
  {\bibfield  {journal} {\bibinfo  {journal} {Phys. Rev. C}\ }\textbf {\bibinfo
  {volume} {103}},\ \bibinfo {pages} {054310} (\bibinfo {year}
  {2021})}\BibitemShut {NoStop}%
\bibitem [{\citenamefont {Miller}\ \emph {et~al.}(2019)\citenamefont {Miller},
  \citenamefont {Minamisono}, \citenamefont {Klose}, \citenamefont {Garand},
  \citenamefont {Kujawa}, \citenamefont {Lantis}, \citenamefont {Liu},
  \citenamefont {Maa{\ss}}, \citenamefont {Mantica}, \citenamefont
  {Nazarewicz}, \citenamefont {N$\ddot{\rm o}$rtersh$\ddot{\rm a}$user},
  \citenamefont {Pineda}, \citenamefont {Reinhard}, \citenamefont {Rossi},
  \citenamefont {Sommer}, \citenamefont {Sumithrarachchi}, \citenamefont
  {Teigelh$\ddot{\rm o}$fer},\ and\ \citenamefont
  {Watkins}}]{Miller2019_NP15-432}%
  \BibitemOpen
  \bibfield  {author} {\bibinfo {author} {\bibfnamefont {A.~J.}\ \bibnamefont
  {Miller}}, \bibinfo {author} {\bibfnamefont {K.}~\bibnamefont {Minamisono}},
  \bibinfo {author} {\bibfnamefont {A.}~\bibnamefont {Klose}}, \bibinfo
  {author} {\bibfnamefont {D.}~\bibnamefont {Garand}}, \bibinfo {author}
  {\bibfnamefont {C.}~\bibnamefont {Kujawa}}, \bibinfo {author} {\bibfnamefont
  {J.~D.}\ \bibnamefont {Lantis}}, \bibinfo {author} {\bibfnamefont
  {Y.}~\bibnamefont {Liu}}, \bibinfo {author} {\bibfnamefont {B.}~\bibnamefont
  {Maa{\ss}}}, \bibinfo {author} {\bibfnamefont {P.~F.}\ \bibnamefont
  {Mantica}}, \bibinfo {author} {\bibfnamefont {W.}~\bibnamefont {Nazarewicz}},
  \bibinfo {author} {\bibfnamefont {W.}~\bibnamefont {N$\ddot{\rm
  o}$rtersh$\ddot{\rm a}$user}}, \bibinfo {author} {\bibfnamefont {S.~V.}\
  \bibnamefont {Pineda}}, \bibinfo {author} {\bibfnamefont {P.-G.}\
  \bibnamefont {Reinhard}}, \bibinfo {author} {\bibfnamefont {D.~M.}\
  \bibnamefont {Rossi}}, \bibinfo {author} {\bibfnamefont {F.}~\bibnamefont
  {Sommer}}, \bibinfo {author} {\bibfnamefont {C.}~\bibnamefont
  {Sumithrarachchi}}, \bibinfo {author} {\bibfnamefont {A.}~\bibnamefont
  {Teigelh$\ddot{\rm o}$fer}},\ and\ \bibinfo {author} {\bibfnamefont
  {J.}~\bibnamefont {Watkins}},\ }\href
  {https://doi.org/10.1038/s41567-019-0416-9} {\bibfield  {journal} {\bibinfo
  {journal} {Nature Phys.}\ }\textbf {\bibinfo {volume} {15}},\ \bibinfo
  {pages} {432} (\bibinfo {year} {2019})}\BibitemShut {NoStop}%
\bibitem [{\citenamefont {{D. Adhikari, H. Albataineh, D. Androic \emph{et
  al.}}}(2021)}]{Adhikari2021_PRL126-172502}%
  \BibitemOpen
  \bibfield  {author} {\bibinfo {author} {\bibnamefont {{D. Adhikari, H.
  Albataineh, D. Androic \emph{et al.}}}} (\bibinfo {collaboration} {PREX
  Collaboration}),\ }\href {https://doi.org/10.1103/PhysRevLett.126.172502}
  {\bibfield  {journal} {\bibinfo  {journal} {Phys. Rev. Lett.}\ }\textbf
  {\bibinfo {volume} {126}},\ \bibinfo {pages} {172502} (\bibinfo {year}
  {2021})}\BibitemShut {NoStop}%
\bibitem [{\citenamefont {Long}\ \emph {et~al.}(2006)\citenamefont {Long},
  \citenamefont {{Van Giai}},\ and\ \citenamefont
  {Meng}}]{Long2006_PLB640-150}%
  \BibitemOpen
  \bibfield  {author} {\bibinfo {author} {\bibfnamefont {W.-H.}\ \bibnamefont
  {Long}}, \bibinfo {author} {\bibfnamefont {N.}~\bibnamefont {{Van Giai}}},\
  and\ \bibinfo {author} {\bibfnamefont {J.}~\bibnamefont {Meng}},\ }\href
  {https://doi.org/10.1016/j.physletb.2006.07.064} {\bibfield  {journal}
  {\bibinfo  {journal} {Phys. Lett. B}\ }\textbf {\bibinfo {volume} {640}},\
  \bibinfo {pages} {150} (\bibinfo {year} {2006})}\BibitemShut {NoStop}%
\bibitem [{\citenamefont {Long}\ \emph {et~al.}(2007)\citenamefont {Long},
  \citenamefont {Sagawa}, \citenamefont {Giai},\ and\ \citenamefont
  {Meng}}]{Long2007_PRC76-034314}%
  \BibitemOpen
  \bibfield  {author} {\bibinfo {author} {\bibfnamefont {W.-H.}\ \bibnamefont
  {Long}}, \bibinfo {author} {\bibfnamefont {H.}~\bibnamefont {Sagawa}},
  \bibinfo {author} {\bibfnamefont {N.~V.}\ \bibnamefont {Giai}},\ and\
  \bibinfo {author} {\bibfnamefont {J.}~\bibnamefont {Meng}},\ }\href
  {https://doi.org/10.1103/PhysRevC.76.034314} {\bibfield  {journal} {\bibinfo
  {journal} {Phys. Rev. C}\ }\textbf {\bibinfo {volume} {76}},\ \bibinfo
  {pages} {034314} (\bibinfo {year} {2007})}\BibitemShut {NoStop}%
\bibitem [{\citenamefont {Shen}\ \emph {et~al.}(2017)\citenamefont {Shen},
  \citenamefont {Liang}, \citenamefont {Meng}, \citenamefont {Ring},\ and\
  \citenamefont {Zhang}}]{Shen2017_PRC96-014316}%
  \BibitemOpen
  \bibfield  {author} {\bibinfo {author} {\bibfnamefont {S.}~\bibnamefont
  {Shen}}, \bibinfo {author} {\bibfnamefont {H.}~\bibnamefont {Liang}},
  \bibinfo {author} {\bibfnamefont {J.}~\bibnamefont {Meng}}, \bibinfo {author}
  {\bibfnamefont {P.}~\bibnamefont {Ring}},\ and\ \bibinfo {author}
  {\bibfnamefont {S.}~\bibnamefont {Zhang}},\ }\href
  {https://doi.org/10.1103/PhysRevC.96.014316} {\bibfield  {journal} {\bibinfo
  {journal} {Phys. Rev. C}\ }\textbf {\bibinfo {volume} {96}},\ \bibinfo
  {pages} {014316} (\bibinfo {year} {2017})}\BibitemShut {NoStop}%
\bibitem [{\citenamefont {Binder}\ \emph {et~al.}(2016)\citenamefont {Binder},
  \citenamefont {Calci}, \citenamefont {Epelbaum}, \citenamefont {Furnstahl},
  \citenamefont {Golak}, \citenamefont {Hebeler}, \citenamefont {Kamada},
  \citenamefont {Krebs}, \citenamefont {Langhammer}, \citenamefont {Liebig},
  \citenamefont {Maris}, \citenamefont {{Ulf-G. Mei\ss{}ner}}, \citenamefont
  {Minossi}, \citenamefont {Nogga}, \citenamefont {Potter}, \citenamefont
  {Roth}, \citenamefont {Skibi\ifmmode~\acute{n}\else \'{n}\fi{}ski},
  \citenamefont {Topolnicki}, \citenamefont {Vary},\ and\ \citenamefont
  {Wita\l{}a}}]{Binder2016_PRC93-044002}%
  \BibitemOpen
  \bibfield  {author} {\bibinfo {author} {\bibfnamefont {S.}~\bibnamefont
  {Binder}}, \bibinfo {author} {\bibfnamefont {A.}~\bibnamefont {Calci}},
  \bibinfo {author} {\bibfnamefont {E.}~\bibnamefont {Epelbaum}}, \bibinfo
  {author} {\bibfnamefont {R.~J.}\ \bibnamefont {Furnstahl}}, \bibinfo {author}
  {\bibfnamefont {J.}~\bibnamefont {Golak}}, \bibinfo {author} {\bibfnamefont
  {K.}~\bibnamefont {Hebeler}}, \bibinfo {author} {\bibfnamefont
  {H.}~\bibnamefont {Kamada}}, \bibinfo {author} {\bibfnamefont
  {H.}~\bibnamefont {Krebs}}, \bibinfo {author} {\bibfnamefont
  {J.}~\bibnamefont {Langhammer}}, \bibinfo {author} {\bibfnamefont
  {S.}~\bibnamefont {Liebig}}, \bibinfo {author} {\bibfnamefont
  {P.}~\bibnamefont {Maris}}, \bibinfo {author} {\bibnamefont {{Ulf-G.
  Mei\ss{}ner}}}, \bibinfo {author} {\bibfnamefont {D.}~\bibnamefont
  {Minossi}}, \bibinfo {author} {\bibfnamefont {A.}~\bibnamefont {Nogga}},
  \bibinfo {author} {\bibfnamefont {H.}~\bibnamefont {Potter}}, \bibinfo
  {author} {\bibfnamefont {R.}~\bibnamefont {Roth}}, \bibinfo {author}
  {\bibfnamefont {R.}~\bibnamefont {Skibi\ifmmode~\acute{n}\else
  \'{n}\fi{}ski}}, \bibinfo {author} {\bibfnamefont {K.}~\bibnamefont
  {Topolnicki}}, \bibinfo {author} {\bibfnamefont {J.~P.}\ \bibnamefont
  {Vary}},\ and\ \bibinfo {author} {\bibfnamefont {H.}~\bibnamefont
  {Wita\l{}a}} (\bibinfo {collaboration} {LENPIC Collaboration}),\ }\href
  {https://doi.org/10.1103/PhysRevC.93.044002} {\bibfield  {journal} {\bibinfo
  {journal} {Phys. Rev. C}\ }\textbf {\bibinfo {volume} {93}},\ \bibinfo
  {pages} {044002} (\bibinfo {year} {2016})}\BibitemShut {NoStop}%
\bibitem [{\citenamefont {Navr$\acute{\rm
  a}$til}(2007)}]{Navratil2007_FS41-117}%
  \BibitemOpen
  \bibfield  {author} {\bibinfo {author} {\bibfnamefont {P.}~\bibnamefont
  {Navr$\acute{\rm a}$til}},\ }\href
  {https://doi.org/10.1007/s00601-007-0193-3} {\bibfield  {journal} {\bibinfo
  {journal} {Few-Body Systems}\ }\textbf {\bibinfo {volume} {41}},\ \bibinfo
  {pages} {117} (\bibinfo {year} {2007})}\BibitemShut {NoStop}%
\bibitem [{\citenamefont {Rodr\'{\i}guez-Guzm\'an}\ \emph
  {et~al.}(2000)\citenamefont {Rodr\'{\i}guez-Guzm\'an}, \citenamefont
  {Egido},\ and\ \citenamefont {Robledo}}]{Rodriguez-Guzman2000_PRC62-054319}%
  \BibitemOpen
  \bibfield  {author} {\bibinfo {author} {\bibfnamefont {R.~R.}\ \bibnamefont
  {Rodr\'{\i}guez-Guzm\'an}}, \bibinfo {author} {\bibfnamefont {J.~L.}\
  \bibnamefont {Egido}},\ and\ \bibinfo {author} {\bibfnamefont {L.~M.}\
  \bibnamefont {Robledo}},\ }\href {https://doi.org/10.1103/PhysRevC.62.054319}
  {\bibfield  {journal} {\bibinfo  {journal} {Phys. Rev. C}\ }\textbf {\bibinfo
  {volume} {62}},\ \bibinfo {pages} {054319} (\bibinfo {year}
  {2000})}\BibitemShut {NoStop}%
\bibitem [{\citenamefont {Nik\ifmmode \check{s}\else
  \v{s}\fi{}i\ifmmode~\acute{c}\else \'{c}\fi{}}\ \emph
  {et~al.}(2006)\citenamefont {Nik\ifmmode \check{s}\else
  \v{s}\fi{}i\ifmmode~\acute{c}\else \'{c}\fi{}}, \citenamefont {Vretenar},\
  and\ \citenamefont {Ring}}]{Niksic2006_PRC73-034308}%
  \BibitemOpen
  \bibfield  {author} {\bibinfo {author} {\bibfnamefont {T.}~\bibnamefont
  {Nik\ifmmode \check{s}\else \v{s}\fi{}i\ifmmode~\acute{c}\else \'{c}\fi{}}},
  \bibinfo {author} {\bibfnamefont {D.}~\bibnamefont {Vretenar}},\ and\
  \bibinfo {author} {\bibfnamefont {P.}~\bibnamefont {Ring}},\ }\href
  {https://doi.org/10.1103/PhysRevC.73.034308} {\bibfield  {journal} {\bibinfo
  {journal} {Phys. Rev. C}\ }\textbf {\bibinfo {volume} {73}},\ \bibinfo
  {pages} {034308} (\bibinfo {year} {2006})}\BibitemShut {NoStop}%
\bibitem [{\citenamefont {Sun}\ \emph {et~al.}(2020)\citenamefont {Sun},
  \citenamefont {Zhao},\ and\ \citenamefont {Zhou}}]{Sun2020_NPA1003-122011}%
  \BibitemOpen
  \bibfield  {author} {\bibinfo {author} {\bibfnamefont {X.-X.}\ \bibnamefont
  {Sun}}, \bibinfo {author} {\bibfnamefont {J.}~\bibnamefont {Zhao}},\ and\
  \bibinfo {author} {\bibfnamefont {S.-G.}\ \bibnamefont {Zhou}},\ }\href
  {https://doi.org/https://doi.org/10.1016/j.nuclphysa.2020.122011} {\bibfield
  {journal} {\bibinfo  {journal} {Nucl. Phys. A}\ }\textbf {\bibinfo {volume}
  {1003}},\ \bibinfo {pages} {122011} (\bibinfo {year} {2020})}\BibitemShut
  {NoStop}%
\bibitem [{\citenamefont {Bohr}\ and\ \citenamefont
  {Mottelson}(1969)}]{Bohr1969}%
  \BibitemOpen
  \bibfield  {author} {\bibinfo {author} {\bibfnamefont {A.}~\bibnamefont
  {Bohr}}\ and\ \bibinfo {author} {\bibfnamefont {B.~R.}\ \bibnamefont
  {Mottelson}},\ }\href@noop {} {\emph {\bibinfo {title} {Nuclear
  structure}}},\ edited by\ \bibinfo {editor} {\bibfnamefont {W.~A.}\
  \bibnamefont {Benjamin}},\ Vol.~\bibinfo {volume} {I}\ (\bibinfo  {publisher}
  {World Scientific Publishing Co. Pte. Ltd.},\ \bibinfo {year}
  {1969})\BibitemShut {NoStop}%
\bibitem [{\citenamefont {Itagaki}\ \emph {et~al.}(2000)\citenamefont
  {Itagaki}, \citenamefont {Okabe},\ and\ \citenamefont
  {Ikeda}}]{Itagaki2000_PRC62-034301}%
  \BibitemOpen
  \bibfield  {author} {\bibinfo {author} {\bibfnamefont {N.}~\bibnamefont
  {Itagaki}}, \bibinfo {author} {\bibfnamefont {S.}~\bibnamefont {Okabe}},\
  and\ \bibinfo {author} {\bibfnamefont {K.}~\bibnamefont {Ikeda}},\ }\href
  {https://doi.org/10.1103/PhysRevC.62.034301} {\bibfield  {journal} {\bibinfo
  {journal} {Phys. Rev. C}\ }\textbf {\bibinfo {volume} {62}},\ \bibinfo
  {pages} {034301} (\bibinfo {year} {2000})}\BibitemShut {NoStop}%
\bibitem [{\citenamefont {Kanada-En'yo}\ \emph {et~al.}(2012)\citenamefont
  {Kanada-En'yo}, \citenamefont {Kimura},\ and\ \citenamefont
  {Ono}}]{KanadaEnyo2012_PTEP2012-01A202}%
  \BibitemOpen
  \bibfield  {author} {\bibinfo {author} {\bibfnamefont {Y.}~\bibnamefont
  {Kanada-En'yo}}, \bibinfo {author} {\bibfnamefont {M.}~\bibnamefont
  {Kimura}},\ and\ \bibinfo {author} {\bibfnamefont {A.}~\bibnamefont {Ono}},\
  }\href {https://doi.org/10.1093/ptep/pts001} {\bibfield  {journal} {\bibinfo
  {journal} {Prog. Theor. Exp. Phys.}\ }\textbf {\bibinfo {volume} {2012}},\
  \bibinfo {pages} {01A202} (\bibinfo {year} {2012})}\BibitemShut {NoStop}%
\end{thebibliography}%
\end{document}